\documentclass[letterpaper,prc,aps,showpacs,showkeys,superscriptaddress]{revtex4}
\usepackage{ae,aecompl}
\usepackage[T1]{fontenc}
\usepackage[latin9]{inputenc}
\usepackage{amsmath}
\usepackage{amssymb}
\usepackage{graphicx}
\usepackage{float}
\usepackage{subfigure}
\usepackage{multirow}
\usepackage{mathtools}
\usepackage{tabularx}

\makeatletter

\pdfpageheight\paperheight
\pdfpagewidth\paperwidth

\begin{document}
\title{Double differential fragmentation cross sections measurements of 95 MeV/u $^{12}$C on thin targets for hadrontherapy}

\author{J. Dudouet}
\affiliation{LPC Caen, ENSICAEN, Universit\'e de Caen, CNRS/IN2P3, Caen, France}
\author{D. Juliani}  
\affiliation{Institut Pluridisciplinaire Hubert Curien Strasbourg, France}
\author{J.C. Ang\'elique}
\affiliation{LPC Caen, ENSICAEN, Universit\'e de Caen, CNRS/IN2P3, Caen, France}
\author{B. Braunn}
\affiliation{CEA/Saclay, DSM/Irfu/SPhN, Gif-sur-Yvette, France}
\author{J. Colin}
\author{D. Cussol}
\affiliation{LPC Caen, ENSICAEN, Universit\'e de Caen, CNRS/IN2P3, Caen, France}
\author{Ch. Finck}
\affiliation{Institut Pluridisciplinaire Hubert Curien Strasbourg, France}
\author{J.M. Fontbonne}
\author{H. Gu\'erin}
\affiliation{LPC Caen, ENSICAEN, Universit\'e de Caen, CNRS/IN2P3, Caen, France}
\author{P. Henriquet}
\author{J. Krimmer}
\affiliation{IPN Lyon, Universit\'e de Lyon, Universit\'e Lyon 1, CNRS/IN2P3, Villeurbanne, France}
\author{M. Labalme}
\affiliation{LPC Caen, ENSICAEN, Universit\'e de Caen, CNRS/IN2P3, Caen, France}
\author{M. Rousseau} 
\affiliation{Institut Pluridisciplinaire Hubert Curien Strasbourg, France}
\author{M.G. Saint-Laurent}
\affiliation{GANIL, Caen, France}
\author{S. Salvador}
\affiliation{LPC Caen, ENSICAEN, Universit\'e de Caen, CNRS/IN2P3, Caen, France}

\date{\today}

\begin{abstract}

During therapeutic treatment with heavy ions like carbon, the beam undergoes nuclear fragmentation and secondary light charged particles, in particular protons and alpha particles, are produced. To estimate the dose deposited into the tumors and the surrounding healthy tissues, an accurate prediction on the fluences of these secondary fragments is necessary. Nowadays, a very limited set of double differential carbon fragmentation cross sections are being measured in the energy range used in hadrontherapy (40 to 400~MeV/u). Therefore, new measurements are performed to determine the double differential cross section of carbon on different thin targets. This work describes the experimental results of an experiment performed on May 2011 at GANIL. The double differential cross sections and the angular distributions of secondary fragments produced in the $^{12}$C fragmentation at 95~MeV/u on thin targets (C, CH$_2$, Al, Al$_2$O$_3$, Ti and PMMA) have been measured. The experimental setup will be precisely described, the systematic error study will be explained and all the experimental data will be presented.

\end{abstract}



\maketitle

\section{Introduction}

The use of carbon ion beams in hadrontherapy to treat cancerous tumors is motivated by the highly localized dose distribution. The high carbon mass (compared to proton) leads to a smaller angular scattering and a higher dose deposition at the end of the radiation range (i.e. at the Bragg peak). Moreover, the biological efficiency, which is strongly correlated to the linear energy transfer (LET), is higher for carbon ions in the Bragg peak region. Carbon ions allow thus to better target the tumor while preserving the surrounding healthy tissues. However, to keep the benefits of carbon ions in radiotherapy, a very high accuracy on the dose deposition is required ($\pm$3\% and $\pm$1 mm). The physical dose deposition is affected by the fragmentation of the ions along their penetration path in the human tissues~\cite{Schardt96}. As a consequence, the number of incident ions reaching the tumor is reduced for example up to 70\%  for 400~MeV/u $^{12}$C in water. The carbon beam fragmentation in the human body leads to a mixed radiation field composed of lighter fragments of higher ranges and angular distributions with respect to the primary ions. These lighter fragments have different relative biological effectivenesses (RBE) which contribute to the deposited dose all along the carbon path. These effects, due to the carbon fragmentation, result in a new spatial dose distribution, particularly on healthy tissues. This must be taken into account for the evaluation of the biological dose~\cite{Scholz00} by accurately evaluating the fragmentation processes.

Simulation codes are used to compute the transportation of ions in the matter but the constraints on nuclear models and fragmentation cross sections in the energy range used in hadrontherapy (up to 400 MeV/u) are not yet sufficient to reproduce the fragmentation processes with the required accuracy for clinical treatments~\cite{Bohlen10,Braunn12,Catane12}. Nuclear cross sections are critical inputs for these simulation frameworks. In particular there is a lack of experimental cross section data available for light ions on light targets in the energy range from 30 to 400 MeV/u. These experimental data are necessary to benchmark the Monte Carlo codes for their use in hadrontherapy.

To improve the knowledge on the $^{12}$C fragmentation processes, experiments have been performed in Japan and in Europe for more than 15 years. Measurements of light charged fragment production in water and PMMA have been performed by the Japanese treatment centers (Chiba and Hyogo)~\cite{Matsufuji03,Matsufuji05,Toshito07} and by the GSI biophysics department~\cite{Schall96,Gunzert04,Gunzert08} in the energy range from 200 to 400 MeV/u. All these measurements were used to determine the integrated flux and energy distributions of the fragments relative to the penetration depth in water equivalent matter. To extend these data to the lowest energies, a first integral experiment on thick PMMA targets has been performed by our collaboration in May 2008 at GANIL (France)~\cite{Braunn11}. Energy and angular distributions of the fragments produced by nuclear reactions of a 95 MeV/u $^{12}$C beam on thick PMMA targets have been obtained. Comparisons between Monte Carlo simulation codes (FLUKA~\cite{fluka}, INCL~\cite{INCL} and GEANT4~\cite{geant}) have been performed~\cite{Bohlen10,Braunn10,Braunn12} using different physics processes (Binary Cascade or Quantum Molecular Dynamic). These simulations have shown discrepancies up to one order of magnitude for the production rates of fragments. The angular and energy distributions were also not well reproduced. 

Nevertheless, it is difficult to constrain the nuclear reaction models by a direct comparison to thick target experiments.
To improve the models and reach the accuracy required for a reference simulation code for hadrontherapy, experiments on thin targets were planned. An experiment has been performed at 62 MeV/u on carbon target in Catania~\cite{Catane12}. An other one has been performed at 400 MeV/u on carbon target at GSI by the collaboration FIRST~\cite{FIRST12}. A third experiment has been performed by our collaboration on May 2011 at GANIL to study carbon reaction on C, H, O, Al and $^{nat}$Ti targets at 95 MeV/u. The data analysis method has already been described in a previous paper~\cite{Dudouet12}. In the following part, the experimental set-up will be recalled, in a second part, the study of the systematic errors will be detailed and finally, in the last part, the experimental results will be presented.

\section{Experimental setup}

The experiment used the ECLAN reaction chamber at the GANIL G22 beam line. In order to obtain the double differential fragmentation cross sections of carbon on carbon, oxygen, hydrogen and calcium targets (which represent around 95\% of the human body composition), a $^{12}$C beam at 94.6$\pm$0.09~MeV/u was sent on different thin targets (C, CH$_2$, Al, Al$_2$O$_3$, $^{nat}$Ti, PMMA (C$_5$H$_8$O$_2$)). We made here the approximation that the cross sections for the calcium target must be close to those of titanium ($^{40}_{20}$Ca $\sim$ $^{48}_{22}$Ti). The target area densities are about $\sim$50~mg/cm$^2$. The area densities (density ($\rho) \times$ thickness (th)) given values have been measured with a high precision balance leading to a relative error on the area densities of 1\%. The target characteristics are described in Table \ref{targets_char}.

\begin{table}[h]
\begin{center}
\begin{tabular}{|c|c|c|}
\hline
Targets		&th (in $\mu$m)	 & $\rho \times$th (in g.cm$^{-2}$) 	\\
\hline
\hline
C 		&  $250$		 		 & 0,0411			\\
\hline
CH$_2$ 		&  $500$				 &0,0481			\\
\hline
Ti 		&  $125$				 &0,0576			\\
\hline
Al 		&  $200$				 &0,0540			\\
\hline
 Al$_2$O$_3$ 	&  $150$				 &0,0550			\\
\hline
PMMA (C$_5$H$_8$O$_2$) &  $500$				 &0,0642			\\
\hline
\end{tabular}
\end{center}
\caption{Target characteristics}
\label{targets_char}
\end{table}

\begin{figure}[H]
\subfigure[]{\label{eclan1}{\includegraphics[width=0.49\linewidth]{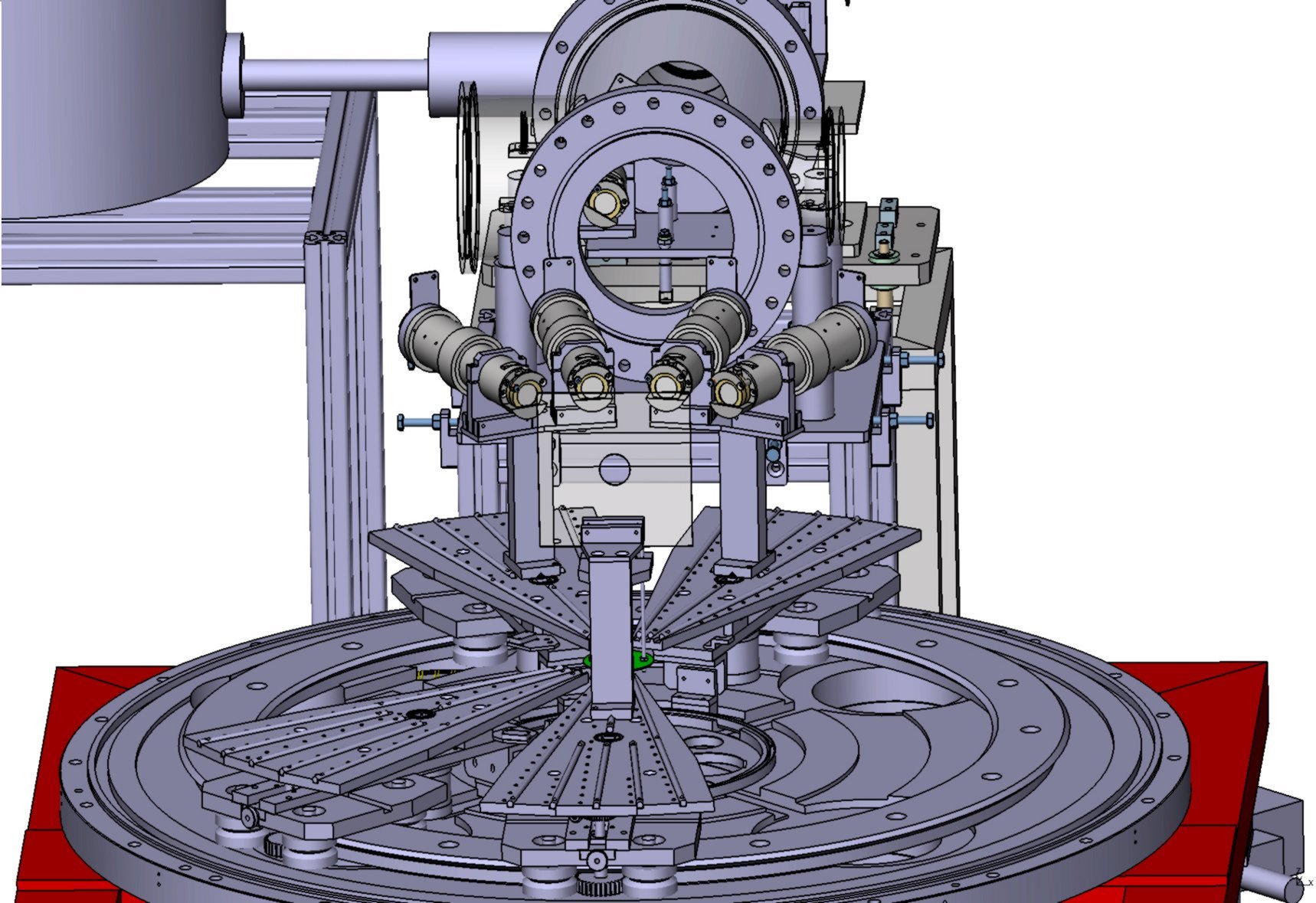}}}
\subfigure[]{\label{eclan2}{\includegraphics[width=0.49\linewidth]{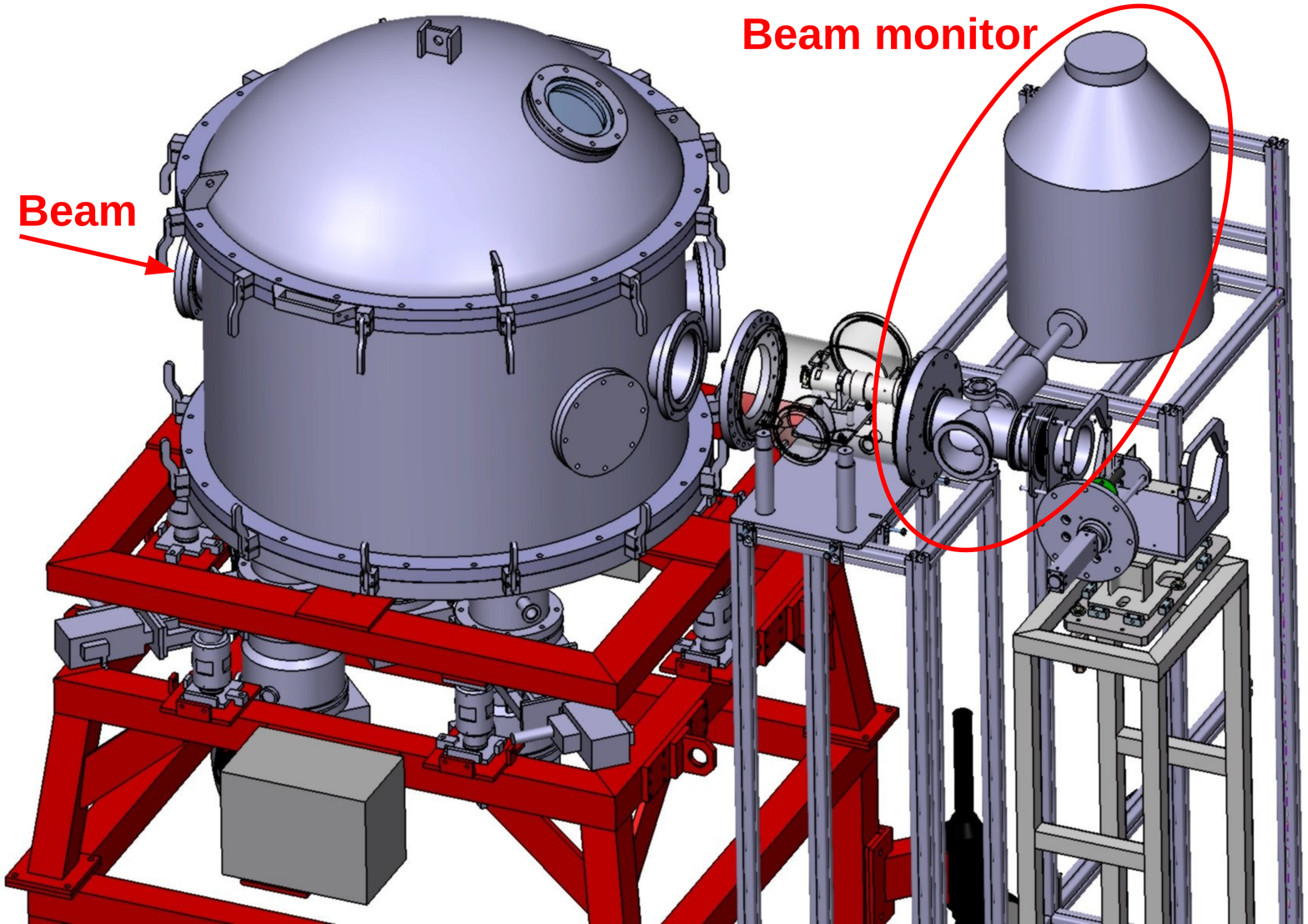}}}
\caption{On the Fig.\ref{eclan1}, representation of inside the vacuum chamber with the 4 rotating telescopes and the fixed telescope backward. On the Fig.\ref{eclan2}, a larger view including the beam monitor.}
\label{eclan}
\end{figure}

For the charged particles detection, the setup consists of four $\Delta E_{thin} /\Delta E_{thick}/E$ telescopes mounted two by two on two stages that allows rotation inside the chamber from $0^\circ$ to $43^\circ$ with $2^\circ$ steps at a distance of 204~mm behind the target. A fifth telescope was mounted downstream, at a fixed  angle  of $4^\circ$ and a distance of 820~mm behind the target (cf. Fig.~\ref{eclan}). The geometrical properties of the experimental setup are described in Table \ref{geometrical_property}.

The five $\Delta E_{thin} /\Delta E_{thick}/E$ telescopes consist in two silicon detectors followed by a 10~cm thick CsI(Tl) scintillator. The thicknesses of the silicon detectors were $\sim$150 $\mu$m and $\sim$1000~$\mu$m respectively. To prevent interaction of the scintillation light emitted by the CsI with the thick silicon detector, a thin Mylar sheet of 12 $\mu$m is placed between these two detectors. The silicon detectors and the CsI crystal have a diameter of 1.9 cm and 3 cm respectively. The higher diameter of the scintillating crystal is used to prevent particles with high incidence angles to escape out of the crystal by scattering. The detector dimensions are summarized in Tables~\ref{geometrical_property} and \ref{detectors_char}. Acceptances given in the Table~\ref{geometrical_property} are those of the thick silicon detectors. Indeed, the acquisition was triggered only if the thick silicon detector was hit. As the silicon detector thicknesses were not the same, an energy threshold has been used. This threshold has been calculated for each detected isotope and for the thickest of the thin silicon detectors (telescope n$^\circ$1). The resulting energy thresholds are given in Table~\ref{energy_threshold}.

\begin{table}[H]
\begin{center}
{\setlength{\tabcolsep}{1mm}
{\renewcommand{\arraystretch}{1.5}
\begin{tabular}{|c|c|c|c|c||c|}
\cline{2-6}
\multicolumn{1}{c|}{} &Telescope			& Thin Si	  & Thick Si		& CsI 			&$\Omega$ (msr)	\\
\cline{2-6}
\hline
\multirow{2}{2.2cm}{Distance to the target (mm)}&1 			&  820 (5)		  & 828 (5)               & 834 (5) &0.43 (3)\\
\cline{2-6}
&2-3-4-5				& 204 (5)		  & 212 (5)               & 218 (5)&6.6 (5)\\
\hline
\multicolumn{6}{c}{\vspace{-5mm}}\\
\cline{1-5}
diameter&	1$\rightarrow$5	&  1.954 cm		&       1.954 cm	&	3 cm		\\
\cline{1-5}
\end{tabular}}}
\end{center}
\caption{Geometrical properties of the detectors}
\label{geometrical_property}
\end{table}

\begin{table}[H]
\begin{center}
{\setlength{\tabcolsep}{1mm}
{\renewcommand{\arraystretch}{1.5}
\begin{tabular}{|c|c|c|c|}
\hline
Telescope			& Thin Si	  & Thick Si		& CsI 				\\
\hline
1 			&  161	$\mu$m		 & 	1011 $\mu$m	&\multirow{5}{*}{10 cm}	\\
\cline{1-3}
2 			&  159	$\mu$m		 & 	1014 $\mu$m	&				\\
\cline{1-3}
3 			&  160	$\mu$m		 &	1014	$\mu$m	&				\\
\cline{1-3}
4 			&  152	$\mu$m		 & 	1011	$\mu$m	&				\\
\cline{1-3}
5 			&  101 $\mu$m	 & 	1018	$\mu$m	&				\\
\hline
\end{tabular}}}
\end{center}
\caption{Charged particles detectors thicknesses.}
\label{detectors_char}
\end{table}

\begin{table}[H]
\begin{center}
{\setlength{\tabcolsep}{1mm}
{\renewcommand{\arraystretch}{1.5}
\begin{tabular}{|c|c|c|c|c|c|c|c|c|c|}
\cline{1-9}
isotope $^{\text{A}}$X  &$^1$H&$^2$H&$^3$H&$^3$He&$^4$He&$^6$He&$^6$Li&$^7$Li \\	
\cline{1-9}
E$_{\text{seuil}}$ (MeV)&4.0&5.2&6.1&14.2&16.0&18.6&29.9&31.7\\	
\cline{1-9}
\multicolumn{10}{c}{\vspace{-5mm}}\\
\hline
isotope $^{\text{A}}$X  &$^7$Be&$^9$Be&$^{10}$Be&$^8$B&$^{10}$B&$^{11}$B&$^{10}$C&$^{11}$C&$^{12}$C \\	
\hline
E$_{\text{seuil}}$ (MeV)&44.3&48.6&50.5&60.6&65.8&68.1&81.3&84.2&86.9\\	
\hline
\end{tabular}}}
\caption{Energy threshold.}
\label{energy_threshold}
\end{center}
\end{table}

To determine the number of incident ions, a beam monitor is placed after the reaction chamber. It is based on the measurement of fluorescence X-rays emitted by a thin Ag foil (7~$\mu$m thick) which is set in the beam. X-rays are detected by means of a Si(Li) detector located at $90^\circ$ with respect to the beam direction. This detector is calibrated at different beam intensities ranging from 10$^3$ to 10$^6$ ions/s using a plastic scintillator intercepting the beam. The calibration is then extrapolated on the whole intensity range of the experiment (from 10$^5$ to 10$^7$ ions/s). The error on the beam intensity has been estimated of about 5\%~\cite{Dudouet12}.

For the electronic part, a digital acquisition named FASTER developed in-house was used. A FASTER module contains one QDC (charge to digital converter) used for the CsI and an ADC (analog to digital converter) used for the Silicon detectors.
 
\section{Systematic errors study}
\label{systematic_section}

A detailed description of the analysis method is described in Dudouet et al.~\cite{Dudouet12}. It has been performed using the KaliVeda framework developed by the INDRA collaboration~\cite{Pouthas95,kaliveda}. However, systematic errors due to detector effects still have to be determined to finalize the analysis.

The $^{12}$C fragmentation cross sections for a $^\text{A}_\text{Z}X$ fragment are obtained as follows :

\begin{equation}
\frac{d\sigma}{d \Omega} (^\text{A}_\text{Z}\text{X}) = \frac{N_{^\text{A}_\text{Z}\text{X}} \times A_{target}}{N_{^{12}\text{C}} \times \Omega \times \rho \times \text{th} \times \mathcal{N}_{\text{A}}},
\label{cross_sections}
\end{equation}

where $N_{^\text{A}_\text{Z}\text{X}}$ is the number of $^\text{A}_\text{Z}\text{X}$ fragments detected, A and Z are respectively the mass and charge of the fragment X, $A_{target}$ is the target mass, $N_{^{12}\text{C}}$ is the number of incident carbon nuclei, $\Omega$ is the solid angle of the detector, $\rho \times$th is the target area density and $\mathcal{N}_{\text{A}}$ is the Avogadro's number.

Errors on $N_{^{12}\text{C}}$, $\Omega$ and $\rho \times$th have been discussed in the first part, therefore, it remains to determine the error on $N_{^\text{A}_\text{Z}\text{X}}$. This last term is coming from detector effects generating wrong identifications. As statistical errors are low e.g. from 1\% to 5\% for the rare isotopes, the systematic errors are the largest sources of uncertainties and must be determined. A first source of systematic errors comes from the identification method. This error was already estimated to be of the order of 5\% for most of the measurements~\cite{Dudouet12}. In order to estimate other sources of systematic errors, Monte~Carlo simulations have been conducted, taking into account the experimental setup.

The experimental setup has been simulated by using the version 9.4 of the GEANT4 Monte Carlo simulation~\cite{geant}. Electromagnetic interactions were described by using the ``electromagnetic standard package option 3''. Regarding the inelastic processes, two GEANT4 models have been tested: the Binary Cascade for light ions~\cite{BIC} and then the G4QMD model~\cite{QMD}. The simulated data were then analyzed with exactly the same method as for the experimental data. Overall, two sources of errors were observed.

The first one is due to inelastic processes in the CsI detector. Some detected particles might encounter inelastic processes in the CsI. This leads to energy losses coming on one hand from the Q-Value of the reaction and neutrons which are not detected and on the other hand, from lighter fragments emitted with high incidences and leaving the detector. Moreover, even if all the fragments are stopped in the detector, their resulting emitted scintillation light is changed due to quenching effects in the scintillator~\cite{Bir64} which depend on the charge and mass of the particles. Such cases will induce a part of wrong identification.

The second source of errors comes from the pile-up of two particles in a telescope. If two particles from the same $^{12}$C event are detected in a single detector, it will appear as one particle which has deposited the energy of two detected particles. The identification of the two different particles is thus no longer feasible. It appears in the simulations that $\sim$4\% of the events are due to a pile-up of two or more particles. According to simulation results, the main contribution is due to the detection of two $\alpha$ particles at the same time. Although this phenomenon only represents a few percent of the whole statistics, it must be taken into account. As shown in Fig.~\ref{2a_de_e}, the events coming from the pile-up of two alpha particles appears on the $^{6}$He and on the $^{6}$Li-$^{7}$Li identification lines in the $\Delta$E-E matrix. As the CsI is larger than the thick silicon detector, it is possible for a particle to reach the CsI without crossing the thick Silicon detector. The pile-up component on the $^{6}$He corresponds to events with only one $\alpha$ detected in the thick silicon and the sum of two in the CsI. The component on the lithium isotopes corresponds to the detection of both $\alpha$ in all detectors. The measured cross sections of these three rare isotopes ($^{6}$He, $^{6}$Li and $^{7}$Li) are then modified by the pile-up.  

\begin{figure}[H]
\centerline{\includegraphics[width=0.5\linewidth]{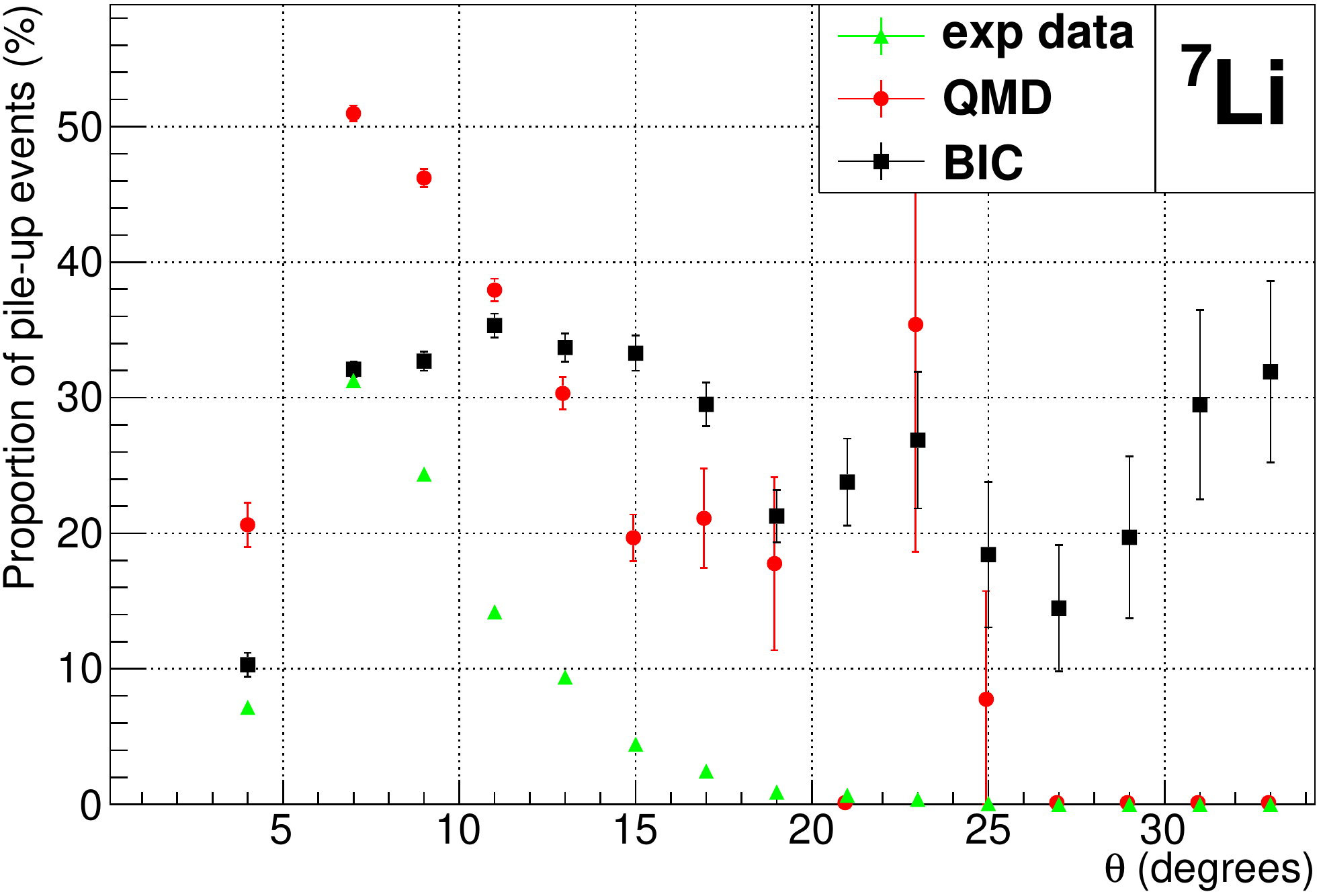}}
\caption{Proportion of pile-up events in the $^{7}$Li statistic. The results have been estimated by GEANT4 simulations with BIC and QMD models. They are compared to the experimental results obtained with the CsI fast/slow matrix. Only Statistical errors are shown.}
\label{err_id_7Li}
\end{figure}

\begin{figure}[H]
\centerline{\includegraphics[width=0.7\linewidth]{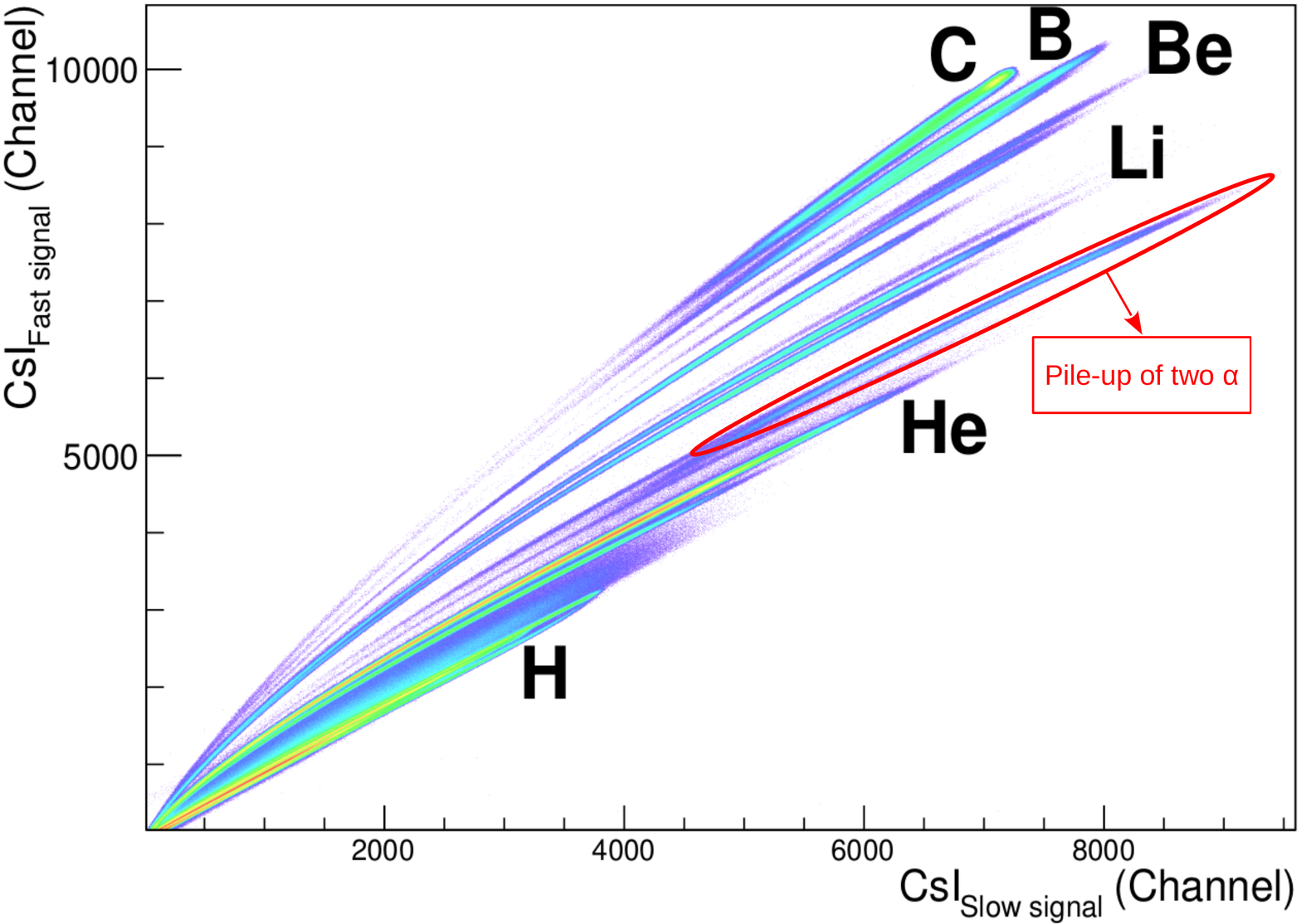}}
\caption{CsI(fast/slow) matrix. The different isotopes from protons to carbons are separated. The pile-up contribution is surrounded by red.}
\label{slow_fast_map}
\end{figure}

\begin{figure}[H]
\subfigure[]{\label{simu_de_e}{\includegraphics[width=0.49\linewidth]{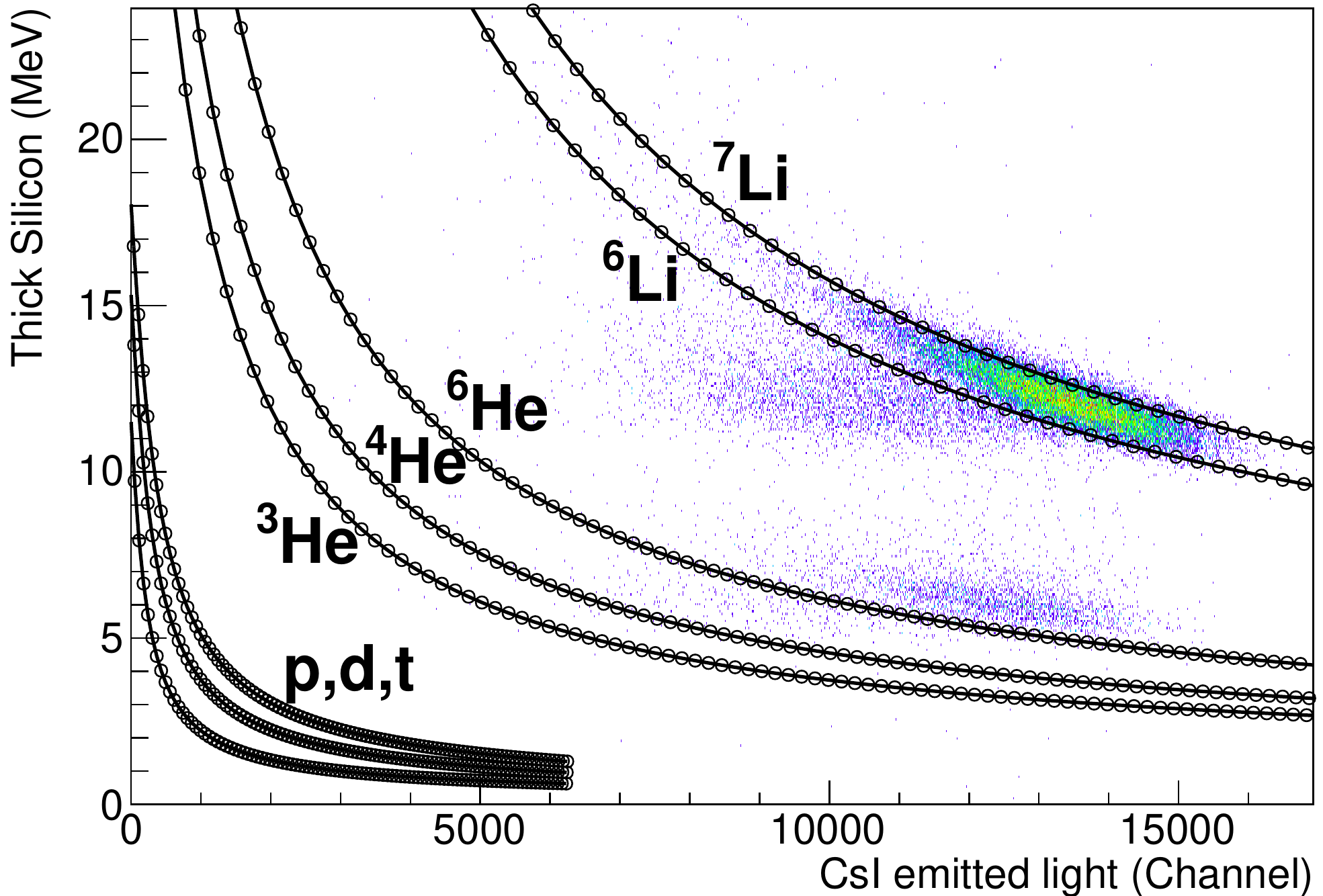}}}
\subfigure[]{\label{exp_de_e}{\includegraphics[width=0.49\linewidth]{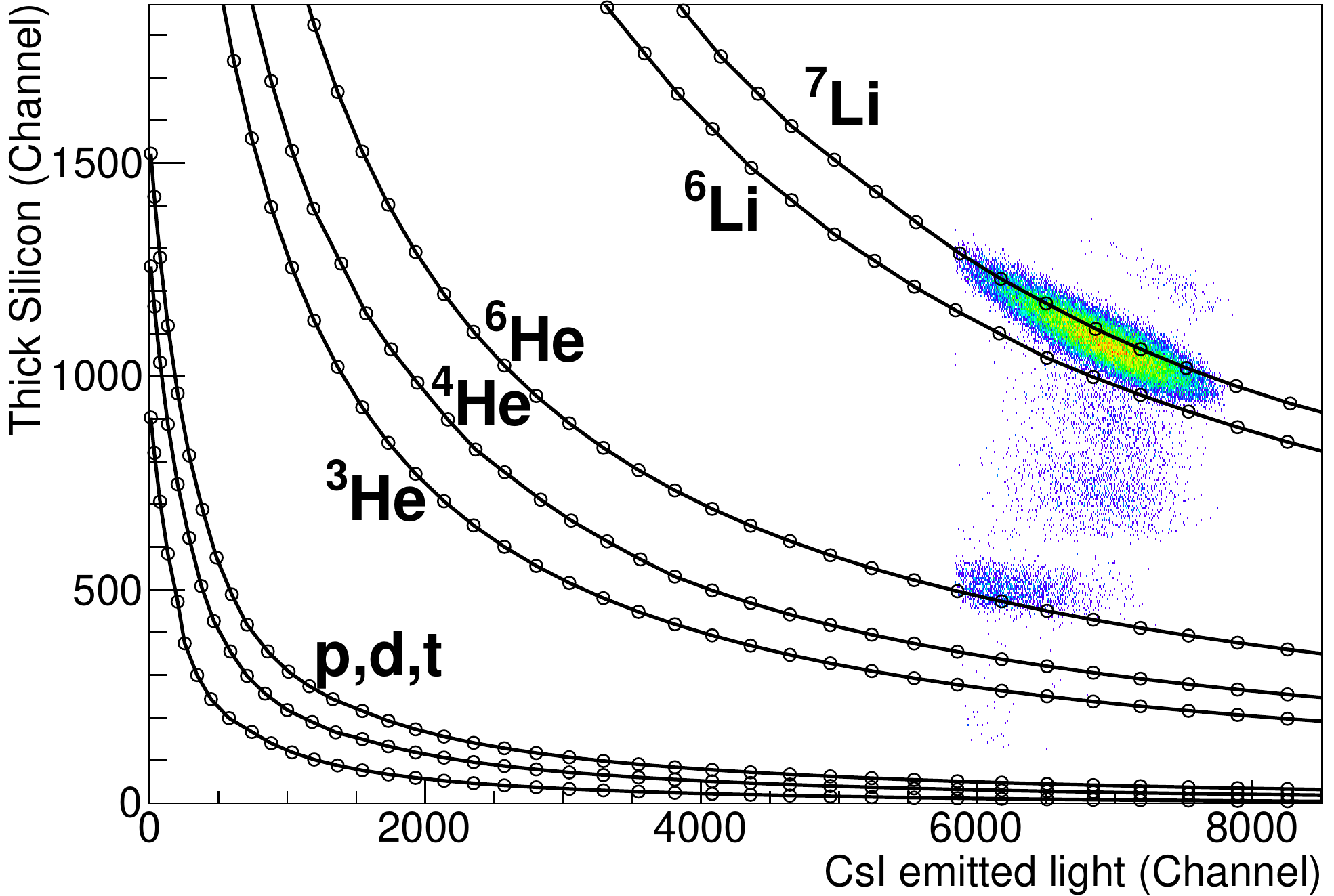}}}
\caption{Background of the $\Delta$E-E matrix due to pile-up of two alpha in a telescope. On the Fig.\ref{simu_de_e} : results obtained by GEANT4 simulations. On the Fig.\ref{exp_de_e} : Experimental pollution due to events selected as pile-up from the fast/slow matrix. The surrounded selection on Fig.~\ref{slow_fast_map} has been projected on the $\Delta$E-E matrix.}
\label{2a_de_e}
\end{figure}

Although this phenomenon was observed in simulations with the two tested models, their respective quantities are strongly model dependent (cf. Fig.~\ref{err_id_7Li}) and then, cannot be applied as an accurate correction on experimental data. Therefore, in order to remove the pile-up events from the $\Delta$E-E matrix, an experimental method has been chosen by analyzing the CsI light output signal shape. Using the digital acquisition, different gates in time have been chosen on the QDC (fast,slow,total) to integrate the light emitted by the scintillating crystal. By representing the fast component of the signal versus the slow one, a matrix on which each isotope is separated has been obtained (cf. Fig.~\ref{slow_fast_map}). It turns out that the events that are coming from the pile-up of two $\alpha$ particles are clearly separated on this matrix. These pile-up events are located in a line which is between the $^{4}$He and the $^{6}$Li lines. It is thus possible to make a 
cut with the graphical tools of ROOT~\cite{ROOT} on this pile-up contribution to remove it from the $\Delta$E-E matrix. Fig.~\ref{err_id_7Li} represents the proportion of pile-up events in the $^{7}$Li statistics obtained by the experimental method and by GEANT4 simulations using BIC and QMD nuclear models. The QMD model seems to be more realistic but neither of the two models are able to correctly estimate the pile-up contribution.

Fig.~\ref{2a_de_e} represents an example of the pollution due to pile-up events on the $\Delta$E-E matrix. On the Fig.\ref{simu_de_e} are represented the pile-up events obtained by simulation with the QMD model of GEANT4 which seems to be the least far from the data. The light emitted by the CsI crystal has been simulated by using quenching formula described in Parlog et al.\cite{Parlog02}. On the Fig.\ref{exp_de_e} are represented the experimental $\Delta$E-E matrix for the events selected as pile-up from the fast/slow matrix.

\begin{figure}[H]
\subfigure[]{\label{7Li_case1}{\includegraphics[width=0.49\linewidth]{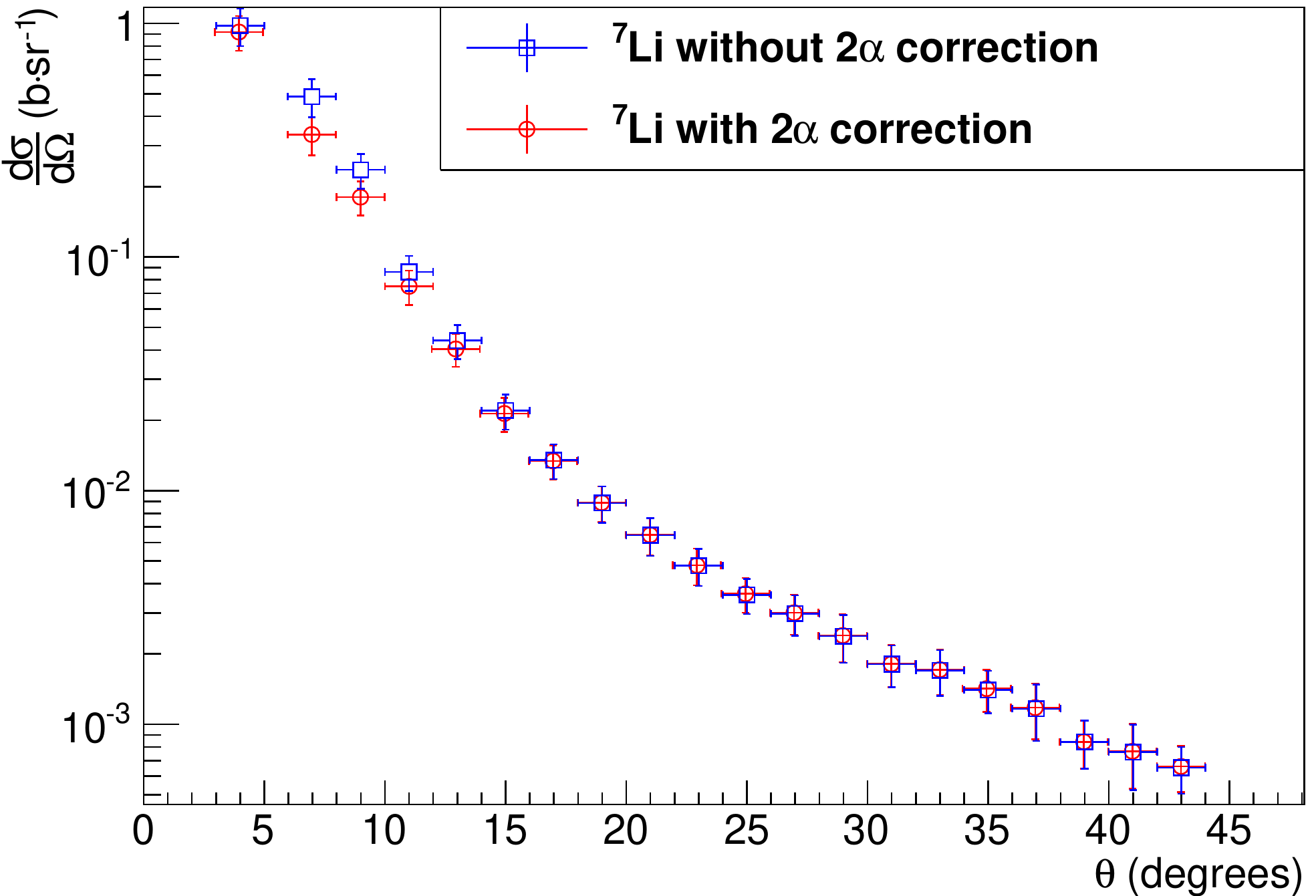}}}
\subfigure[]{\label{7Li_case2}{\includegraphics[width=0.49\linewidth]{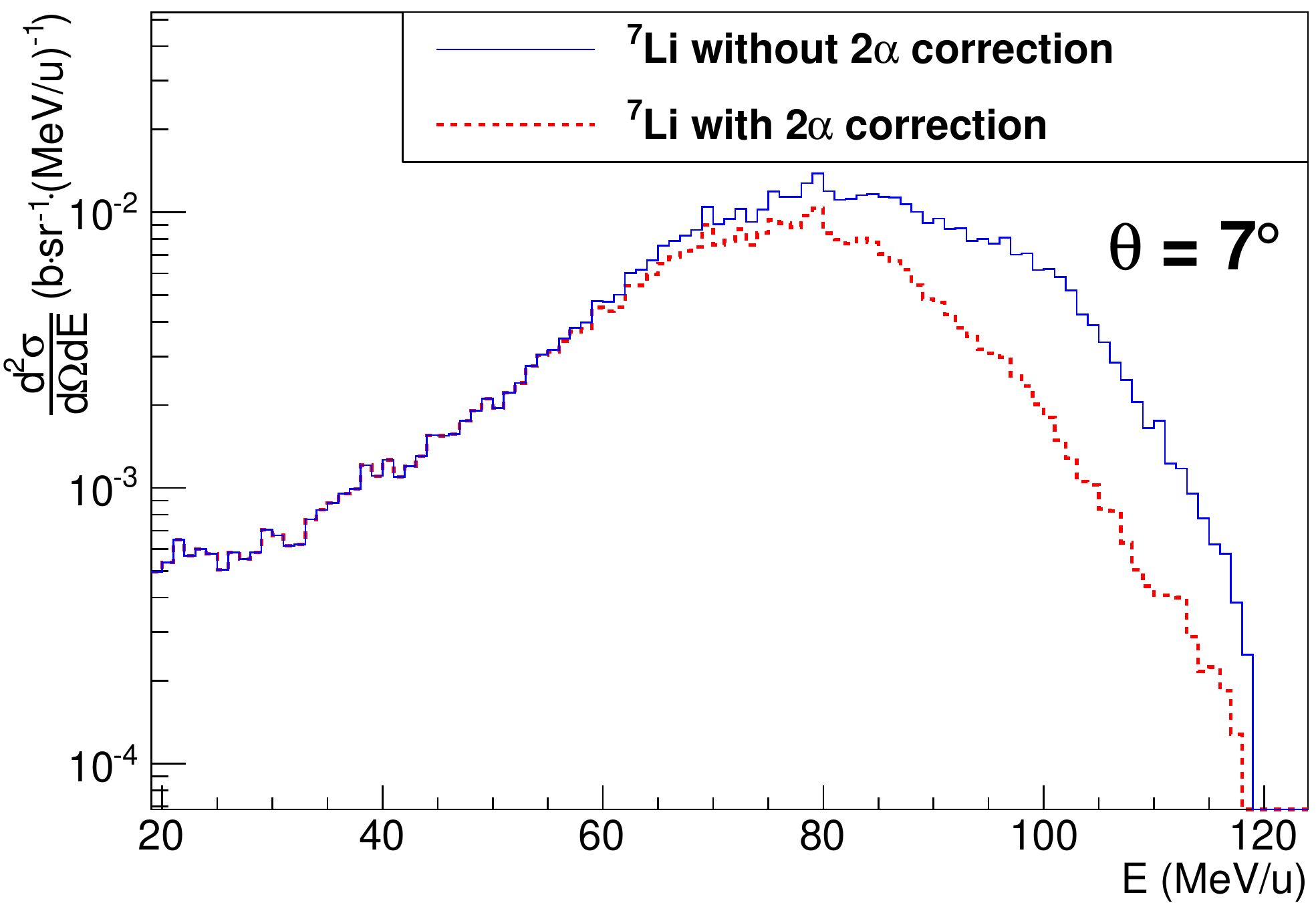}}}
\caption{Example of the ``2$\alpha$'' pile-up correction for the $^{7}$Li distributions. The angular distribution is represented on the left and the energy distribution for $\theta=7^\circ$ is on the right.}
\label{7Li_case}
\end{figure}

Due to the proximity of the pile-up and $^{6}$He events in the fast/slow matrix, some pile-up events could have been missed or some $^{6}$He events could have been removed. The error bars for the $^{6}$He distributions shown in the results were thus calculated as the difference between with and without pile-up correction,resulting conservative values. An example of the correction applied to the $^{7}$Li distributions is given in Fig.~\ref{7Li_case}. The Fig.\ref{7Li_case1} represents the angular distributions with and without corrections. The Fig.\ref{7Li_case2} represents the energy distributions of $^{7}$Li fragments emitted at $7^\circ$ with and without corrections. The pile-up correction is mainly visible at low angles and high energies.

GEANT4 simulations have given a proportion of particles which fragment in the CsI of the order of several percents. In such cases, the inelastic processes lead to a loss of a part of the fragment's energy released in the CsI. It appears on the $\Delta$E-E matrix as horizontal shift toward the lower energies. This phenomenon is affecting all the isotopes but is in general negligible. This is not the case for helium isotopes for which the $\alpha$ particles are produced ten times more than $^{3}$He. Some $\alpha$ particles will then be identified as an $^{3}$He. The proportion of $\alpha$ that have been identified as $^{3}$He has been estimated to be about 5\% which represents an error of 30\% on the $^{3}$He statistic. A correction has been done to the $^{3}$He distributions by using the fast/slow components of the CsI light output.

Apart for these major error sources, the simulations were used to determine the percentage of well identified particles induced by the analysis method. The conclusion of this study is that the systematic errors are too model-dependent to be accurately estimated. To be conservative, the errors obtained at forward angles with the QMD model (which seems to be the closest to experimental data) have been used for all detection angles. Indeed, the identification errors are larger for the higher energies which are mostly detected at forward angles. The systematic errors have thus been overestimated, ranging from 2\% to 15\% depending on the isotope. To precisely determine these systematic errors, this study will need to be refined with a nuclear model which gives results closer to experimental data.

The systematic error study was realized by combining Monte Carlo simulations and experimental method. The simulation results have shown that the nuclear models available in GEANT4 are not sufficiently accurate to be used to correct the data from detection effects as pile-up or fragmentation in the CsI. Thus, an experimental method using the CsI light shape analysis has been used to correct the distributions of $^{3}$He, $^{6}$He, $^{6}$Li and $^{7}$Li from these effects. Overall, the simulation results were used, in a conservative way, to estimate the error bars on the identification due to the analysis method. The main consequence is a probable overestimation of the error bars in the following results.

\section{Experimental results}

\subsection{hydrogen and oxygen target cross sections}

In order to obtain the double differential cross sections for oxygen and hydrogen, composite targets (CH$_2$ and Al$_2$O$_3$) have been used. The hydrogen cross section has been obtained by combining the cross sections of CH$_2$ and C targets. The same method has been used to obtain the cross section of the oxygen by combining the cross sections of the Al$_2$O$_3$ and Al targets as follow :

\begin{align}
 \frac{d \sigma}{d\Omega}(\text{H}) &= \frac{1}{2} \times \left(\frac{d \sigma}{d\Omega}(\text{CH}_2) - \frac{d \sigma}{d\Omega}(\text{C})\right),\\
 \frac{d \sigma}{d\Omega}(\text{O}) &= \frac{1}{3} \times \left(\frac{d \sigma}{d\Omega}(\text{Al$_2$O$_3$}) - 2 \times \frac{d \sigma}{d\Omega}(\text{Al})\right).
\end{align}

\begin{figure}[H]
\centerline{\includegraphics[width=0.7\linewidth]{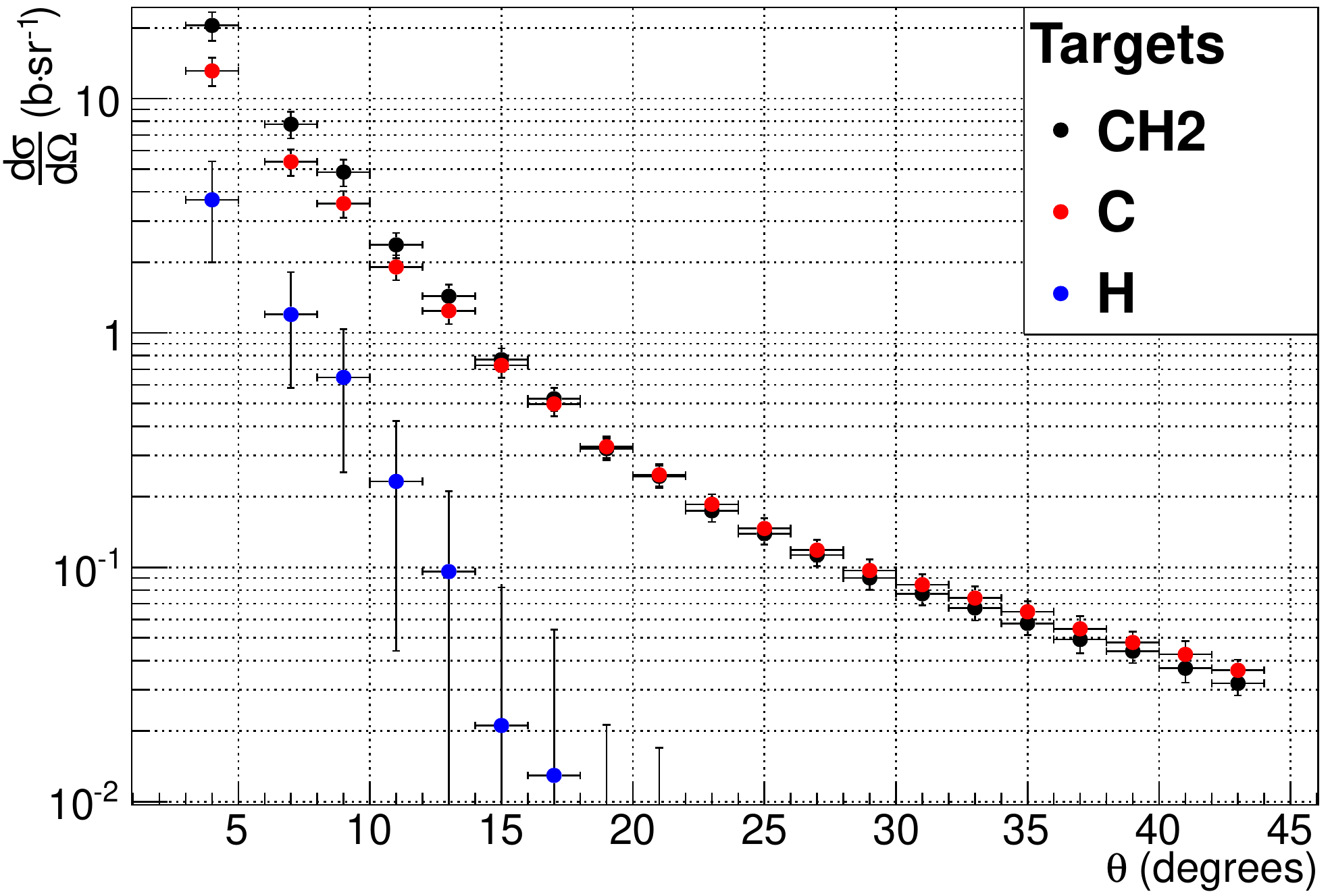}}
\caption{Combination of the carbon and CH$_2$ angular distribution to determine the hydrogen angular distribution for alpha fragments. The angular distribution for the hydrogen target is the difference between the CH$_2$ and carbon target, divided by two.}
\label{H_from_C_CH2}
\end{figure}

Fig.~\ref{H_from_C_CH2} represents an example of the method used to extract the alpha angular differential cross sections of hydrogen from CH$_2$ and C cross sections. A disadvantage of such a method is that the resulting cross sections are obtained by subtracting the cross sections of the two targets but the uncertainties are the quadratic sum of the uncertainties of the both individual targets. Hydrogen cross sections, which are small compared to carbon cross sections (cf. Fig.~\ref{H_from_C_CH2}), have been obtained with larger error bars. 

\subsection{Results}

All the elements to determine the double differential cross sections are now available. Firstly, the angular distributions will be presented, then, the energy distributions and secondly, the comparison between the real PMMA target and the reconstructed PMMA target cross sections will be shown.

\subsubsection{Angular distributions}

Concerning the angular distributions, all the measurements are represented on the positive angle side. It has to be noted that one measurement out of two in the presented distribution was made at the opposite angle and thus on an other telescope (cf. Fig.~\ref{eclan}). All the results for each target (real target or target obtained by composite target reconstruction) are summarized in the appendices (cf Tab.~\ref{Table_cross_section_H_target}, ~\ref{Table_cross_section_C_target}, ~\ref{Table_cross_section_O_target}, ~\ref{Table_cross_section_Al_target}, and \ref{Table_cross_section_Ti_target}).

\begin{figure}[H]
\subfigure[Z=1]{\label{C_H}{\includegraphics[width=0.49\linewidth]{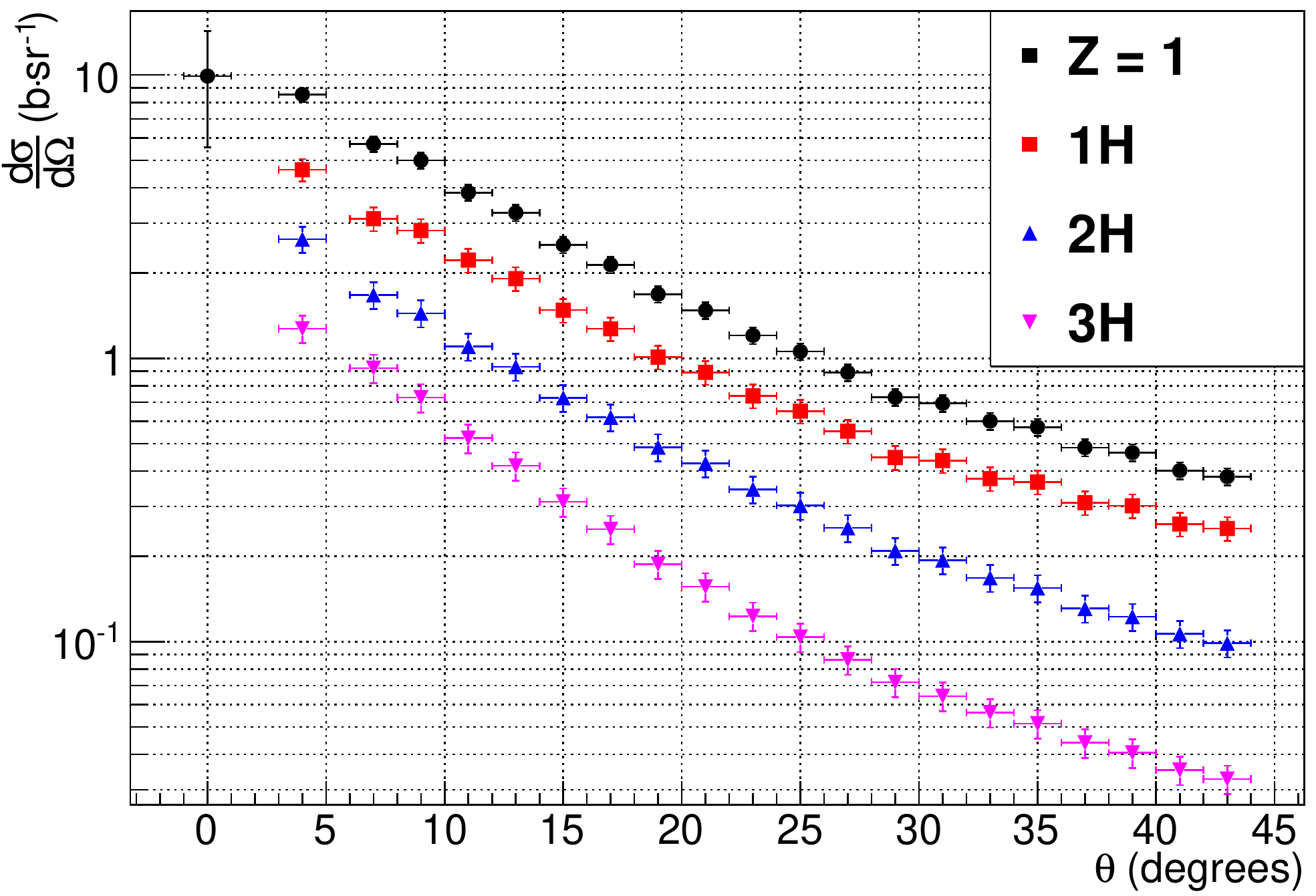}}}
\subfigure[Z=2]{\label{C_He}{\includegraphics[width=0.49\linewidth]{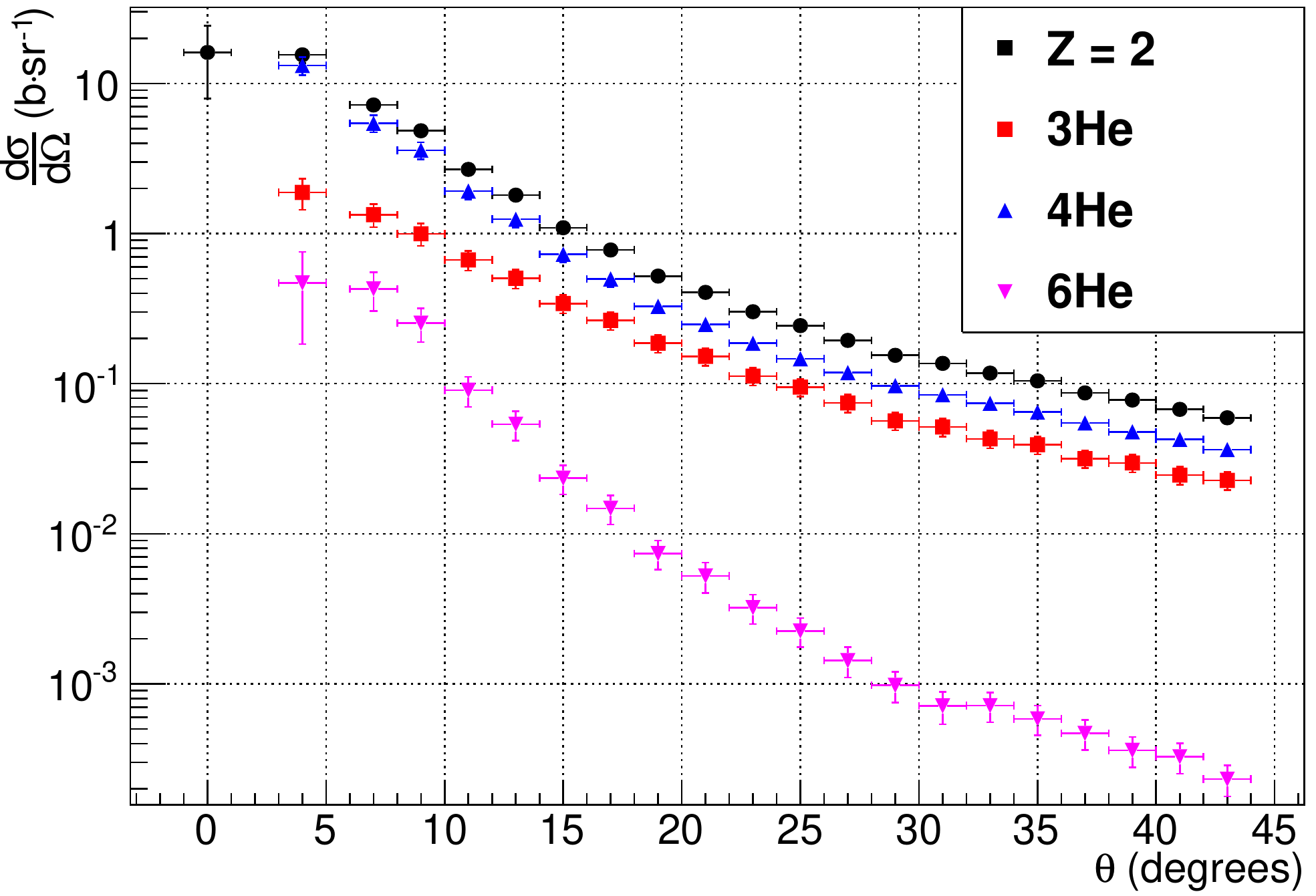}}}\\
\subfigure[Z=3]{\label{C_Li}{\includegraphics[width=0.49\linewidth]{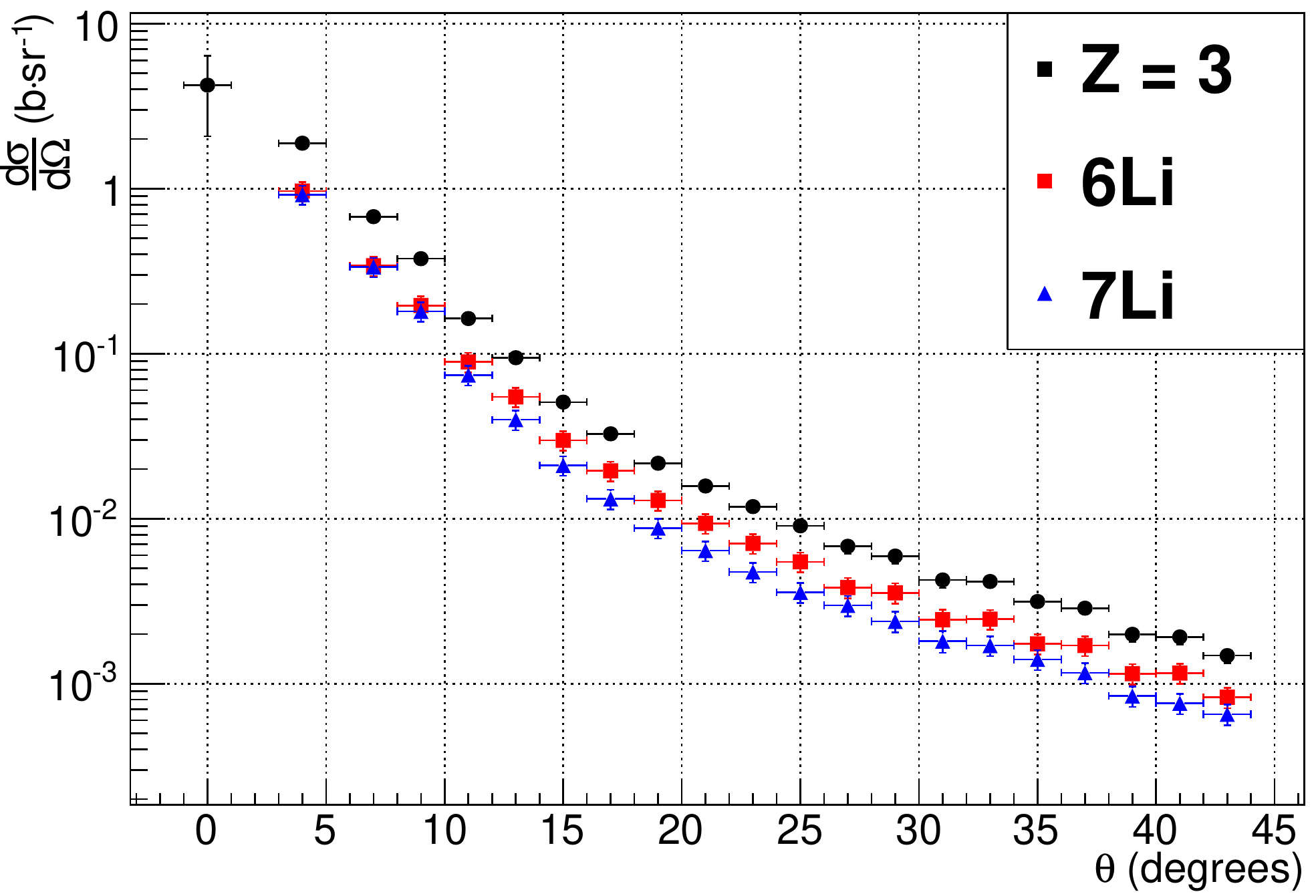}}}
\subfigure[Z=4]{\label{C_Be}{\includegraphics[width=0.49\linewidth]{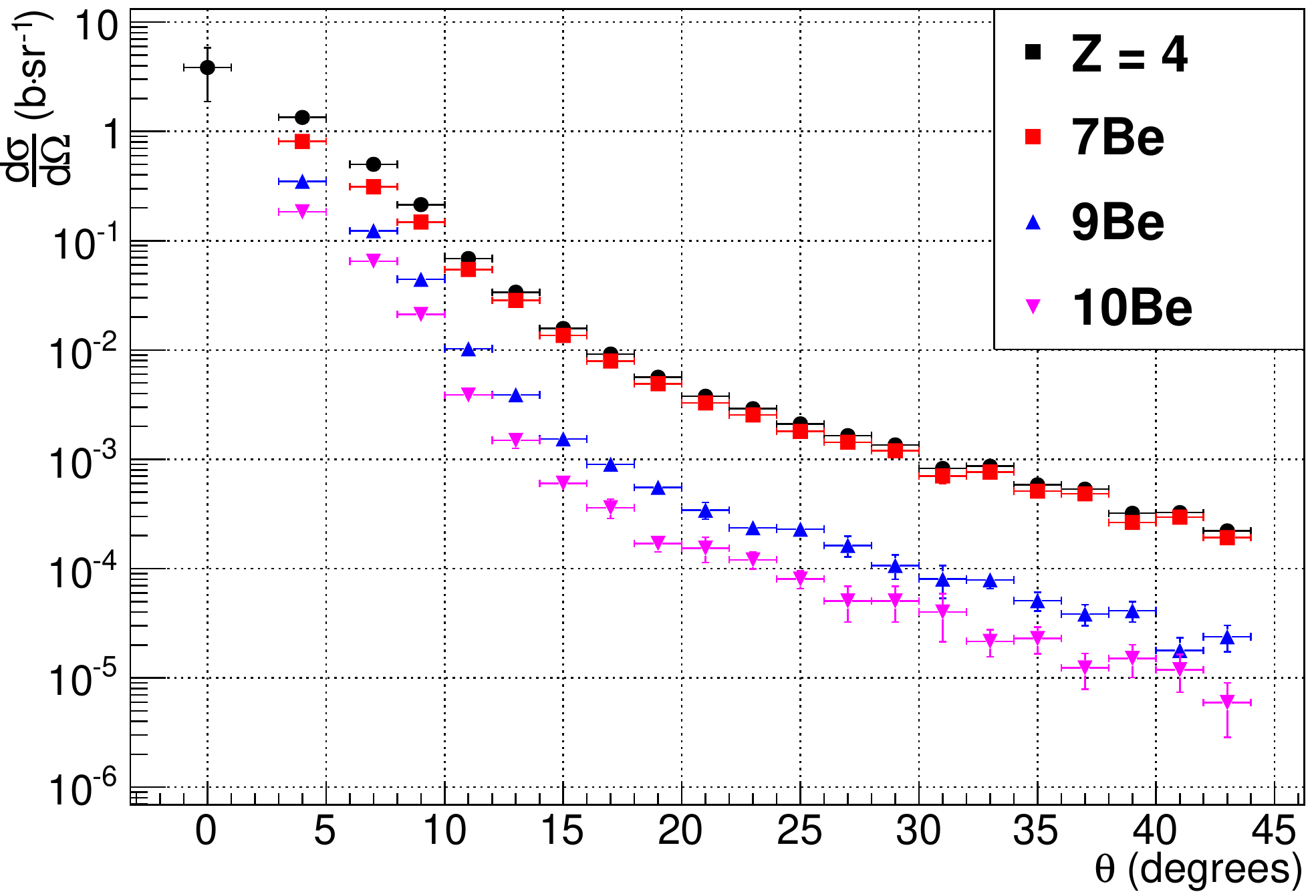}}}\\
\subfigure[Z=5]{\label{C_B}{\includegraphics[width=0.49\linewidth]{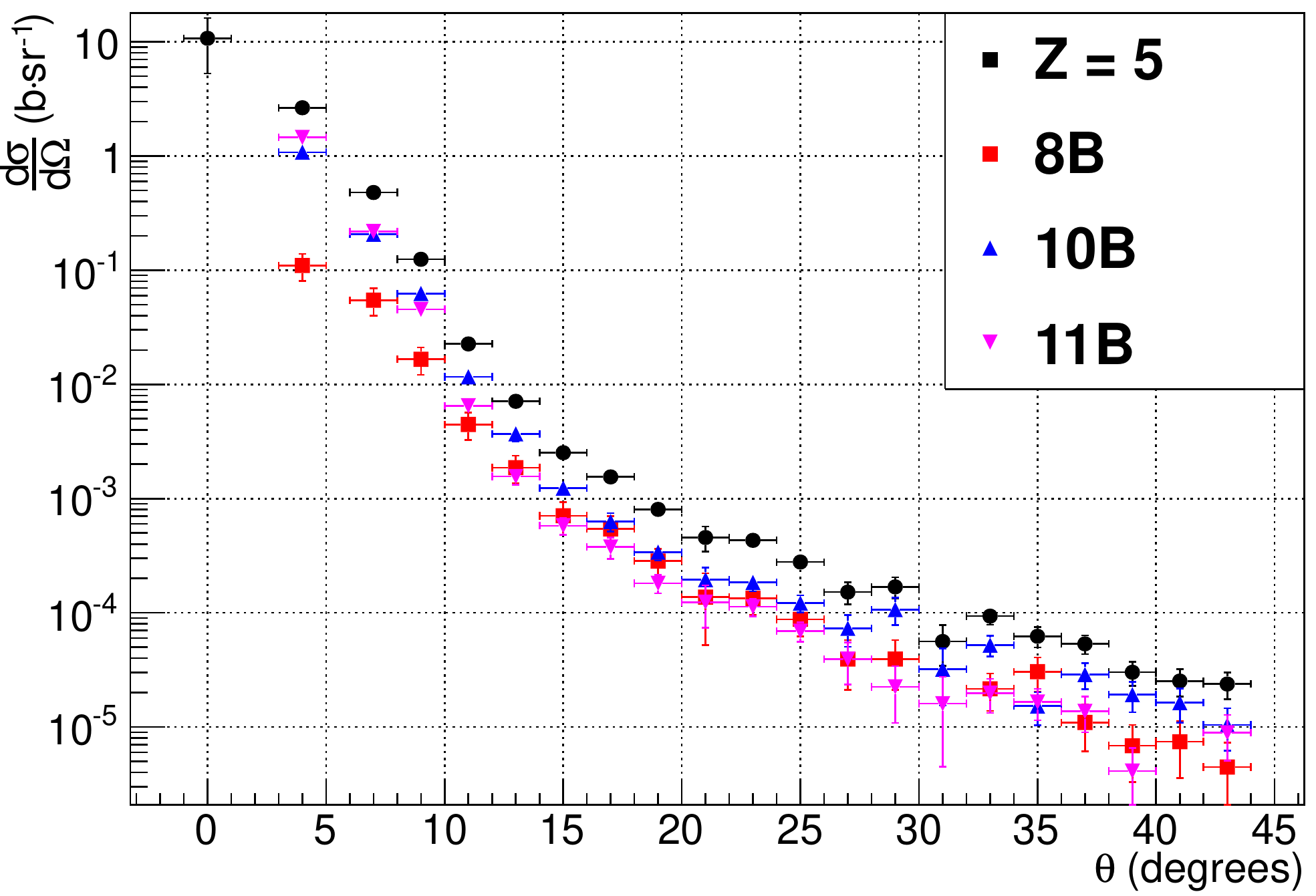}}}
\subfigure[Z=6]{\label{C_C}{\includegraphics[width=0.49\linewidth]{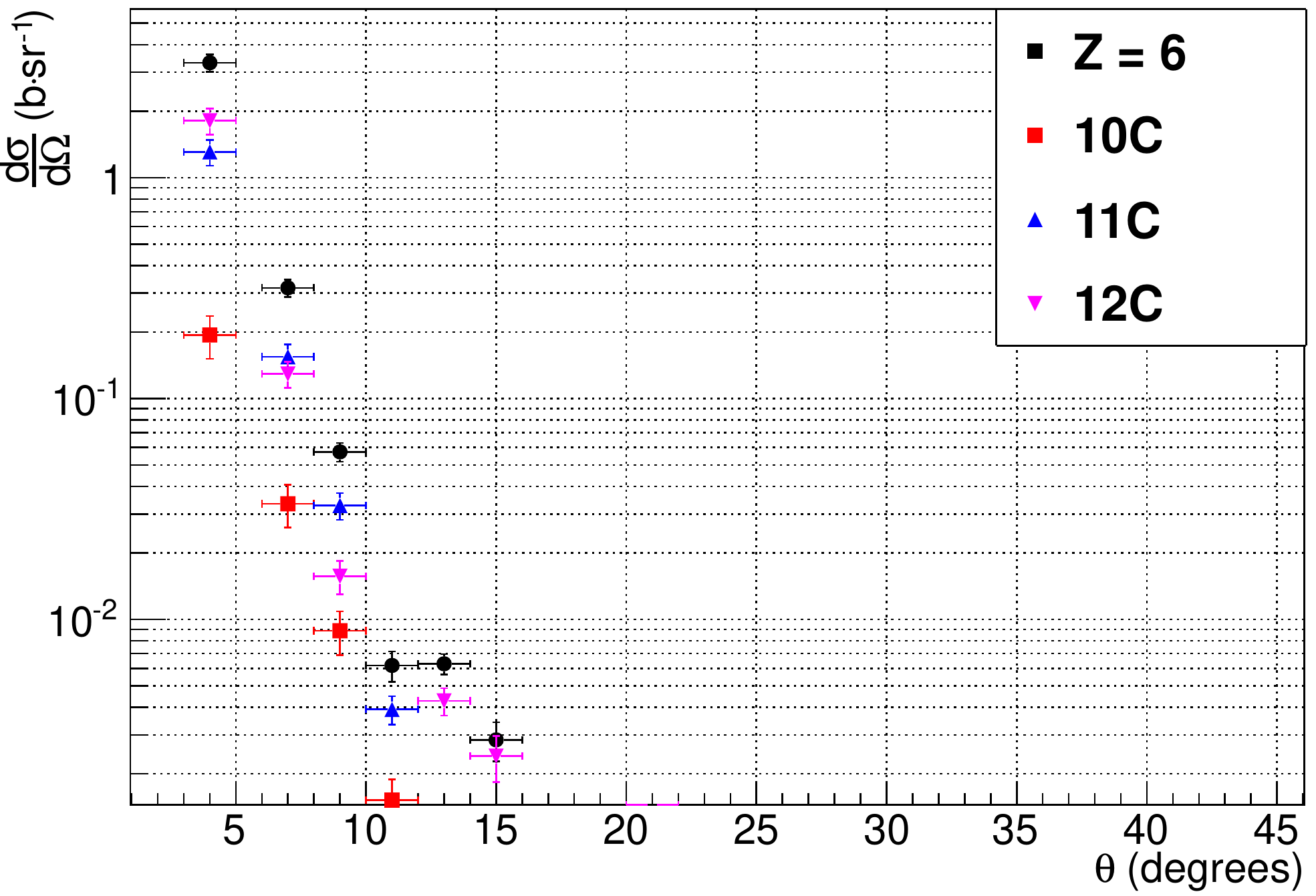}}}
\caption{Angular distributions for fragments resulting from the fragmentation on carbon target. Each graph represents the distribution per Z value with the measure at zero degree (except for Z=6 for which fragments are not dissociable from the carbon coming from the beam). The distributions of the different isotopes are superimposed.}
\label{C_distrib}
\end{figure}

The angular distributions for the carbon target are  shown in Fig.~\ref{C_distrib}. One graph per Z is drawn, on which the different detected isotopes and their sum, including a measurement at zero degree angle value, are superimposed. 

Regarding zero degree angle value measurements, the normalization versus the number of carbon ions has been done by integrating the number of Z=6 on the $\Delta$E-E matrix. Due to the direct detection of the beam particles, these measurements were done at low beam intensity ($\sim10^{3}$s$^{-1}$). Moreover, the detection devices were not adapted for zero degree angle measurements and have not permitted a mass identification. This is why the error bars are very large for the zero degree angle value measurements.

It has to be noticed that the production rates are dominated by the hydrogen and helium isotopes with a predominance of $\alpha$  at small angles (below $10^\circ$) which is compatible with the alpha cluster structure of the $^{12}$C. The results also show an angular emission most forward peaked for heavier fragments. For most of the isotopes, the overall error is of about 5 to 15\%, but is dominated by systematic errors that should be reduced by estimating them with simulations done with more realistic nuclear models (cf. section \ref{systematic_section}).

\begin{figure}[H]
\subfigure[$^{1}$H]{\label{1H}{\includegraphics[width=0.49\linewidth]{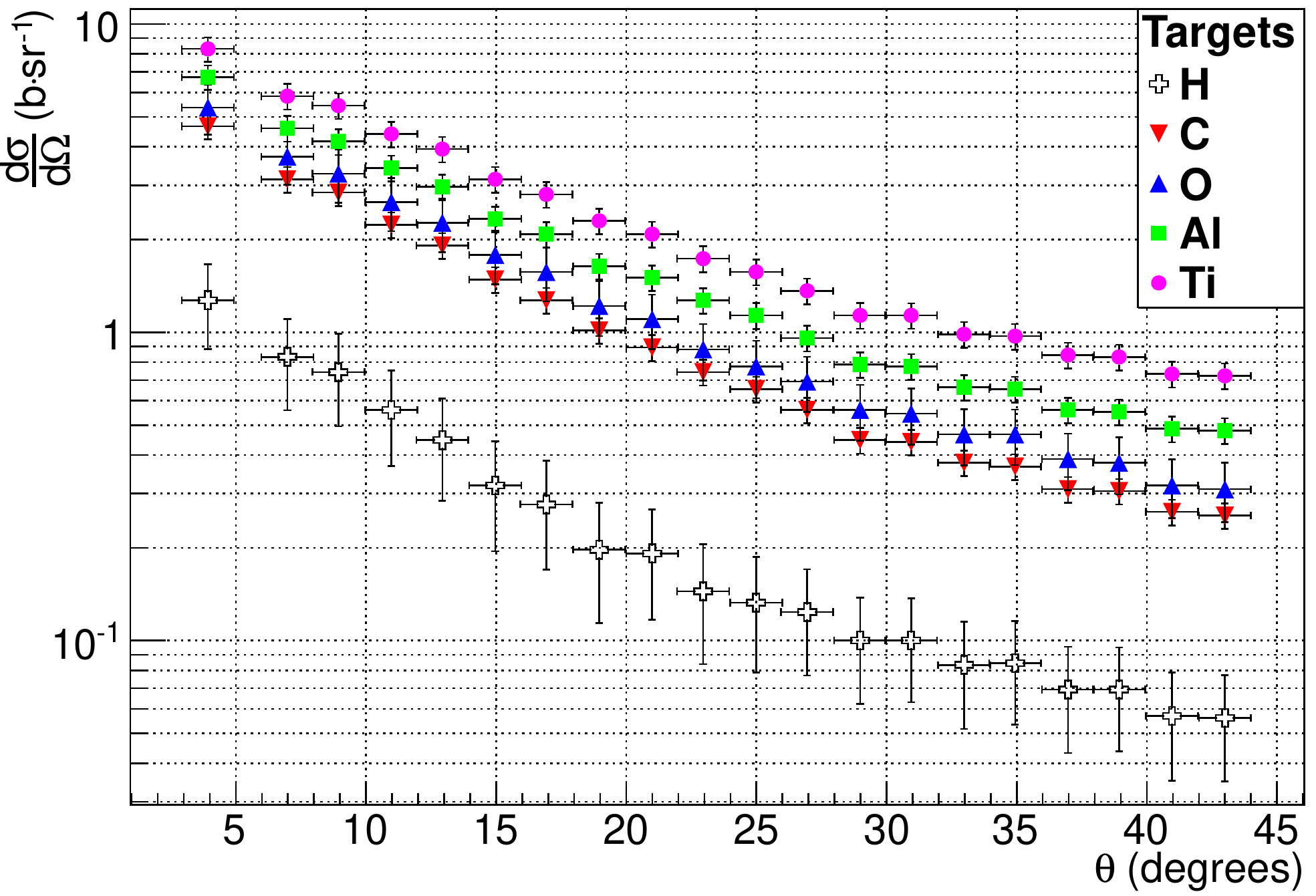}}}
\subfigure[$^{4}$He]{\label{4He}{\includegraphics[width=0.49\linewidth]{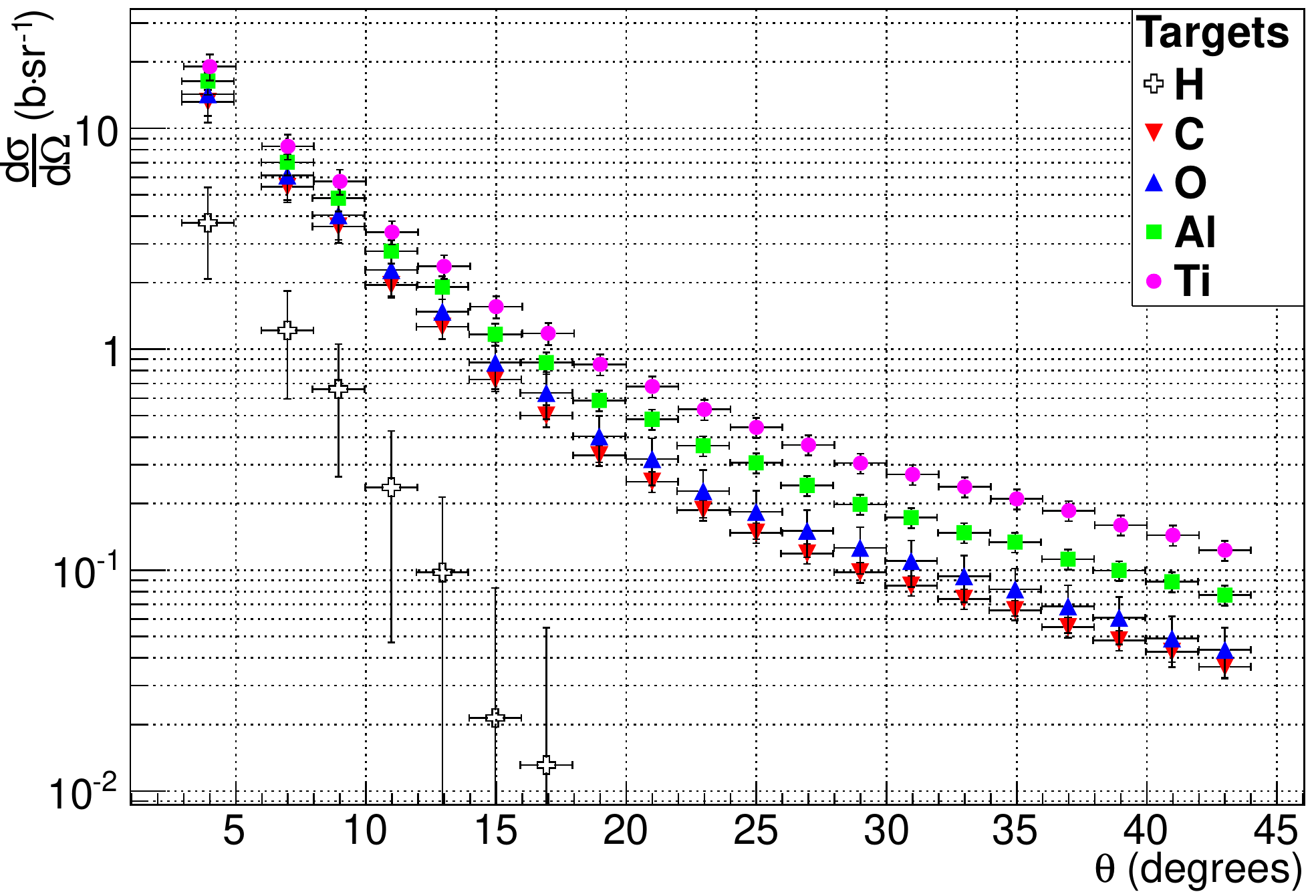}}}\\
\subfigure[$^{6}$Li]{\label{6Li}{\includegraphics[width=0.49\linewidth]{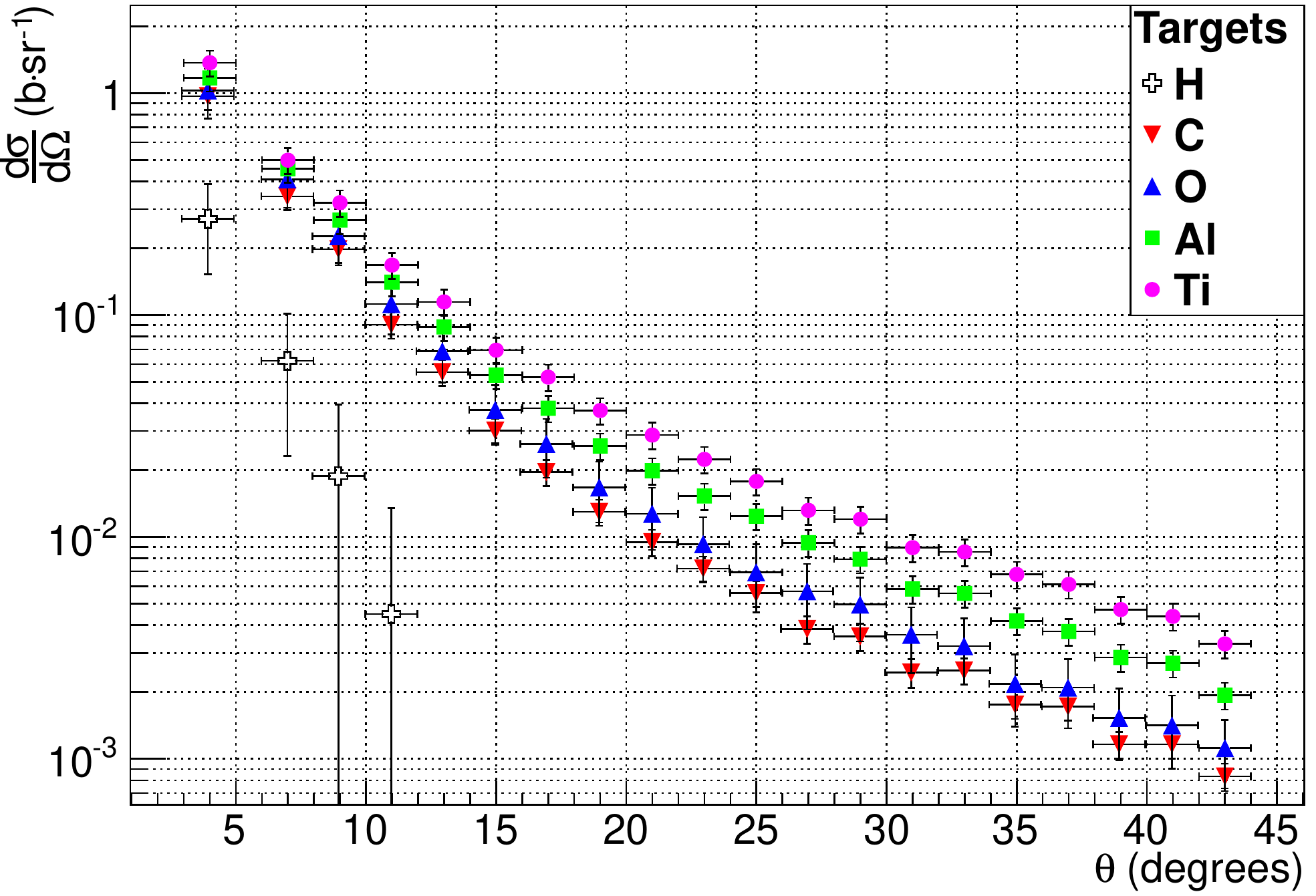}}}
\subfigure[$^{7}$Be]{\label{7Be}{\includegraphics[width=0.49\linewidth]{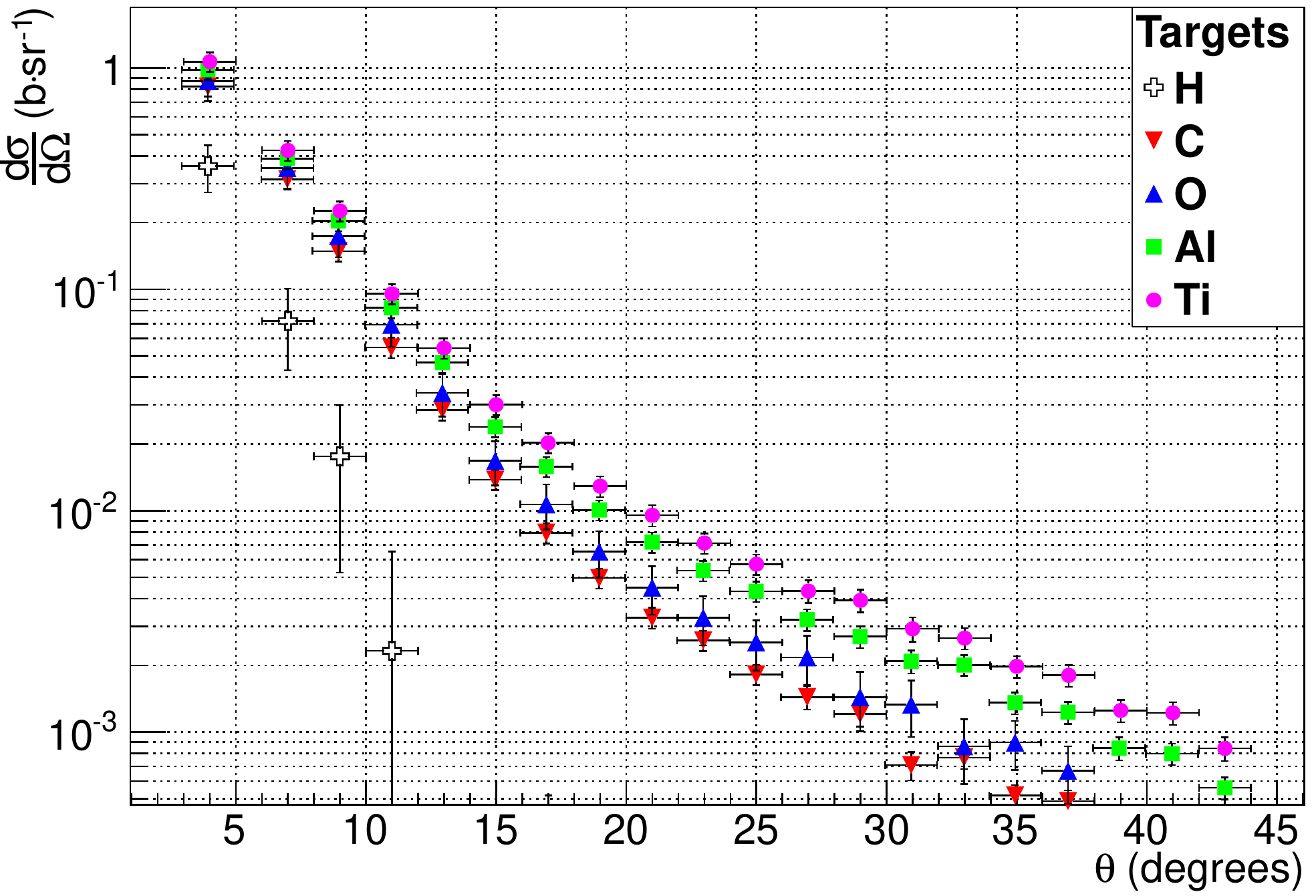}}}\\
\subfigure[$^{11}$B]{\label{11B}{\includegraphics[width=0.49\linewidth]{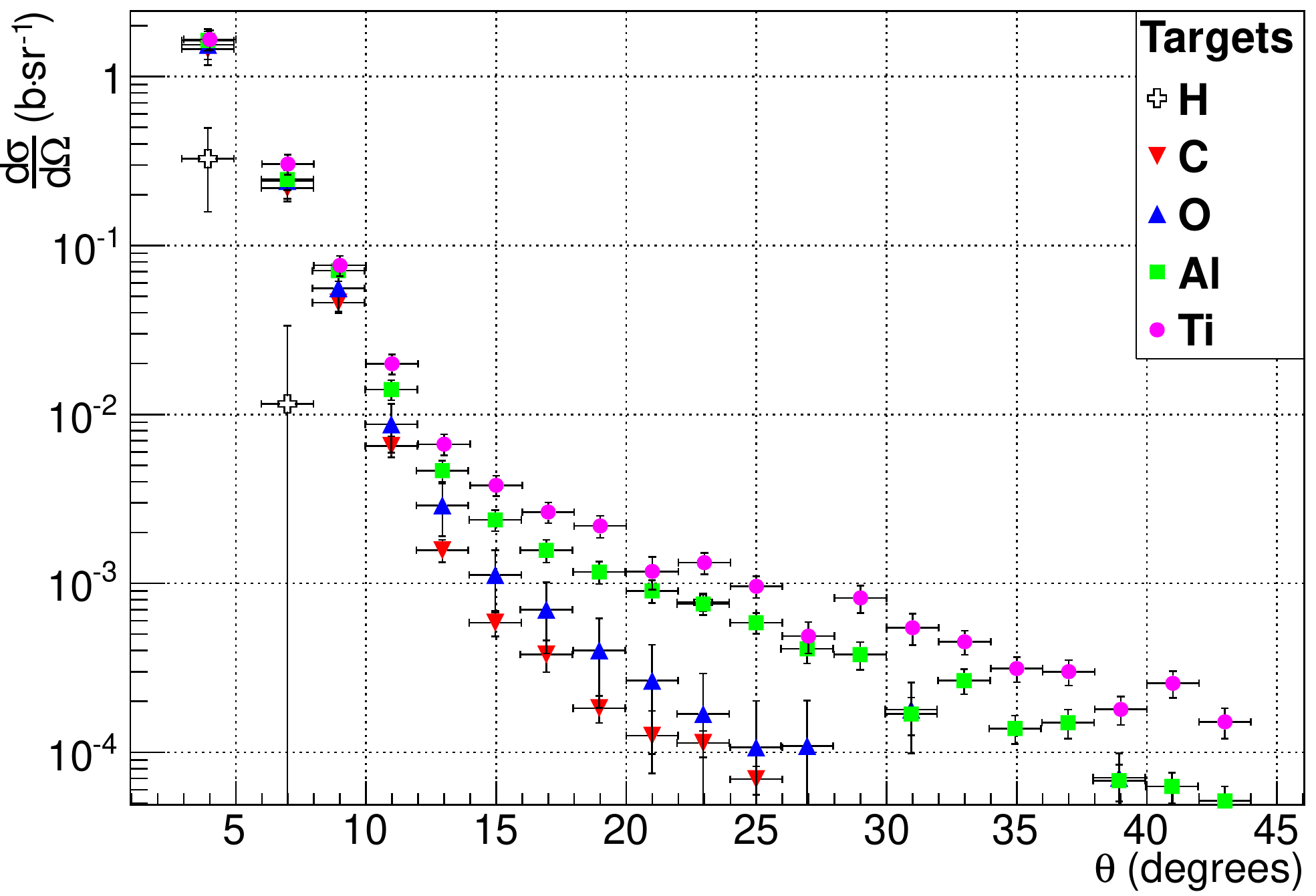}}}
\subfigure[$^{11}$C]{\label{11C}{\includegraphics[width=0.49\linewidth]{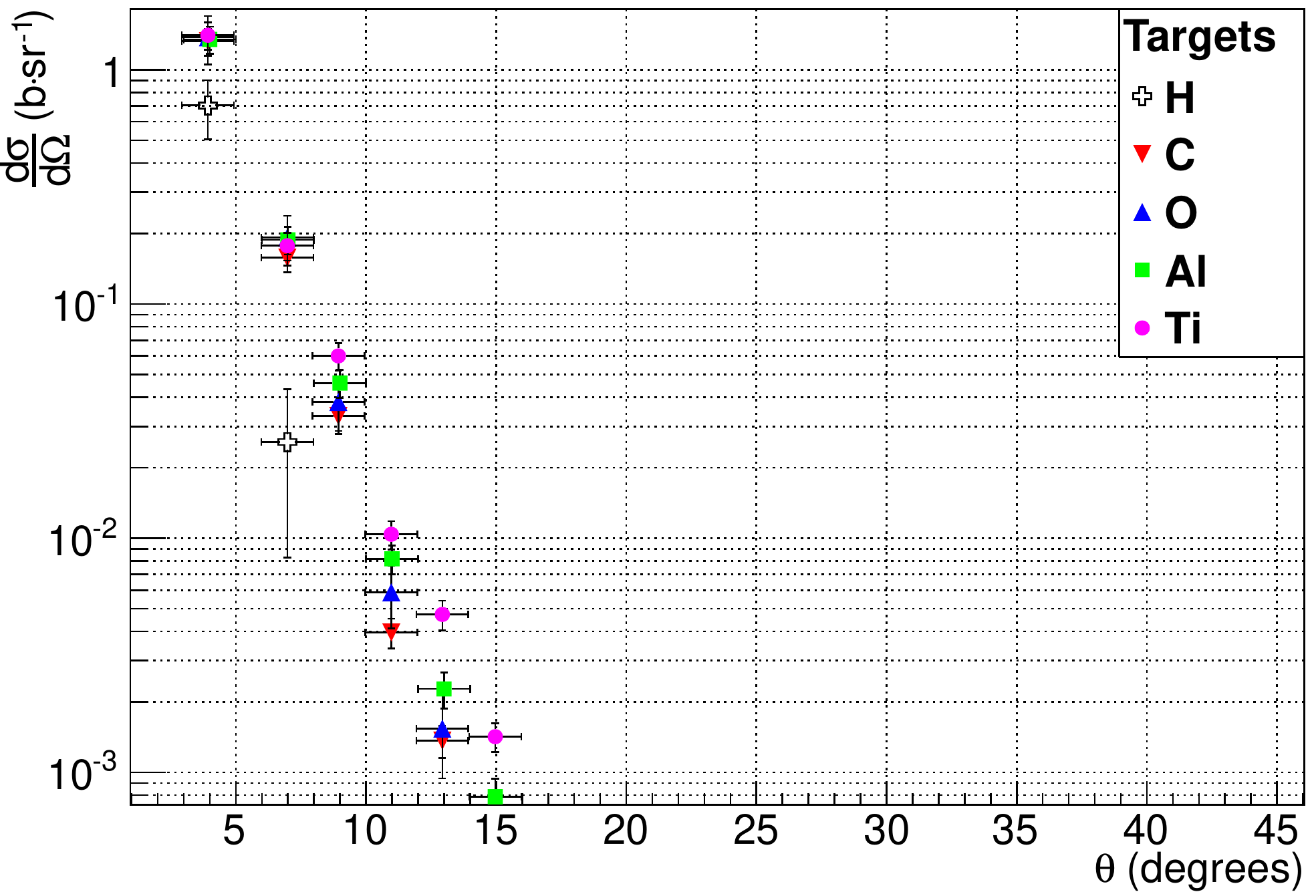}}}
\caption{Angular distributions for the most produced fragments of each Z and each elemental targets.}
\label{C_all_targets_distrib}
\end{figure}

Fig.~\ref{C_all_targets_distrib} presents the angular differential cross sections of several isotopes (the most produced of each Z) obtained for the different elemental targets: H, C, O, Al and $^{nat}$Ti. 

\subsubsection{Production cross sections per isotope}

The angular distributions are in a first approximation well described by a gaussian distribution at forward angles corresponding to fragments emitted by the quasi projectile. For larger angles, the data are better reproduced by an exponential distribution corresponding to a mid-rapidity emission. The need for this second component has been observed before by Golovkov and Matsufuji\cite{Golovkov97,Matsufuji05}. The angular distribution cross sections obtained for all isotopes have been fitted by a function resulting of the sum of a gaussian and an exponential function. The resulting function is described as follows :

\begin{align}
\frac{d\sigma}{d \Omega} = a \times exp{\left(-\frac{(\theta-b)^{2}}{2\cdot c^{2}}\right)} + 
\begin{cases}
d\times exp(e\times \theta)  ~ &\text{if } (\theta \geq 0)\\
d\times exp(-e\times \theta) ~ &\text{if } (\theta <    0)
\end{cases}
\label{fit_func}
\end{align}

where a, b, c, d and e are free parameters corresponding respectively to the gaussian constant, mean and sigma and to the exponential constant and slope.

\begin{figure}[H]
\subfigure[$^{1}$H distribution for the carbon target.]{\label{C_1H}{\includegraphics[width=0.49\linewidth]{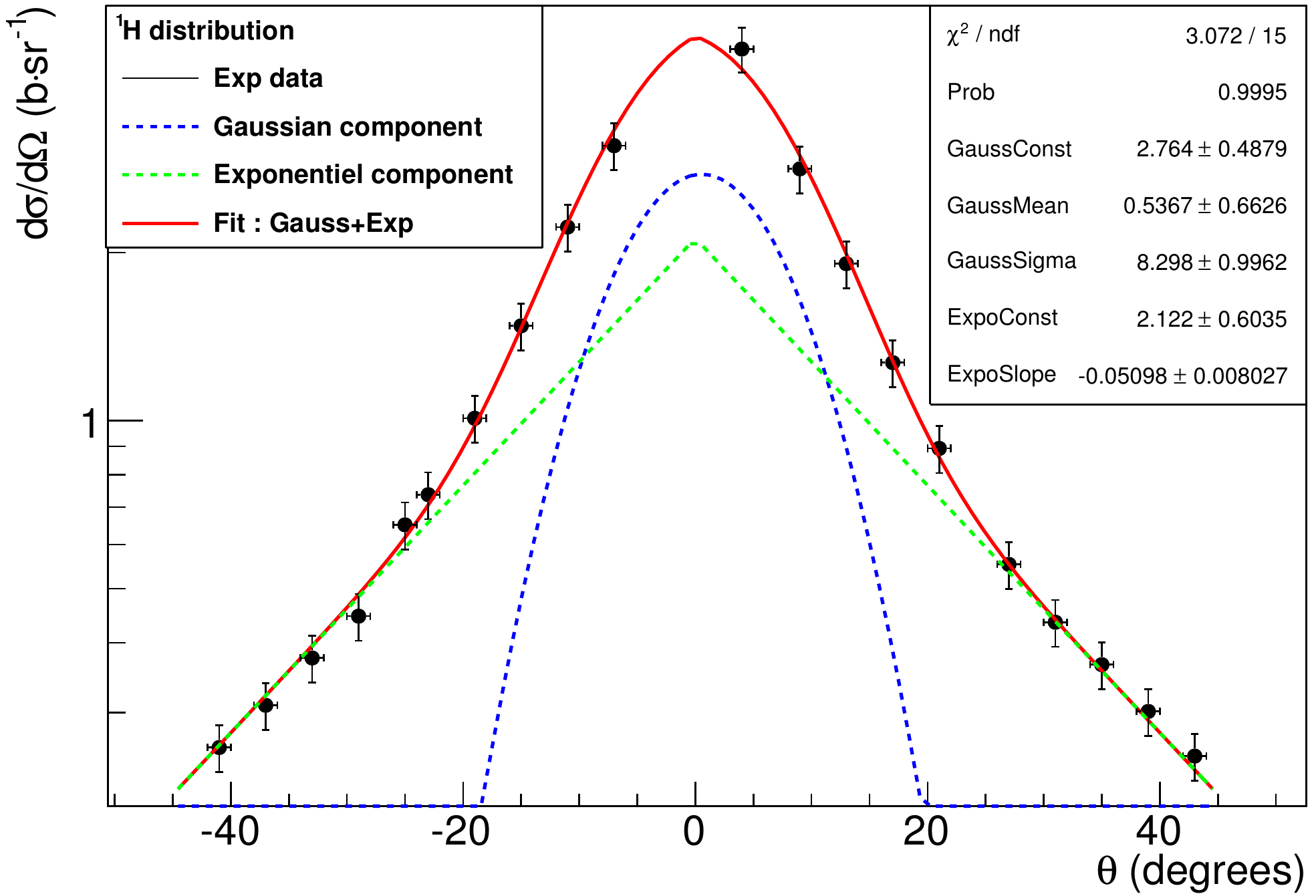}}}
\subfigure[$^{4}$He distribution for the carbon target.]{\label{C_4He}{\includegraphics[width=0.49\linewidth]{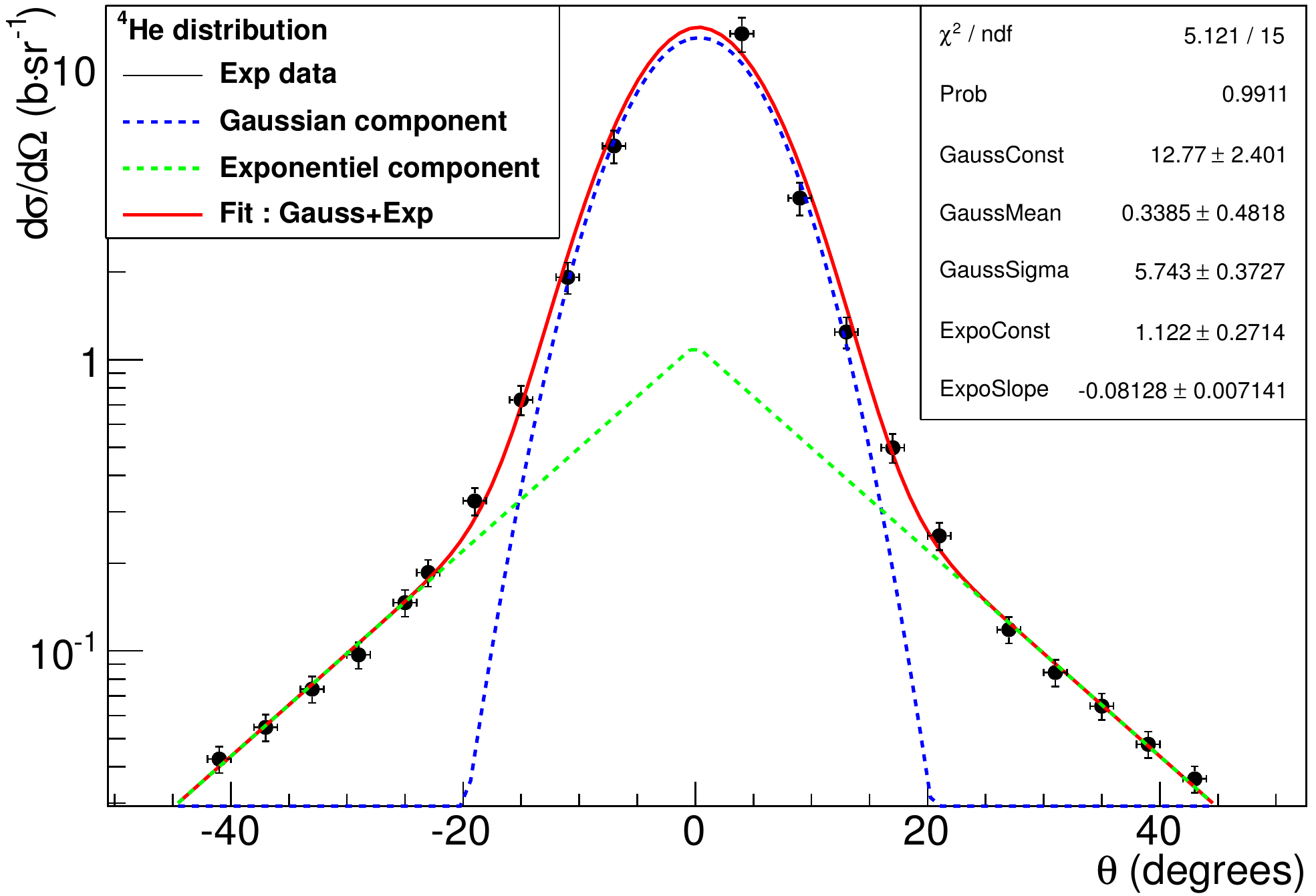}}}\\
\subfigure[$^{1}$H distribution for the hydrogen target.]{\label{H_1H}{\includegraphics[width=0.49\linewidth]{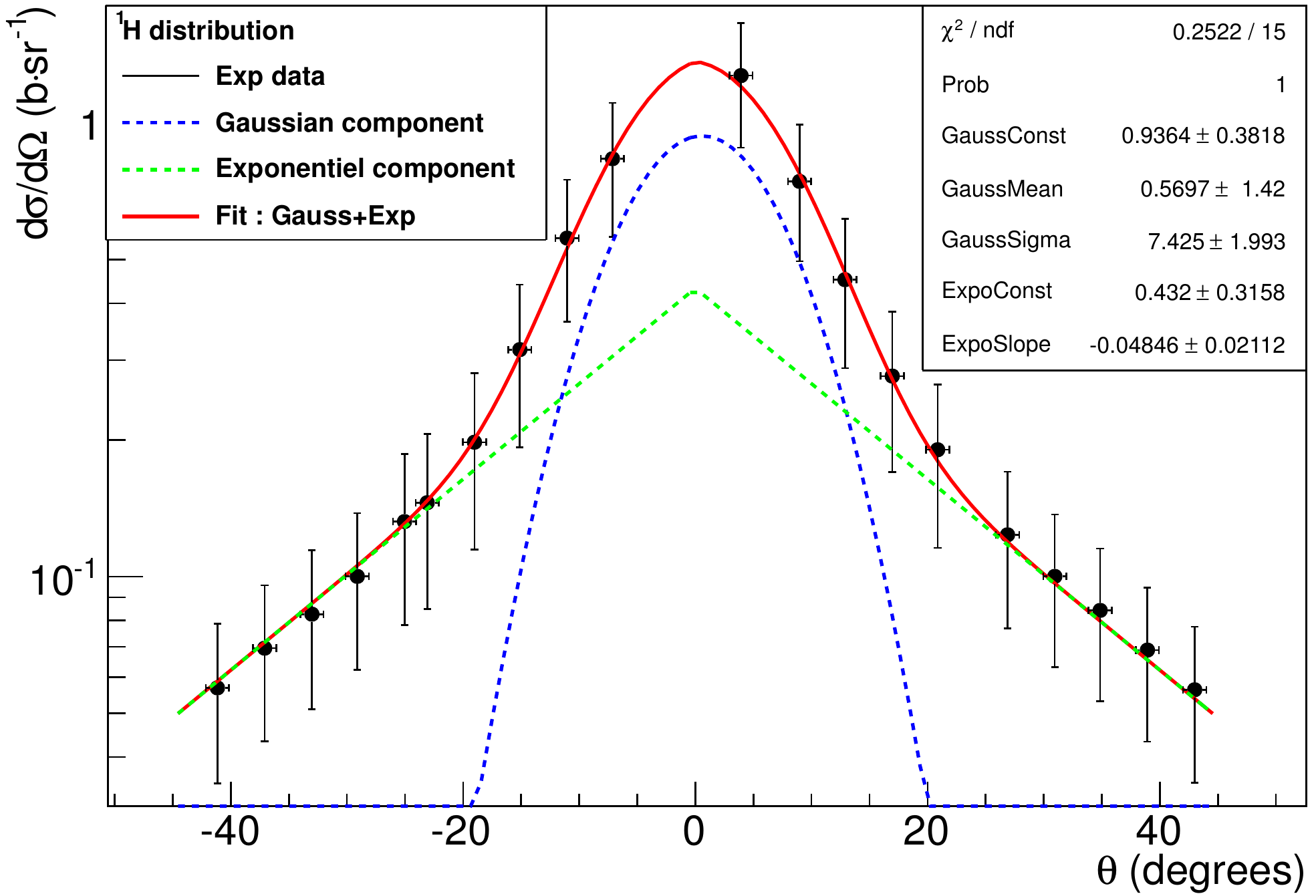}}}
\subfigure[$^{4}$He distribution for the hydrogen target.]{\label{H_4He}{\includegraphics[width=0.49\linewidth]{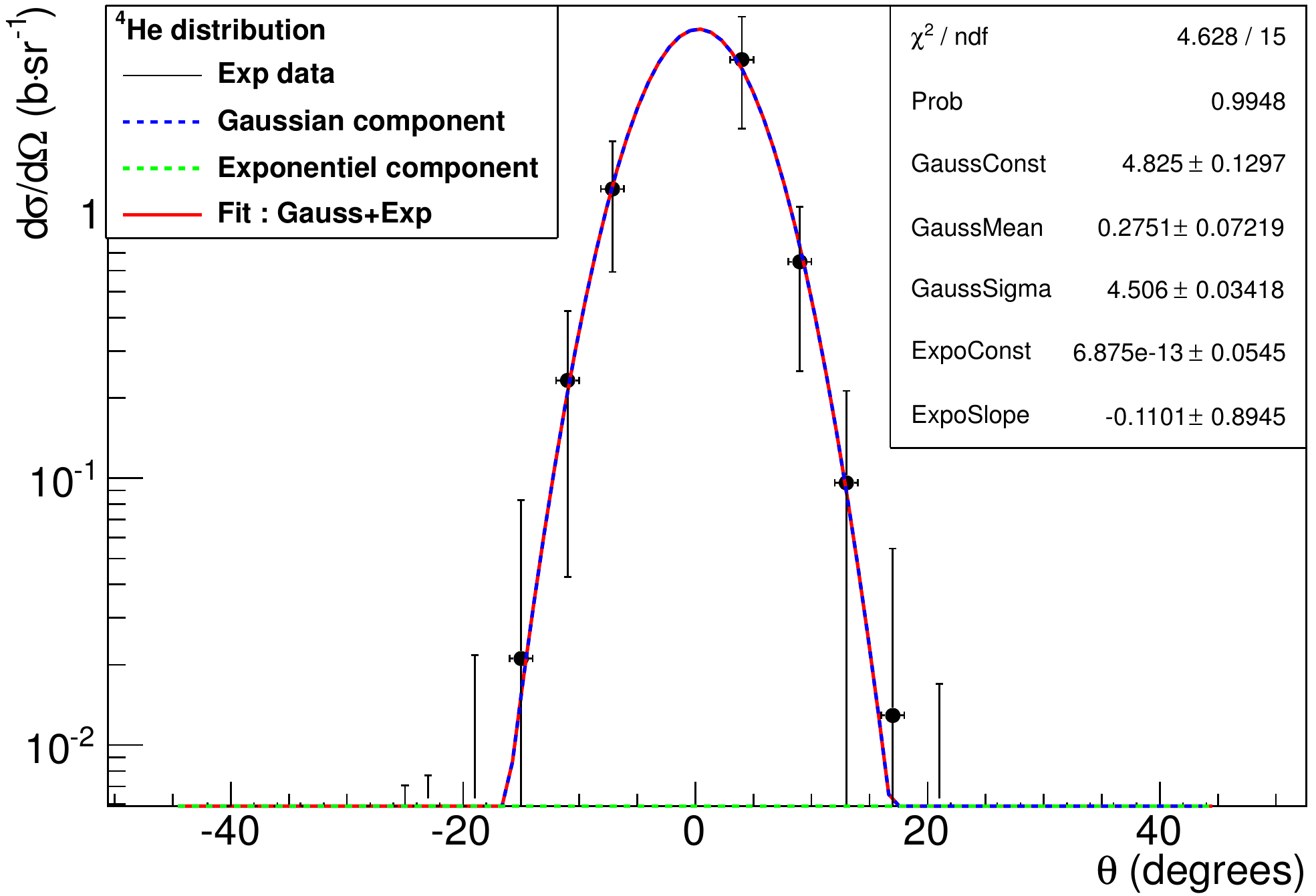}}}
\caption{Angular distributions of the protons and $\alpha$ production for the carbon and hydrogen targets. The distributions have been fitted by a sum of a gaussian and an exponential function.}
\label{fit_dist}
\end{figure}

Fig \ref{fit_dist} represents an example of this fit for proton and $\alpha$ production for the carbon and hydrogen targets. The agreement between the fits and the data is very good for all isotopes and targets. The heavier the fragments are, the most important the contribution of the gaussian term is (quasi-projectile contribution) and the heavier the targets are, the most important the contribution of the exponential term is (mid-rapidity contribution). It has to be noted that the mid-rapidity component is not present for Z>1 fragments in the interactions with the hydrogen target.

By integrating the fitted distribution (cf. Equ. \ref{int_fit_func}), it is then possible to determine the production cross sections of each fragment and and for each target. The results of these integrations are summarized in Table \ref{cross_sec_per_isotope}. Regarding the hydrogen target, it is not possible to integrate the distributions for fragments heavier than $^{7}$Be. Indeed, in such cases, the gaussian width become too small to be fitted with our experimental measurements.

\begin{align}
\sigma = \int_\Omega \frac{d\sigma}{d\Omega} \, \mathrm d\Omega = 2\pi\times\int_0^\pi \frac{d\sigma}{d\Omega} \, sin(\theta) \, \mathrm  d\theta
\label{int_fit_func}
\end{align}

\begin{table}[H]
\begin{center}
{\setlength{\tabcolsep}{1mm}
{\renewcommand{\arraystretch}{1.5}
\begin{tabular}{|c||c|c|c|c|c|}
\cline{2-6}
\multicolumn{1}{c||}{}& \multicolumn{5}{c|}{$\sigma$ (barns)}\\
\multicolumn{1}{c||}{} 	& \multicolumn{5}{c|}{Targets} \\
\hline
Fragments & 		H	&	C		&	O		&	Al		&	Ti	\\
\hline
\hline
$^{1}$H	&$4.1(0.6).10^{-1}$	&$1.7(0.1)$		&$2.1(0.2)$		&$3.0(0.2)$		&$4.4(0.3)$		\\
\hline
$^{2}$H &$7.3(1.5).10^{-2}$	&$7.9(0.4).10^{-1}$	&$9.3(0.8).10^{-1}$	&$1.3(0.1)$		&$1.9(0.1)$		\\
\hline
$^{3}$H	&$1.6(0.3).10^{-2}$	&$3.2(0.2).10^{-1}$	&$3.6(0.4).10^{-1}$	&$5.2(0.3).10^{-1}$	&$7.7(0.4).10^{-1}$		\\
\hline
$^{3}$He&$4.1(0.4).10^{-2}$	&$3.3(0.6).10^{-1}$	&$4.0(0.5).10^{-1}$	&$5.1(0.8).10^{-1}$ 	&$6.7(1.0).10^{-1}$		\\
\hline
$^{4}$He&$2.0(0.8).10^{-1}$	&$1.2(0.3)$		&$1.2(0.4)$		&$1.7(0.1)$		&$2.0(0.2)$		\\
\hline
$^{6}$He&$1.0(0.1).10^{-2}$	&$4.9(1.2).10^{-2}$	&$5.7(2.2).10^{-2}$	&$7.0(1.4).10^{-2}$	&$8.7(1.5).10^{-2}$		\\
\hline
$^{6}$Li&$1.2(0.6).10^{-2}$	&$6.8(1.1).10^{-2}$	&$7.4(1.5).10^{-2}$	&$9.4(1.3).10^{-2}$	&$1.2(0.2).10^{-1}$		\\
\hline
$^{7}$Li&$8.0(1.8).10^{-3}$	&$6.1(1.0).10^{-2}$	&$6.6(1.4).10^{-2}$	&$8.6(1.3).10^{-2}$	&$1.1(0.1).10^{-1}$		\\
\hline
$^{7}$Be&$1.6(0.3).10^{-2}$	&$5.0(0.8).10^{-2}$	&$5.5(1.0).10^{-2}$	&$6.7(1.0).10^{-2}$	&$7.6(1.1).10^{-2}$		\\
\hline
$^{9}$Be&-		     	&$1.8(0.7).10^{-2}$	&$1.9(0.4).10^{-2}$	&$2.3(0.5).10^{-2}$	&$3.0(0.5).10^{-2}$		\\
\hline
$^{10}$Be&-			&$9.3(2.0).10^{-3}$	&$1.0(0.3).10^{-2}$	&$1.2(0.3).10^{-2}$	&$1.5(0.3).10^{-2}$		\\
\hline
$^{8}$B &-			&$6.1(1.8).10^{-3}$	&$6.9(2.7).10^{-3}$	&$7.8(2.1).10^{-3}$	&$8.5(6.3).10^{-3}$		\\
\hline
$^{10}$B&-			&$4.7(1.5).10^{-2}$	&$5.0(3.3).10^{-2}$	&$5.3(1.6).10^{-2}$	&$6.2(1.8).10^{-2}$		\\
\hline
$^{11}$B&-			&$6.0(2.4).10^{-2}$	&$6.3(6.2).10^{-2}$	&$6.8(3.9).10^{-2}$	&$7.1(2.4).10^{-2}$		\\
\hline
$^{10}$C&-			&$8.2(3.0).10^{-3}$	&$8.5(5.4).10^{-3}$	&$9.3(3.3).10^{-3}$	&$1.1(0.4).10^{-2}$		\\
\hline	
$^{11}$C&-			&$5.3(2.2).10^{-2}$	&$5.5(3.7).10^{-2}$	&$5.5(2.1).10^{-2}$	&$5.8(3.5).10^{-2}$		\\
\hline
$^{12}$C&-			&$7.6(4.4).10^{-2}$	&$8.1(5.0).10^{-2}$	&$8.0(3.9).10^{-2}$	&$7.6(3.3).10^{-2}$		\\
\hline
\end{tabular}}}
\end{center}
\caption{Production cross sections per isotope and for each elemental target.}
\label{cross_sec_per_isotope}
\end{table}

\subsubsection{Energy distributions}

\begin{figure}[H]
\centerline{\includegraphics[width=0.45\linewidth]{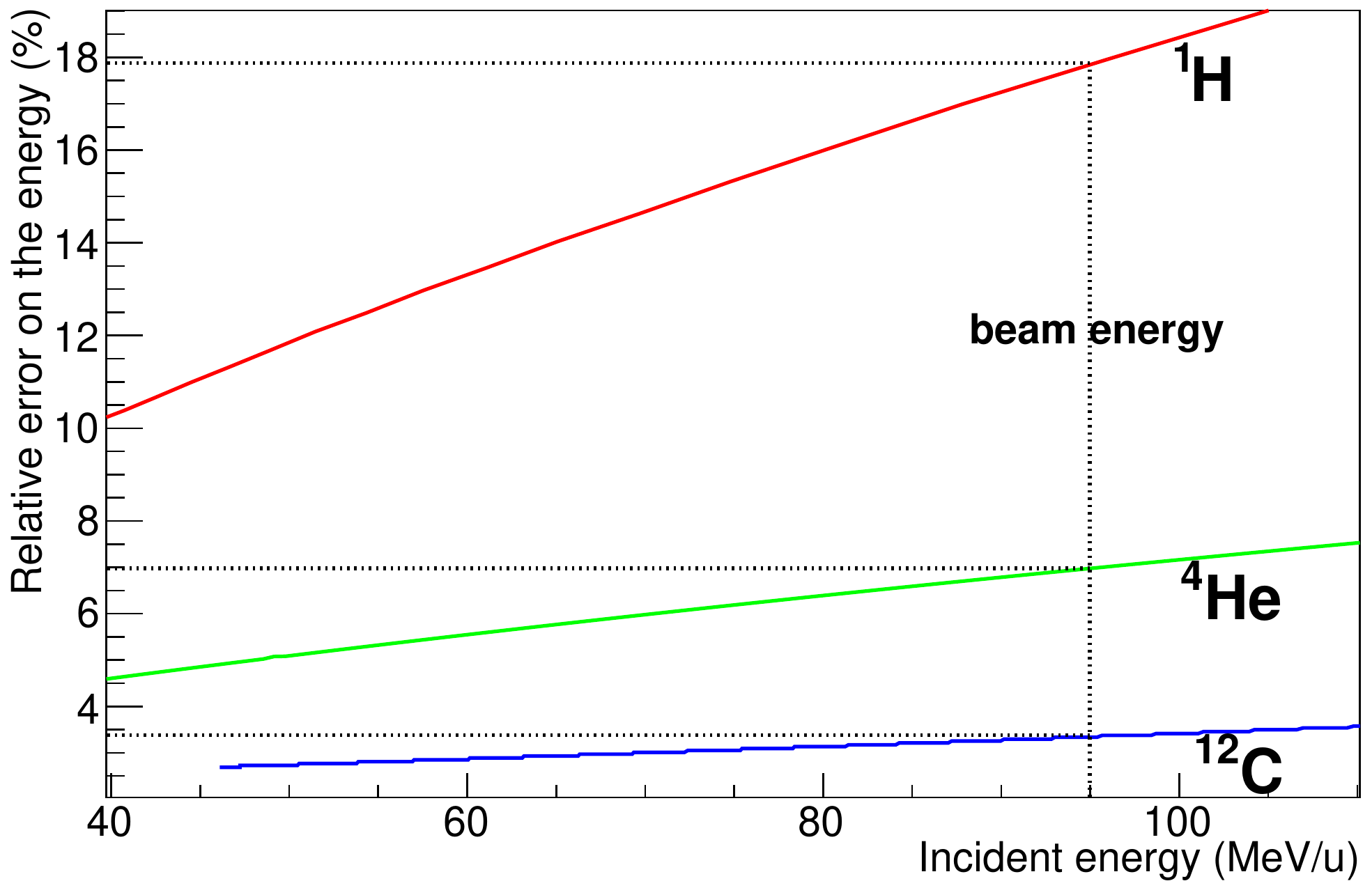}}
\caption{Estimated errors on the energy due to the calibration method used with the thick silicon detector.}
\label{Err_E_CsI}
\end{figure}

To obtain the double differential cross section, the energy distributions need to be determined at each point of the previous angular distributions. The energies of the fragments are obtained from the energy deposited in the thick silicon stage. The calibration method was made with the KaliVeda toolkit as described previously~\cite{Dudouet12}.  The advantage of such a method is that the energy calibration of the non-linear response for the CsI crystal is not needed, but propagates the error off the thick silicon energy calibration (particularly for low energies in the thick silicon which correspond to high energy protons). The errors made on the energy calibration depend on the nature of the particle. They are summarized in Fig.~\ref{Err_E_CsI} for several isotopes. They have been estimated from about 20\% for 95~MeV protons to $\sim$3-4\% for 95~MeV/u $^{12}$C.

\begin{figure}[H]
\subfigure[$^{4}$He energy distribution for the carbon target at different angles.]{{\includegraphics[width=0.49\linewidth]{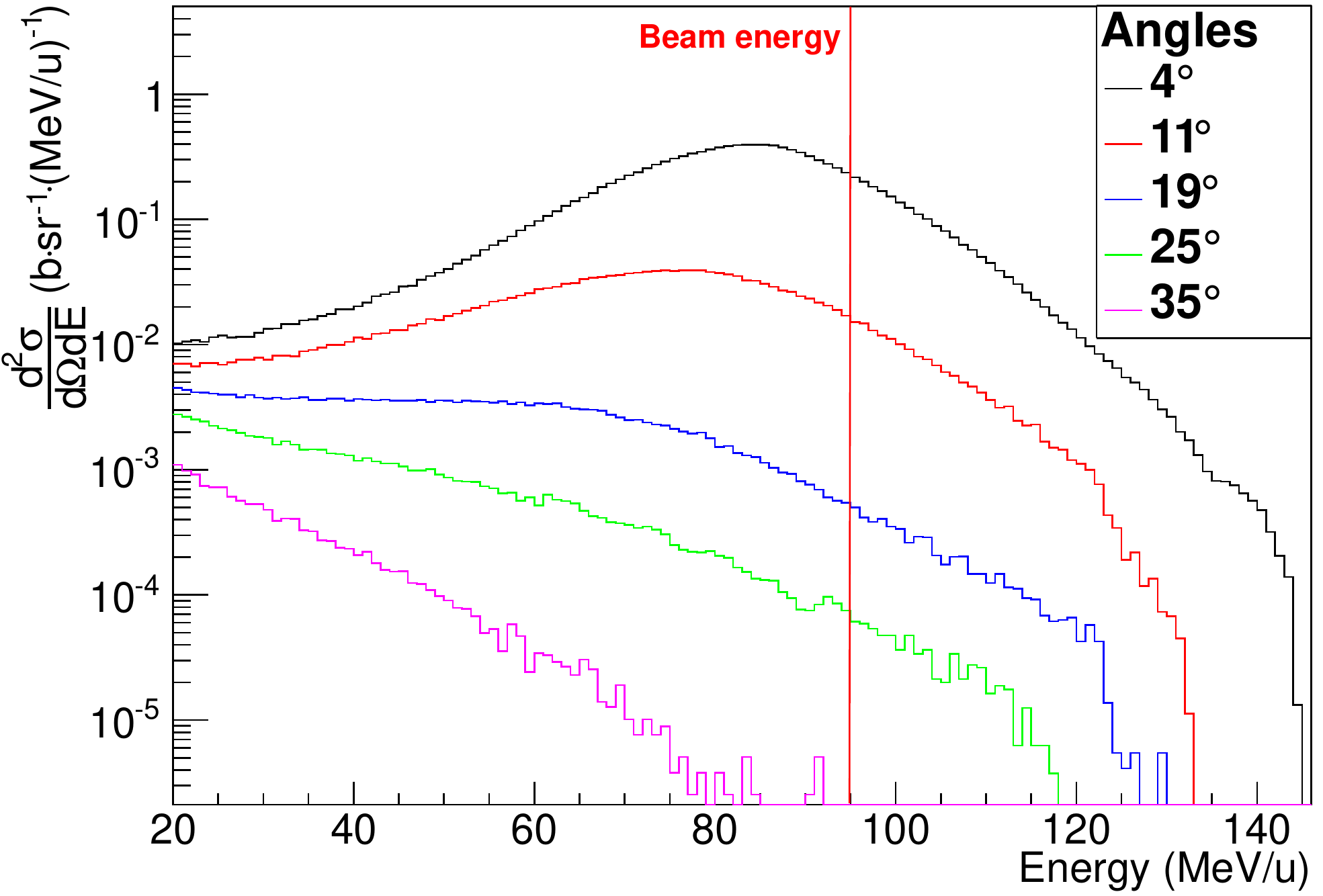}}}
\subfigure[$^{4}$He energy distribution at 4 degrees for different targets.]{{\includegraphics[width=0.49\linewidth]{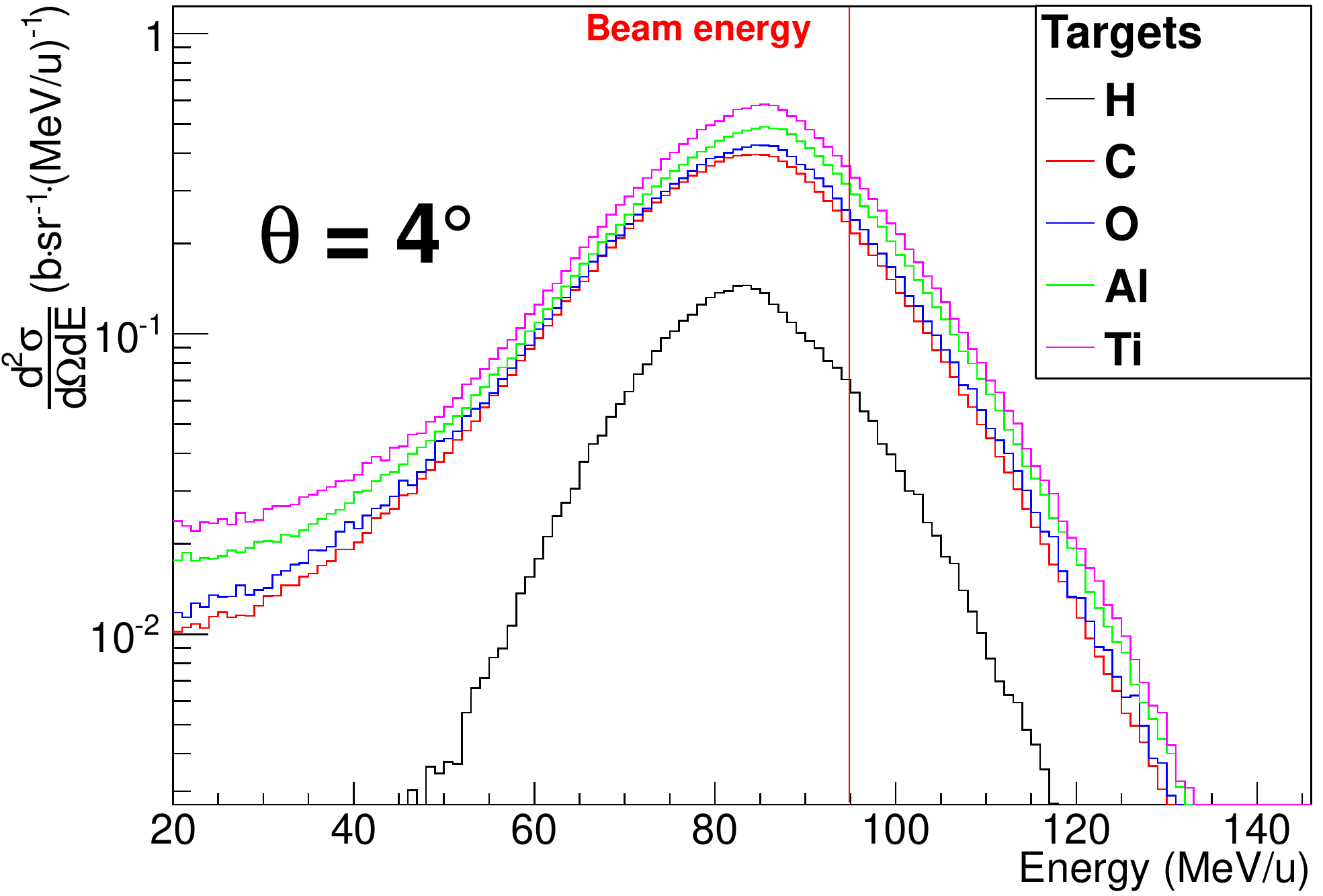}}}\\
\subfigure[$^{6}$Li energy distribution for the carbon target at different angles.]{{\includegraphics[width=0.49\linewidth]{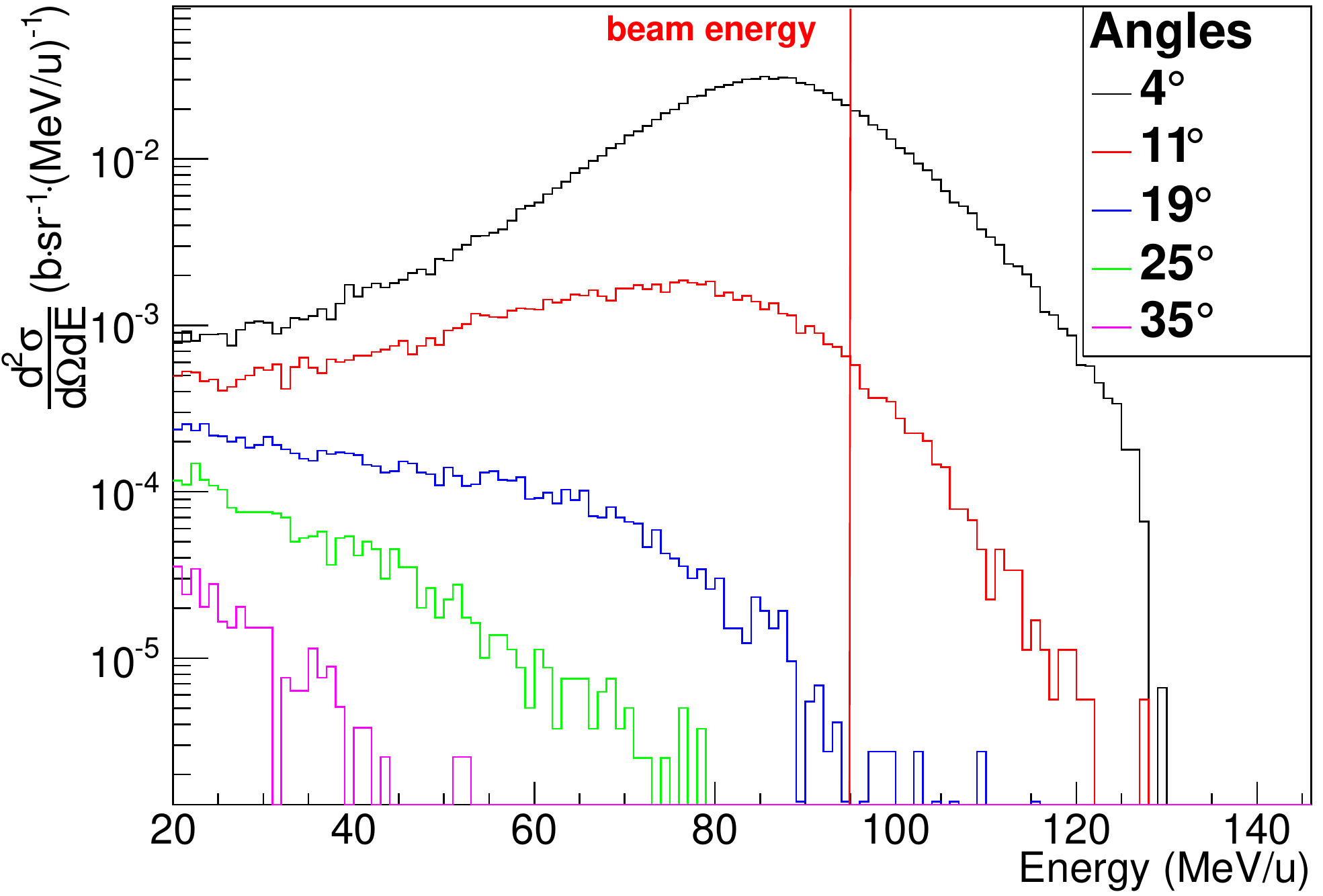}}}
\subfigure[$^{6}$Li energy distribution at 4 degrees for different targets.]{{\includegraphics[width=0.49\linewidth]{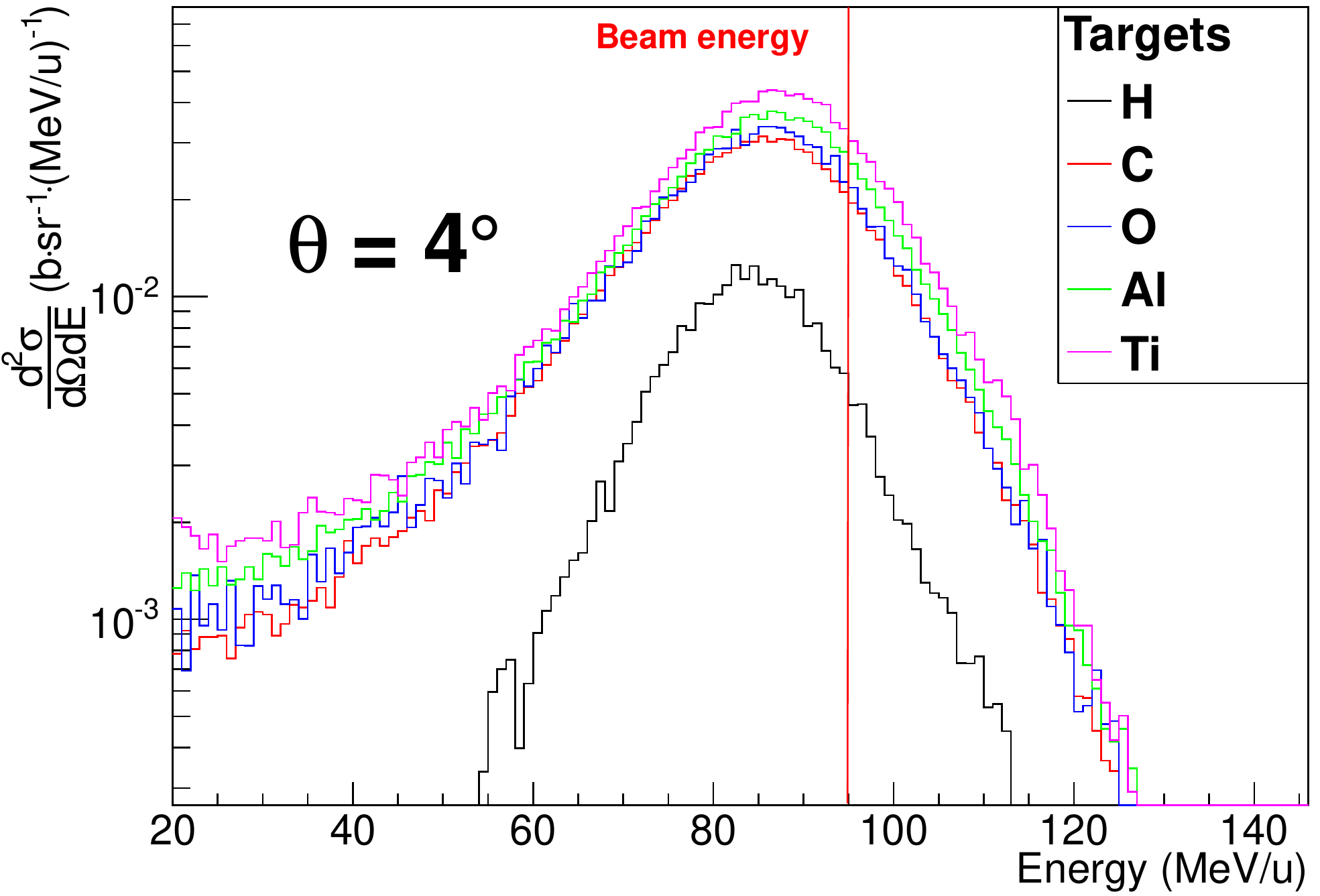}}}
\caption{$^{4}$He and $^{6}$Li energy distributions.}
\label{E_dist}
\end{figure}

An example of energy distributions is shown in Fig.~\ref{E_dist}. On the top are represented the $^{4}$He energy distributions, at different angles for the carbon target on the left hand side and, on different targets at 4$^\circ$ on the right hand side. The same figures are represented below for $^{6}$Li fragments.

These distributions are dominated at small angles by a peak centered close to the beam energy (94.6 MeV/u). The energy and amplitude of this peak decrease with the emission angle and the width of this peak decreases with the increasing mass of the fragments. This observation confirms that most of the detected fragments are coming from the projectile fragmentation. 

Concerning the different targets, as for the angular distributions, the cross sections increase with the charge and mass of the target. Moreover, the heavier the target, the higher the cross sections at low energy are.

\subsubsection{PMMA reconstruction}

To check the validity of the target cross sections combination method, the results obtained with the PMMA target will be used. The PMMA chemical composition is C$_5$H$_8$O$_2$, by combining the obtained carbon, hydrogen and oxygen target cross sections, we should be able to reproduce the cross sections for the PMMA target as follow :

\begin{align}
 \frac{d \sigma}{d\Omega}(\text{PMMA}) &= 5 \times \frac{d \sigma}{d\Omega}(\text{C}) + 8 \times \frac{d \sigma}{d\Omega}(\text{H}) + 2 \times \frac{d \sigma}{d\Omega}(\text{O}).
\end{align}

\begin{figure}[H]
\subfigure[Angular distributions for different isotopes]{{\includegraphics[width=0.49\linewidth]{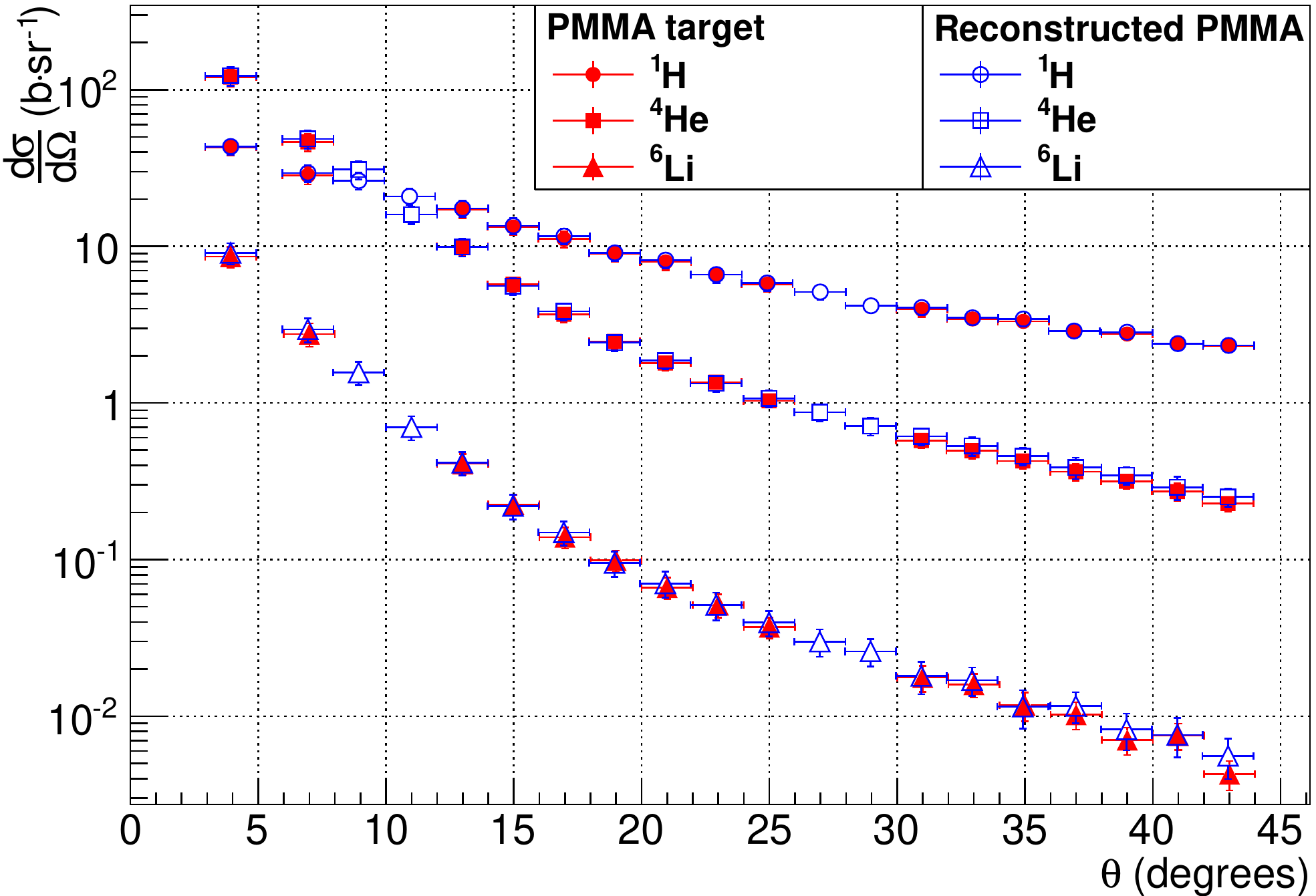}}}
\subfigure[$^{4}$He energy distributions at different angles.]{{\includegraphics[width=0.49\linewidth]{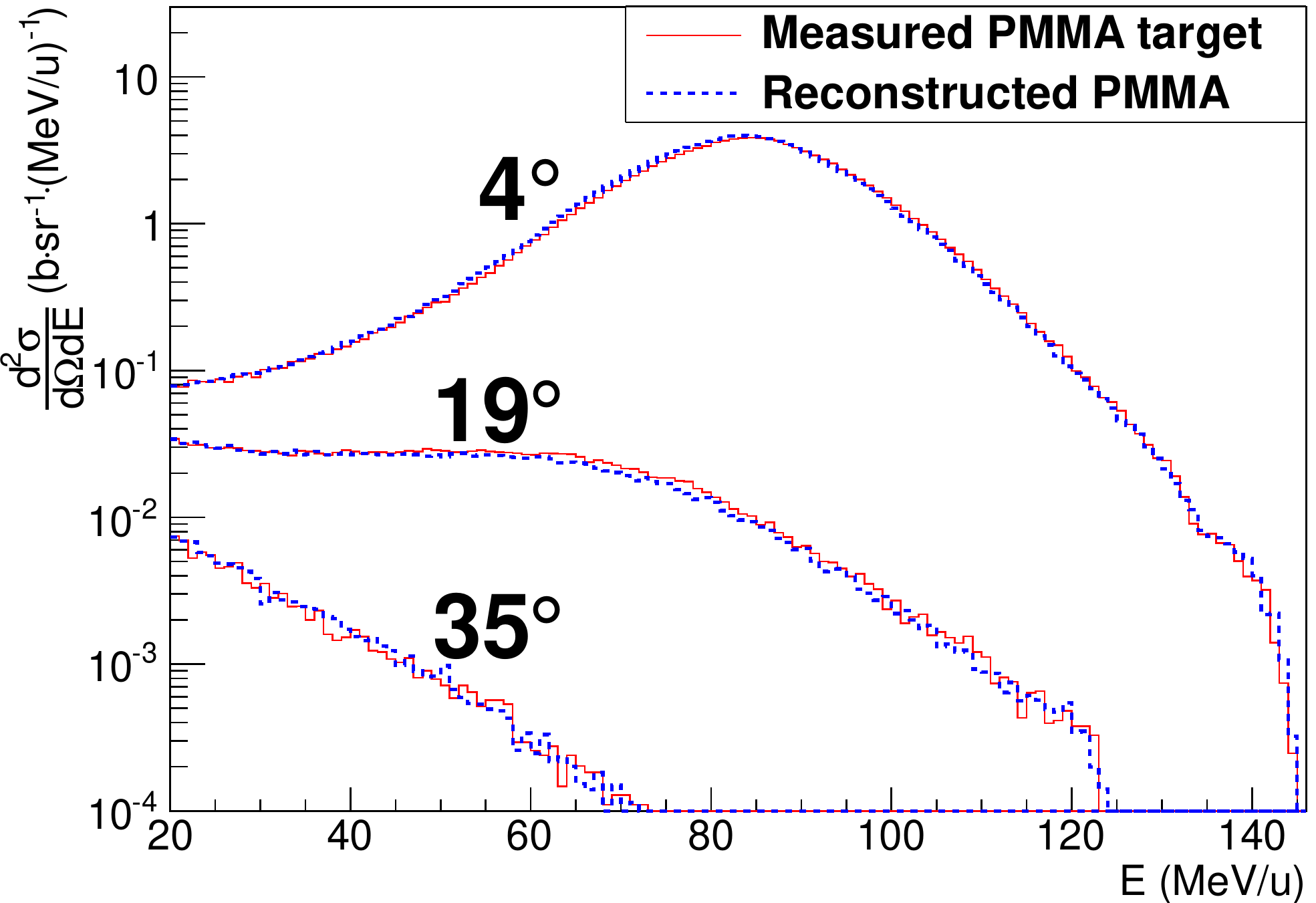}}}
\caption{Comparisons of the differential cross sections for the PMMA target and the reconstructed PMMA target.}
\label{PMMA_compare}
\end{figure}

Comparisons between the cross sections measured with the PMMA target and the calculated ones from elemental target cross sections are shown in Fig.~\ref{PMMA_compare}. On the left hand side are represented the angular distributions for protons, $^{4}$He and $^{6}$Li fragments. On the right hand side are represented the $^{4}$He energy distributions at 4, 19 and 35$^\circ$. The data at 9, 11, 27 and 39$^\circ$ are missing for the PMMA target, explaining the holes in the PMMA angular distributions.

The results between the real target and the reconstructed one are very close to each other. The agreement between these two targets is about 1 to 5\% both for angular and energy distributions.

This last result confirms that the data are of good quality and overall, validate the target combination method. Thus, with the cross sections presented in this article, most of the human organic tissues cross sections can be reproduced at 95 MeV/u.

\section{Conclusion}

Measurements for twenty different angles and five different targets led to the double differential fragmentation cross sections of $^{12}$C on hydrogen, carbon, oxygen, aluminum and titanium in fragmented particles, ranging from protons to carbon ion isotopes.

The angular distributions are dominated by the emission of light fragments (Z $<$ 3), more especially by $\alpha$  particles at forward angles, which is consistent with the $\alpha$ cluster structure of the $^{12}$C ion. These  results are in agreement  with previous experiments with thick targets~\cite{Braunn11} and at lower energies~\cite{Catane12}. The production rates and the emission angles are increasing when the charge and mass of the target are increasing. 

Energy distributions are dominated at low angles by a peak close to the beam energy, while the amplitude and energy mean value of the peak decrease when the angle of emission increases. The production of low energy particles increases with the target charge and mass. These observations indicate that most of the emitted particles result from the projectile fragmentation during nuclear reactions.

The presented experimental results will contribute directly to constrain the nuclear reaction models used in different Monte-Carlo simulation toolkits as GEANT4 or FLUKA.

Moreover, the method used to combine the cross sections of composite targets to extract an elemental target (CH$_2$ and C to extract H) has been validated. By comparing the double differential cross sections of the real PMMA target with the one of the reconstructed PMMA target (result of combination of Al$_2$O$_3$, Al, CH$_2$ and C target), it gives us the information that most of the human body tissue cross sections are reproducible using this method.

In order to complete these data, new measurements at zero degree for the different thin targets are planned.

\medskip

All the experimental results including all angular and energy distributions for each target are in free access on the web-site \url{http://hadrontherapy-data.in2p3.fr}.

\newpage 

\appendix

\section{Cross section for the hydrogen target}

\begin{table}[H]
\begin{center}
\begin{tabular}{|c|c|c|c|c|c|c|}
\hline
$\theta$ & $^{1}$H & $^{2}$H & $^{3}$H & $^{3}$He & $^{4}$He & $^{6}$He \\ 
(Degrees)  & $d\sigma/d\Omega$ (b.sr$^{-1}$) & $d\sigma/d\Omega$ (b.sr$^{-1}$) & $d\sigma/d\Omega$ (b.sr$^{-1}$) & $d\sigma/d\Omega$ (b.sr$^{-1}$) & $d\sigma/d\Omega$ (b.sr$^{-1}$) & $d\sigma/d\Omega$ (b.sr$^{-1}$) \\ 
\hline
\hline
4(1)        & 1.26(0.39)        & 4.3(2.3).10$^{-1}$        & 1.6(1.1).10$^{-1}$        & 4.5(4.0).10$^{-1}$        & 3.7(1.7)        & 1.6(3.0).10$^{-1}$ \\ 
\hline
7(1)        & 8.3(2.7).10$^{-1}$        & 2.6(1.5).10$^{-1}$        & 1.09(0.85).10$^{-1}$        & 2.8(2.0).10$^{-1}$        & 1.20(0.62)        & 8.66(11.0).10$^{-2}$ \\ 
\hline
9(1)        & 7.4(2.5).10$^{-1}$        & 2.2(1.3).10$^{-1}$        & 7.4(6.5).10$^{-2}$        & 1.9(1.5).10$^{-1}$        & 6.5(4.0).10$^{-1}$        & 3.2(5.2).10$^{-2}$ \\ 
\hline
11(1)        & 5.6(1.9).10$^{-1}$        & 1.47(0.97).10$^{-1}$        & 4.4(4.6).10$^{-2}$        & 1.04(0.82).10$^{-1}$        & 2.3(1.9).10$^{-1}$        & 3.7(15.0).10$^{-3}$ \\ 
\hline
13(1)        & 4.5(1.6).10$^{-1}$        & 1.12(0.81).10$^{-1}$        & 2.5(3.6).10$^{-2}$        & 5.9(5.8).10$^{-2}$        & 9.61(12.0).10$^{-2}$        & 1.1(8.6).10$^{-3}$ \\ 
\hline
15(1)        & 3.1(1.2).10$^{-1}$        & 8.1(6.3).10$^{-2}$        & 1.1(2.6).10$^{-2}$        & 2.9(3.7).10$^{-2}$        & 2.1(6.2).10$^{-2}$        &         -         \\ 
\hline
17(1)        & 2.7(1.1).10$^{-1}$        & 7.6(5.4).10$^{-2}$        & 9.33(21.0).10$^{-3}$        & 2.1(2.8).10$^{-2}$        & 1.3(4.2).10$^{-2}$        &         -         \\ 
\hline
19(1)        & 1.96(0.83).10$^{-1}$        & 5.3(4.2).10$^{-2}$        & 2.4(15.0).10$^{-3}$        & 7.44(19.0).10$^{-3}$     &         -                &         -         \\ 
\hline
21(1)        & 1.90(0.75).10$^{-1}$        & 5.0(3.7).10$^{-2}$        & 1.94(13.0).10$^{-3}$        & 5.08(15.0).10$^{-3}$     &         -                &         -         \\ 
\hline
23(1)        & 1.44(0.61).10$^{-1}$        & 3.4(2.9).10$^{-2}$        &         -             &              -                &         -               &         -         \\ 
\hline
27(1)        & 1.23(0.47).10$^{-1}$        & 2.5(2.1).10$^{-2}$        &         -             &             -                &         -                &         -         \\ 
\hline
31(1)        & 9.9(3.7).10$^{-2}$        & 1.4(1.6).10$^{-2}$        &         -              &             -                &         -                &         -         \\ 
\hline
35(1)        & 8.4(3.1).10$^{-2}$        & 1.0(1.3).10$^{-2}$        &          -             &            -                &         -                &         -         \\ 
\hline
39(1)        & 6.9(2.6).10$^{-2}$        &         -                &         -                &            -                &         -                &         -         \\ 
\hline
\end{tabular}

\begin{tabular}{|c|c|c|c|c|c|c|}
\hline
$\theta$ & $^{6}$Li & $^{7}$Li & $^{7}$Be & $^{9}$Be & $^{10}$Be & $^{8}$B \\ 
(Degrees)  & $d\sigma/d\Omega$ (b.sr$^{-1}$) & $d\sigma/d\Omega$ (b.sr$^{-1}$) & $d\sigma/d\Omega$ (b.sr$^{-1}$) & $d\sigma/d\Omega$ (b.sr$^{-1}$) & $d\sigma/d\Omega$ (b.sr$^{-1}$) & $d\sigma/d\Omega$ (b.sr$^{-1}$) \\ 
\hline
\hline
4(1)        & 2.7(1.2).10$^{-1}$        & 1.8(1.0).10$^{-1}$        & 3.59(0.86).10$^{-1}$        & 6.4(3.1).10$^{-2}$        & 1.9(1.9).10$^{-2}$        & 3.5(2.8).10$^{-2}$ \\ 
\hline
7(1)        & 6.2(3.9).10$^{-2}$        & 4.4(3.7).10$^{-2}$        & 7.2(2.9).10$^{-2}$        & 1.5(1.1).10$^{-2}$        & 5.2(6.8).10$^{-3}$        & 8.25(12.0).10$^{-3}$ \\ 
\hline
9(1)        & 1.9(2.1).10$^{-2}$        & 9.65(18.0).10$^{-3}$        & 1.8(1.2).10$^{-2}$        & 1.7(3.6).10$^{-3}$        & -3.06(20.0).10$^{-4}$        & 9.04.0(33.0).10$^{-4}$ \\ 
\hline
11(1)        & 4.4(9.0).10$^{-3}$        & 1.0(7.2).10$^{-3}$        & 2.3(4.2).10$^{-3}$        & 2.7(8.4).10$^{-4}$        & 3.4(4.3).10$^{-4}$        & 2.4(9.0).10$^{-4}$ \\ 
\hline
\end{tabular}

\begin{tabular}{|c|c|c|c|c|c|}
\hline
$\theta$ & $^{10}$B & $^{11}$B & $^{10}$C & $^{11}$C & $^{12}$C \\ 
(Degrees)  & $d\sigma/d\Omega$ (b.sr$^{-1}$) & $d\sigma/d\Omega$ (b.sr$^{-1}$) & $d\sigma/d\Omega$ (b.sr$^{-1}$) & $d\sigma/d\Omega$ (b.sr$^{-1}$) & $d\sigma/d\Omega$ (b.sr$^{-1}$) \\ 
\hline
\hline
4(1)        & 4.1(1.4).10$^{-1}$        & 3.2(1.7).10$^{-1}$        & 1.06(0.49).10$^{-1}$        & 7.0(2.0).10$^{-1}$        & 2.0(1.9).10$^{-1}$ \\ 
\hline
7(1)        & 3.5(2.3).10$^{-2}$        & 1.1(2.2).10$^{-2}$        & 7.5(6.5).10$^{-3}$        & 2.6(1.7).10$^{-2}$        &         -         \\ 
\hline
9(1)        & 1.7(6.2).10$^{-3}$        &         -                & 4.4(15.0).10$^{-4}$        &         -                &         -         \\ 
\hline
\end{tabular}
\end{center}
\caption{$^{12}$C fragmentation cross sections for the hydrogen target at different angles. The values in parentheses represent the uncertainties ( 4.55(0.52) is equivalent to 4.55 $\pm$ 0.52)}
\label{Table_cross_section_H_target}
\end{table}

\section{Cross section for the carbon target}

\begin{table}[H]
\begin{center}
\begin{tabular}{|c|c|c|c|c|c|c|}
\hline
$\theta$ & $^{1}$H & $^{2}$H & $^{3}$H & $^{3}$He & $^{4}$He & $^{6}$He \\ 
(Degrees)  & $d\sigma/d\Omega$ (b.sr$^{-1}$) & $d\sigma/d\Omega$ (b.sr$^{-1}$) & $d\sigma/d\Omega$ (b.sr$^{-1}$) & $d\sigma/d\Omega$ (b.sr$^{-1}$) & $d\sigma/d\Omega$ (b.sr$^{-1}$) & $d\sigma/d\Omega$ (b.sr$^{-1}$) \\ 
\hline
\hline
4(1)        & 4.63(0.42)        & 2.63(0.28)        & 1.27(0.14)        & 1.88(0.44)        & 1.32(0.18).10$^{1}$        & 4.7(2.9).10$^{-1}$ \\ 
\hline
7(1)        & 3.11(0.30)        & 1.67(0.18)        & 9.2(1.1).10$^{-1}$        & 1.34(0.23)        & 5.42(0.70)        & 4.3(1.2).10$^{-1}$ \\ 
\hline
9(1)        & 2.82(0.27)        & 1.44(0.16)        & 7.27(0.84).10$^{-1}$        & 10.0(1.7).10$^{-1}$        & 3.59(0.47)        & 2.53(0.64).10$^{-1}$ \\ 
\hline
11(1)        & 2.22(0.21)        & 1.10(0.12)        & 5.23(0.60).10$^{-1}$        & 6.7(1.0).10$^{-1}$        & 1.92(0.24)        & 9.0(2.1).10$^{-2}$ \\ 
\hline
13(1)        & 1.91(0.18)        & 9.3(1.0).10$^{-1}$        & 4.17(0.48).10$^{-1}$        & 5.03(0.73).10$^{-1}$        & 1.24(0.15)        & 5.4(1.2).10$^{-2}$ \\ 
\hline
15(1)        & 1.48(0.14)        & 7.25(0.79).10$^{-1}$        & 3.12(0.36).10$^{-1}$        & 3.42(0.48).10$^{-1}$        & 7.28(0.85).10$^{-1}$        & 2.35(0.52).10$^{-2}$ \\ 
\hline
17(1)        & 1.27(0.12)        & 6.19(0.68).10$^{-1}$        & 2.49(0.29).10$^{-1}$        & 2.64(0.36).10$^{-1}$        & 4.98(0.57).10$^{-1}$        & 1.48(0.33).10$^{-2}$ \\ 
\hline
19(1)        & 1.01(0.097)        & 4.86(0.53).10$^{-1}$        & 1.88(0.22).10$^{-1}$        & 1.86(0.25).10$^{-1}$        & 3.27(0.36).10$^{-1}$        & 7.4(1.6).10$^{-3}$ \\ 
\hline
21(1)        & 8.92(0.86).10$^{-1}$        & 4.26(0.47).10$^{-1}$        & 1.56(0.18).10$^{-1}$        & 1.52(0.21).10$^{-1}$        & 2.48(0.27).10$^{-1}$        & 5.2(1.2).10$^{-3}$ \\ 
\hline
23(1)        & 7.37(0.71).10$^{-1}$        & 3.45(0.38).10$^{-1}$        & 1.23(0.14).10$^{-1}$        & 1.12(0.15).10$^{-1}$        & 1.86(0.20).10$^{-1}$        & 3.22(0.71).10$^{-3}$ \\ 
\hline
27(1)        & 5.53(0.53).10$^{-1}$        & 2.52(0.28).10$^{-1}$        & 8.6(1.0).10$^{-2}$        & 7.4(1.0).10$^{-2}$        & 1.18(0.12).10$^{-1}$        & 1.43(0.33).10$^{-3}$ \\ 
\hline
31(1)        & 4.36(0.42).10$^{-1}$        & 1.94(0.21).10$^{-1}$        & 6.42(0.74).10$^{-2}$        & 5.14(0.70).10$^{-2}$        & 8.43(0.88).10$^{-2}$        & 7.1(1.7).10$^{-4}$ \\ 
\hline
35(1)        & 3.66(0.35).10$^{-1}$        & 1.55(0.17).10$^{-1}$        & 5.14(0.59).10$^{-2}$        & 3.92(0.53).10$^{-2}$        & 6.46(0.67).10$^{-2}$        & 5.8(1.3).10$^{-4}$ \\ 
\hline
39(1)        & 3.02(0.29).10$^{-1}$        & 1.22(0.13).10$^{-1}$        & 4.05(0.47).10$^{-2}$        & 2.96(0.40).10$^{-2}$        & 4.78(0.50).10$^{-2}$        & 3.61(0.82).10$^{-4}$ \\ 
\hline
\end{tabular}

\begin{tabular}{|c|c|c|c|c|c|c|}
\hline
$\theta$ & $^{6}$Li & $^{7}$Li & $^{7}$Be & $^{9}$Be & $^{10}$Be & $^{8}$B \\ 
(Degrees)  & $d\sigma/d\Omega$ (b.sr$^{-1}$) & $d\sigma/d\Omega$ (b.sr$^{-1}$) & $d\sigma/d\Omega$ (b.sr$^{-1}$) & $d\sigma/d\Omega$ (b.sr$^{-1}$) & $d\sigma/d\Omega$ (b.sr$^{-1}$) & $d\sigma/d\Omega$ (b.sr$^{-1}$) \\ 
\hline
\hline
4(1)        & 9.7(1.3).10$^{-1}$        & 9.2(1.2).10$^{-1}$        & 8.09(0.81).10$^{-1}$        & 3.50(0.37).10$^{-1}$        & 1.84(0.24).10$^{-1}$        & 1.10(0.29).10$^{-1}$ \\ 
\hline
7(1)        & 3.41(0.46).10$^{-1}$        & 3.36(0.45).10$^{-1}$        & 3.12(0.32).10$^{-1}$        & 1.23(0.14).10$^{-1}$        & 6.49(0.88).10$^{-2}$        & 5.5(1.5).10$^{-2}$ \\ 
\hline
9(1)        & 1.96(0.27).10$^{-1}$        & 1.81(0.24).10$^{-1}$        & 1.48(0.15).10$^{-1}$        & 4.44(0.49).10$^{-2}$        & 2.11(0.29).10$^{-2}$        & 1.66(0.45).10$^{-2}$ \\ 
\hline
11(1)        & 8.9(1.2).10$^{-2}$        & 7.4(1.0).10$^{-2}$        & 5.46(0.57).10$^{-2}$        & 1.03(0.12).10$^{-2}$        & 3.87(0.56).10$^{-3}$        & 4.5(1.2).10$^{-3}$ \\ 
\hline
13(1)        & 5.47(0.74).10$^{-2}$        & 3.98(0.54).10$^{-2}$        & 2.84(0.30).10$^{-2}$        & 3.89(0.46).10$^{-3}$        & 1.49(0.23).10$^{-3}$        & 1.87(0.51).10$^{-3}$ \\ 
\hline
15(1)        & 2.98(0.40).10$^{-2}$        & 2.10(0.29).10$^{-2}$        & 1.36(0.14).10$^{-2}$        & 1.54(0.18).10$^{-3}$        & 6.04(0.86).10$^{-4}$        & 7.1(2.3).10$^{-4}$ \\ 
\hline
17(1)        & 1.95(0.26).10$^{-2}$        & 1.32(0.18).10$^{-2}$        & 7.92(0.84).10$^{-3}$        & 9.0(1.2).10$^{-4}$        & 3.60(0.73).10$^{-4}$        & 5.4(1.6).10$^{-4}$ \\ 
\hline
19(1)        & 1.29(0.18).10$^{-2}$        & 8.8(1.2).10$^{-3}$        & 4.92(0.52).10$^{-3}$        & 5.54(0.66).10$^{-4}$        & 1.70(0.28).10$^{-4}$        & 2.84(0.78).10$^{-4}$ \\ 
\hline
21(1)        & 9.4(1.3).10$^{-3}$        & 6.40(0.87).10$^{-3}$        & 3.28(0.36).10$^{-3}$        & 3.42(0.60).10$^{-4}$        & 1.54(0.40).10$^{-4}$        & 1.37(0.85).10$^{-4}$ \\ 
\hline
23(1)        & 7.08(0.96).10$^{-3}$        & 4.75(0.65).10$^{-3}$        & 2.55(0.27).10$^{-3}$        & 2.36(0.32).10$^{-4}$        & 1.20(0.21).10$^{-4}$        & 1.34(0.38).10$^{-4}$ \\ 
\hline
27(1)        & 3.83(0.54).10$^{-3}$        & 2.98(0.42).10$^{-3}$        & 1.43(0.17).10$^{-3}$        & 1.63(0.35).10$^{-4}$        & 5.1(1.8).10$^{-5}$        & 3.9(1.8).10$^{-5}$ \\ 
\hline
31(1)        & 2.44(0.36).10$^{-3}$        & 1.81(0.27).10$^{-3}$        & 7.0(1.0).10$^{-4}$        & 8.0(2.7).10$^{-5}$        & 4.0(1.9).10$^{-5}$        & 8.0(8.3).10$^{-6}$ \\ 
\hline
35(1)        & 1.74(0.24).10$^{-3}$        & 1.40(0.19).10$^{-3}$        & 5.11(0.59).10$^{-4}$        & 5.08(0.98).10$^{-5}$        & 2.29(0.62).10$^{-5}$        & 3.0(1.0).10$^{-5}$ \\ 
\hline
39(1)        & 1.15(0.16).10$^{-3}$        & 8.4(1.2).10$^{-4}$        & 2.65(0.33).10$^{-4}$        & 4.11(0.88).10$^{-5}$        & 1.51(0.50).10$^{-5}$        & 6.8(3.6).10$^{-6}$ \\ 
\hline
\end{tabular}

\begin{tabular}{|c|c|c|c|c|c|}
\hline
$\theta$ & $^{10}$B & $^{11}$B & $^{10}$C & $^{11}$C & $^{12}$C \\ 
(Degrees)  & $d\sigma/d\Omega$ (b.sr$^{-1}$) & $d\sigma/d\Omega$ (b.sr$^{-1}$) & $d\sigma/d\Omega$ (b.sr$^{-1}$) & $d\sigma/d\Omega$ (b.sr$^{-1}$) & $d\sigma/d\Omega$ (b.sr$^{-1}$) \\ 
\hline
\hline
4(1)        & 1.08(0.14)        & 1.45(0.19)        & 1.94(0.42).10$^{-1}$        & 1.31(0.17)        & 1.81(0.25) \\ 
\hline
7(1)        & 2.06(0.28).10$^{-1}$        & 2.18(0.30).10$^{-1}$        & 3.34(0.74).10$^{-2}$        & 1.54(0.21).10$^{-1}$        & 1.29(0.18).10$^{-1}$ \\ 
\hline
9(1)        & 6.25(0.85).10$^{-2}$        & 4.55(0.62).10$^{-2}$        & 8.9(2.0).10$^{-3}$        & 3.28(0.45).10$^{-2}$        & 1.57(0.27).10$^{-2}$ \\ 
\hline
11(1)        & 1.16(0.16).10$^{-2}$        & 6.50(0.92).10$^{-3}$        & 1.52(0.37).10$^{-3}$        & 3.91(0.57).10$^{-3}$        & 7.5(7.1).10$^{-4}$ \\ 
\hline
13(1)        & 3.68(0.53).10$^{-3}$        & 1.56(0.24).10$^{-3}$        & 6.6(1.6).10$^{-4}$        & 1.36(0.21).10$^{-3}$        & - \\ 
\hline
15(1)        & 1.24(0.17).10$^{-3}$        & 5.79(0.99).10$^{-4}$        & 1.96(0.46).10$^{-4}$        & 2.50(0.61).10$^{-4}$        & - \\ 
\hline
17(1)        & 6.3(1.2).10$^{-4}$        & 3.77(0.81).10$^{-4}$        & 2.95.0(10.0).10$^{-5}$        & 1.06(0.88).10$^{-4}$        & - \\ 
\hline
19(1)        & 3.40(0.52).10$^{-4}$        & 1.81(0.33).10$^{-4}$        & 4.2(1.2).10$^{-5}$        & 8.0(1.5).10$^{-5}$        & - \\ 
\hline
21(1)        & 1.95(0.53).10$^{-4}$        & 1.24(0.50).10$^{-4}$        & 4.7(3.6).10$^{-5}$        & 3.8(3.0).10$^{-5}$        & - \\ 
\hline
23(1)        & 1.84(0.30).10$^{-4}$        & 1.13(0.20).10$^{-4}$        & 1.19(0.49).10$^{-5}$        & 3.42(0.85).10$^{-5}$        & - \\ 
\hline
27(1)        & 7.3(2.3).10$^{-5}$        & 3.9(1.6).10$^{-5}$        &  -         & 1.7(1.0).10$^{-5}$        &  -  \\ 
\hline
31(1)        & 3.2(1.7).10$^{-5}$        & 1.6(1.2).10$^{-5}$        &  -         & -        &  -  \\ 
\hline
35(1)        & 1.53(0.49).10$^{-5}$        & 1.65(0.51).10$^{-5}$        & -        &  -         &  -  \\ 
\hline
39(1)        & 1.92(0.58).10$^{-5}$        & 4.1(2.4).10$^{-6}$        & -        &  -         &  -  \\ 
\hline
\end{tabular}
\end{center}
\caption{$^{12}$C fragmentation cross sections for the carbon target at different angles. The values in parentheses represent the uncertainties ( 4.55(0.52) is equivalent to 4.55 $\pm$ 0.52)}
\label{Table_cross_section_C_target}
\end{table}

\section{Cross section for the oxygen target}

\begin{table}[H]
\begin{center}
\begin{tabular}{|c|c|c|c|c|c|c|}
\hline
$\theta$ & $^{1}$H & $^{2}$H & $^{3}$H & $^{3}$He & $^{4}$He & $^{6}$He \\ 
(Degrees)  & $d\sigma/d\Omega$ (b.sr$^{-1}$) & $d\sigma/d\Omega$ (b.sr$^{-1}$) & $d\sigma/d\Omega$ (b.sr$^{-1}$) & $d\sigma/d\Omega$ (b.sr$^{-1}$) & $d\sigma/d\Omega$ (b.sr$^{-1}$) & $d\sigma/d\Omega$ (b.sr$^{-1}$) \\ 
\hline
\hline
4(1)        & 5.33(0.98)        & 2.99(0.64)        & 1.41(0.32)        & 2.11(0.92)        & 1.41(0.37).10$^{1}$        & 4.9(5.8).10$^{-1}$ \\ 
\hline
7(1)        & 3.68(0.71)        & 1.93(0.43)        & 1.02(0.24)        & 1.53(0.52)        & 6.0(1.5)        & 4.9(2.5).10$^{-1}$ \\ 
\hline
9(1)        & 3.27(0.64)        & 1.70(0.38)        & 8.4(2.0).10$^{-1}$        & 1.16(0.38)        & 4.0(1.0)        & 2.9(1.4).10$^{-1}$ \\ 
\hline
11(1)        & 2.64(0.52)        & 1.31(0.30)        & 6.2(1.5).10$^{-1}$        & 8.0(2.4).10$^{-1}$        & 2.24(0.55)        & 1.10(0.50).10$^{-1}$ \\ 
\hline
13(1)        & 2.25(0.45)        & 1.12(0.25)        & 5.0(1.2).10$^{-1}$        & 6.0(1.8).10$^{-1}$        & 1.46(0.37)        & 6.5(3.1).10$^{-2}$ \\ 
\hline
15(1)        & 1.78(0.35)        & 8.6(2.0).10$^{-1}$        & 3.69(0.90).10$^{-1}$        & 4.1(1.2).10$^{-1}$        & 8.7(2.1).10$^{-1}$        & 3.0(1.4).10$^{-2}$ \\ 
\hline
17(1)        & 1.55(0.31)        & 7.6(1.7).10$^{-1}$        & 3.07(0.76).10$^{-1}$        & 3.28(0.94).10$^{-1}$        & 6.3(1.5).10$^{-1}$        & 2.02(0.96).10$^{-2}$ \\ 
\hline
19(1)        & 1.21(0.25)        & 5.8(1.4).10$^{-1}$        & 2.23(0.56).10$^{-1}$        & 2.25(0.64).10$^{-1}$        & 4.02(0.96).10$^{-1}$        & 9.7(4.9).10$^{-3}$ \\ 
\hline
21(1)        & 1.09(0.22)        & 5.2(1.2).10$^{-1}$        & 1.90(0.48).10$^{-1}$        & 1.91(0.55).10$^{-1}$        & 3.16(0.77).10$^{-1}$        & 6.5(3.6).10$^{-3}$ \\ 
\hline
23(1)        & 8.7(1.8).10$^{-1}$        & 4.02(0.97).10$^{-1}$        & 1.43(0.37).10$^{-1}$        & 1.33(0.39).10$^{-1}$        & 2.25(0.56).10$^{-1}$        & 4.0(2.3).10$^{-3}$ \\ 
\hline
27(1)        & 6.8(1.4).10$^{-1}$        & 3.11(0.73).10$^{-1}$        & 1.05(0.27).10$^{-1}$        & 9.3(2.6).10$^{-2}$        & 1.49(0.36).10$^{-1}$        & 2.2(1.1).10$^{-3}$ \\ 
\hline
31(1)        & 5.4(1.1).10$^{-1}$        & 2.34(0.56).10$^{-1}$        & 7.5(2.0).10$^{-2}$        & 6.3(1.9).10$^{-2}$        & 1.09(0.26).10$^{-1}$        & 6.1(5.9).10$^{-4}$ \\ 
\hline
35(1)        & 4.60(0.95).10$^{-1}$        & 1.92(0.45).10$^{-1}$        & 6.0(1.5).10$^{-2}$        & 4.7(1.4).10$^{-2}$        & 8.1(2.0).10$^{-2}$        & 6.9(4.2).10$^{-4}$ \\ 
\hline
39(1)        & 3.76(0.80).10$^{-1}$        & 1.50(0.36).10$^{-1}$        & 4.6(1.2).10$^{-2}$        & 3.6(1.0).10$^{-2}$        & 6.0(1.5).10$^{-2}$        & 5.1(2.8).10$^{-4}$ \\ 
\hline
\end{tabular}

\begin{tabular}{|c|c|c|c|c|c|c|}
\hline
$\theta$ & $^{6}$Li & $^{7}$Li & $^{7}$Be & $^{9}$Be & $^{10}$Be & $^{8}$B \\ 
(Degrees)  & $d\sigma/d\Omega$ (b.sr$^{-1}$) & $d\sigma/d\Omega$ (b.sr$^{-1}$) & $d\sigma/d\Omega$ (b.sr$^{-1}$) & $d\sigma/d\Omega$ (b.sr$^{-1}$) & $d\sigma/d\Omega$ (b.sr$^{-1}$) & $d\sigma/d\Omega$ (b.sr$^{-1}$) \\ 
\hline
\hline
4(1)        & 1.02(0.26)        & 9.8(2.5).10$^{-1}$        & 8.7(1.6).10$^{-1}$        & 3.59(0.73).10$^{-1}$        & 1.97(0.50).10$^{-1}$        & 1.16(0.59).10$^{-1}$ \\ 
\hline
7(1)        & 4.0(1.0).10$^{-1}$        & 3.8(1.0).10$^{-1}$        & 3.51(0.69).10$^{-1}$        & 1.53(0.30).10$^{-1}$        & 7.5(1.8).10$^{-2}$        & 6.4(3.1).10$^{-2}$ \\ 
\hline
9(1)        & 2.24(0.60).10$^{-1}$        & 2.06(0.56).10$^{-1}$        & 1.71(0.35).10$^{-1}$        & 5.3(1.1).10$^{-2}$        & 2.49(0.68).10$^{-2}$        & 2.0(1.1).10$^{-2}$ \\ 
\hline
11(1)        & 1.10(0.30).10$^{-1}$        & 9.3(2.6).10$^{-2}$        & 6.9(1.4).10$^{-2}$        & 1.35(0.32).10$^{-2}$        & 5.5(1.7).10$^{-3}$        & 5.3(3.1).10$^{-3}$ \\ 
\hline
13(1)        & 6.8(1.9).10$^{-2}$        & 5.0(1.5).10$^{-2}$        & 3.36(0.75).10$^{-2}$        & 5.4(1.5).10$^{-3}$        & 1.72(0.69).10$^{-3}$        & 2.6(1.5).10$^{-3}$ \\ 
\hline
15(1)        & 3.7(1.1).10$^{-2}$        & 2.72(0.85).10$^{-2}$        & 1.67(0.38).10$^{-2}$        & 2.73(0.67).10$^{-3}$        & 8.4(3.3).10$^{-4}$        & 9.1(6.8).10$^{-4}$ \\ 
\hline
17(1)        & 2.59(0.78).10$^{-2}$        & 1.72(0.57).10$^{-2}$        & 1.05(0.25).10$^{-2}$        & 1.20(0.43).10$^{-3}$        & 6.0(2.5).10$^{-4}$        & 6.8(4.5).10$^{-4}$ \\ 
\hline
19(1)        & 1.64(0.51).10$^{-2}$        & 1.15(0.39).10$^{-2}$        & 6.5(1.5).10$^{-3}$        & 8.2(2.8).10$^{-4}$        & 4.2(1.7).10$^{-4}$        & 3.9(2.9).10$^{-4}$ \\ 
\hline
21(1)        & 1.27(0.40).10$^{-2}$        & 9.0(3.0).10$^{-3}$        & 4.4(1.1).10$^{-3}$        & 5.1(2.4).10$^{-4}$        & 3.7(1.4).10$^{-4}$        & 2.4(2.1).10$^{-4}$ \\ 
\hline
23(1)        & 9.2(3.0).10$^{-3}$        & 6.2(2.3).10$^{-3}$        & 3.27(0.82).10$^{-3}$        & 4.5(1.7).10$^{-4}$        & 2.0(1.1).10$^{-4}$        & 1.9(1.3).10$^{-4}$ \\ 
\hline
27(1)        & 5.6(1.9).10$^{-3}$        & 3.5(1.4).10$^{-3}$        & 2.16(0.55).10$^{-3}$        & 2.6(1.4).10$^{-4}$        &  -         & 1.33(0.74).10$^{-4}$ \\ 
\hline
31(1)        & 3.6(1.2).10$^{-3}$        & 2.5(1.0).10$^{-3}$        & 1.31(0.38).10$^{-3}$        & 1.6(1.1).10$^{-4}$        &  -         & 7.32(51.0).10$^{-6}$ \\ 
\hline
35(1)        & 2.15(0.78).10$^{-3}$        & 2.29(0.72).10$^{-3}$        & 8.9(2.2).10$^{-4}$        & 1.31(0.53).10$^{-4}$        &  -         & 4.8(2.9).10$^{-5}$ \\ 
\hline
39(1)        & 1.51(0.55).10$^{-3}$        & 1.19(0.49).10$^{-3}$        & 4.1(1.4).10$^{-4}$        & 1.39(0.44).10$^{-4}$        &  -         & 1.2(1.7).10$^{-5}$ \\ 
\hline
\end{tabular}

\begin{tabular}{|c|c|c|c|c|c|}
\hline
$\theta$ & $^{10}$B & $^{11}$B & $^{10}$C & $^{11}$C & $^{12}$C \\ 
(Degrees)  & $d\sigma/d\Omega$ (b.sr$^{-1}$) & $d\sigma/d\Omega$ (b.sr$^{-1}$) & $d\sigma/d\Omega$ (b.sr$^{-1}$) & $d\sigma/d\Omega$ (b.sr$^{-1}$) & $d\sigma/d\Omega$ (b.sr$^{-1}$) \\ 
\hline
\hline
4(1)        & 1.12(0.27)        & 1.52(0.37)        & 1.96(0.80).10$^{-1}$        & 1.36(0.32)        & 1.92(0.47) \\ 
\hline
7(1)        & 2.59(0.64).10$^{-1}$        & 2.39(0.59).10$^{-1}$        & 4.3(1.8).10$^{-2}$        & 1.91(0.46).10$^{-1}$        & 1.52(0.38).10$^{-1}$ \\ 
\hline
9(1)        & 7.2(1.9).10$^{-2}$        & 5.6(1.5).10$^{-2}$        & 9.9(4.4).10$^{-3}$        & 3.8(1.0).10$^{-2}$        & 2.12(0.72).10$^{-2}$ \\ 
\hline
11(1)        & 1.66(0.47).10$^{-2}$        & 8.6(2.8).10$^{-3}$        & 2.04(0.99).10$^{-3}$        & 5.8(1.8).10$^{-3}$        & 2.5(1.3).10$^{-3}$ \\ 
\hline
13(1)        & 5.5(1.8).10$^{-3}$        & 2.86(0.99).10$^{-3}$        & 7.8(4.3).10$^{-4}$        &  -         &  -  \\ 
\hline
15(1)        & 1.70(0.66).10$^{-3}$        &  -         &  -         &  -         &  -  \\ 
\hline
\end{tabular}
\end{center}
\caption{$^{12}$C fragmentation cross sections for the oxygen target at different angles. The values in parentheses represent the uncertainties ( 4.55(0.52) is equivalent to 4.55 $\pm$ 0.52)}
\label{Table_cross_section_O_target}
\end{table}

\section{Cross section for the aluminum target}
\begin{table}[H]
\begin{center}
\begin{tabular}{|c|c|c|c|c|c|c|}
\hline
$\theta$ & $^{1}$H & $^{2}$H & $^{3}$H & $^{3}$He & $^{4}$He & $^{6}$He \\ 
(Degrees)  & $d\sigma/d\Omega$ (b.sr$^{-1}$) & $d\sigma/d\Omega$ (b.sr$^{-1}$) & $d\sigma/d\Omega$ (b.sr$^{-1}$) & $d\sigma/d\Omega$ (b.sr$^{-1}$) & $d\sigma/d\Omega$ (b.sr$^{-1}$) & $d\sigma/d\Omega$ (b.sr$^{-1}$) \\ 
\hline
\hline
4(1)        & 6.68(0.61)        & 3.80(0.40)        & 1.83(0.20)        & 2.54(0.56)        & 1.63(0.22).10$^{1}$        & 5.6(3.5).10$^{-1}$ \\ 
\hline
7(1)        & 4.54(0.44)        & 2.49(0.27)        & 1.35(0.15)        & 1.88(0.31)        & 6.93(0.90)        & 5.4(1.5).10$^{-1}$ \\ 
\hline
9(1)        & 4.15(0.40)        & 2.20(0.24)        & 1.11(0.13)        & 1.43(0.23)        & 4.76(0.62)        & 3.43(0.86).10$^{-1}$ \\ 
\hline
11(1)        & 3.38(0.33)        & 1.71(0.19)        & 8.20(0.94).10$^{-1}$        & 9.9(1.5).10$^{-1}$        & 2.75(0.34)        & 1.38(0.31).10$^{-1}$ \\ 
\hline
13(1)        & 2.94(0.28)        & 1.49(0.16)        & 6.80(0.78).10$^{-1}$        & 7.8(1.1).10$^{-1}$        & 1.89(0.23)        & 8.9(2.0).10$^{-2}$ \\ 
\hline
15(1)        & 2.33(0.22)        & 1.16(0.13)        & 5.10(0.59).10$^{-1}$        & 5.31(0.74).10$^{-1}$        & 1.16(0.13)        & 4.12(0.91).10$^{-2}$ \\ 
\hline
17(1)        & 2.05(0.20)        & 1.03(0.11)        & 4.30(0.49).10$^{-1}$        & 4.36(0.60).10$^{-1}$        & 8.65(0.98).10$^{-1}$        & 2.84(0.63).10$^{-2}$ \\ 
\hline
19(1)        & 1.64(0.16)        & 7.98(0.87).10$^{-1}$        & 3.22(0.37).10$^{-1}$        & 3.02(0.41).10$^{-1}$        & 5.84(0.63).10$^{-1}$        & 1.52(0.34).10$^{-2}$ \\ 
\hline
21(1)        & 1.49(0.14)        & 7.26(0.79).10$^{-1}$        & 2.79(0.32).10$^{-1}$        & 2.59(0.35).10$^{-1}$        & 4.73(0.51).10$^{-1}$        & 1.16(0.26).10$^{-2}$ \\ 
\hline
23(1)        & 1.25(0.12)        & 5.90(0.64).10$^{-1}$        & 2.21(0.25).10$^{-1}$        & 1.90(0.26).10$^{-1}$        & 3.62(0.38).10$^{-1}$        & 7.7(1.7).10$^{-3}$ \\ 
\hline
27(1)        & 9.48(0.91).10$^{-1}$        & 4.36(0.48).10$^{-1}$        & 1.55(0.18).10$^{-1}$        & 1.24(0.17).10$^{-1}$        & 2.41(0.25).10$^{-1}$        & 3.31(0.74).10$^{-3}$ \\ 
\hline
31(1)        & 7.70(0.74).10$^{-1}$        & 3.36(0.37).10$^{-1}$        & 1.16(0.13).10$^{-1}$        & 8.9(1.2).10$^{-2}$        & 1.71(0.18).10$^{-1}$        & 2.19(0.50).10$^{-3}$ \\ 
\hline
35(1)        & 6.50(0.63).10$^{-1}$        & 2.70(0.29).10$^{-1}$        & 9.0(1.0).10$^{-2}$        & 6.58(0.89).10$^{-2}$        & 1.32(0.14).10$^{-1}$        & 1.40(0.31).10$^{-3}$ \\ 
\hline
39(1)        & 5.48(0.53).10$^{-1}$        & 2.17(0.24).10$^{-1}$        & 7.03(0.81).10$^{-2}$        & 4.93(0.67).10$^{-2}$        & 9.8(1.0).10$^{-2}$        & 8.5(1.9).10$^{-4}$ \\ 
\hline
\end{tabular}

\begin{tabular}{|c|c|c|c|c|c|c|}
\hline
$\theta$ & $^{6}$Li & $^{7}$Li & $^{7}$Be & $^{9}$Be & $^{10}$Be & $^{8}$B \\ 
(Degrees)  & $d\sigma/d\Omega$ (b.sr$^{-1}$) & $d\sigma/d\Omega$ (b.sr$^{-1}$) & $d\sigma/d\Omega$ (b.sr$^{-1}$) & $d\sigma/d\Omega$ (b.sr$^{-1}$) & $d\sigma/d\Omega$ (b.sr$^{-1}$) & $d\sigma/d\Omega$ (b.sr$^{-1}$) \\ 
\hline
\hline
4(1)        & 1.17(0.15)        & 1.13(0.15)        & 9.68(0.97).10$^{-1}$        & 4.10(0.43).10$^{-1}$        & 2.26(0.30).10$^{-1}$        & 1.31(0.35).10$^{-1}$ \\ 
\hline
7(1)        & 4.56(0.62).10$^{-1}$        & 4.41(0.60).10$^{-1}$        & 3.83(0.40).10$^{-1}$        & 1.56(0.17).10$^{-1}$        & 7.6(1.0).10$^{-2}$        & 6.6(1.8).10$^{-2}$ \\ 
\hline
9(1)        & 2.68(0.36).10$^{-1}$        & 2.55(0.35).10$^{-1}$        & 2(0.21).10$^{-1}$        & 6.37(0.70).10$^{-2}$        & 3.08(0.42).10$^{-2}$        & 2.41(0.64).10$^{-2}$ \\ 
\hline
11(1)        & 1.40(0.19).10$^{-1}$        & 1.21(0.16).10$^{-1}$        & 8.16(0.85).10$^{-2}$        & 1.89(0.21).10$^{-2}$        & 8.2(1.1).10$^{-3}$        & 7.8(2.1).10$^{-3}$ \\ 
\hline
13(1)        & 8.8(1.2).10$^{-2}$        & 7.18(0.98).10$^{-2}$        & 4.60(0.48).10$^{-2}$        & 9.1(1.0).10$^{-3}$        & 3.50(0.50).10$^{-3}$        & 3.58(0.97).10$^{-3}$ \\ 
\hline
15(1)        & 5.35(0.73).10$^{-2}$        & 4.29(0.58).10$^{-2}$        & 2.38(0.25).10$^{-2}$        & 3.74(0.44).10$^{-3}$        & 1.68(0.24).10$^{-3}$        & 1.57(0.48).10$^{-3}$ \\ 
\hline
17(1)        & 3.80(0.52).10$^{-2}$        & 2.95(0.40).10$^{-2}$        & 1.57(0.16).10$^{-2}$        & 2.99(0.34).10$^{-3}$        & 1.28(0.19).10$^{-3}$        & 1.05(0.31).10$^{-3}$ \\ 
\hline
19(1)        & 2.57(0.35).10$^{-2}$        & 2.07(0.28).10$^{-2}$        & 9.9(1.0).10$^{-3}$        & 1.82(0.22).10$^{-3}$        & 7.8(1.2).10$^{-4}$        & 6.7(2.0).10$^{-4}$ \\ 
\hline
21(1)        & 1.99(0.27).10$^{-2}$        & 1.57(0.21).10$^{-2}$        & 7.17(0.77).10$^{-3}$        & 1.67(0.20).10$^{-3}$        & 6.5(1.0).10$^{-4}$        & 5.3(1.6).10$^{-4}$ \\ 
\hline
23(1)        & 1.53(0.21).10$^{-2}$        & 1.26(0.17).10$^{-2}$        & 5.36(0.56).10$^{-3}$        & 1.07(0.12).10$^{-3}$        & 5.81(0.84).10$^{-4}$        & 3.09(0.85).10$^{-4}$ \\ 
\hline
27(1)        & 9.4(1.3).10$^{-3}$        & 7.9(1.1).10$^{-3}$        & 3.18(0.36).10$^{-3}$        & 7.7(1.1).10$^{-4}$        & 3.74(0.70).10$^{-4}$        & 1.06(0.38).10$^{-4}$ \\ 
\hline
31(1)        & 5.82(0.82).10$^{-3}$        & 5.35(0.75).10$^{-3}$        & 2.05(0.25).10$^{-3}$        & 4.41(0.76).10$^{-4}$        & 1.89(0.46).10$^{-4}$        & 1.18(0.44).10$^{-4}$ \\ 
\hline
35(1)        & 4.18(0.58).10$^{-3}$        & 3.52(0.49).10$^{-3}$        & 1.35(0.15).10$^{-3}$        & 2.75(0.41).10$^{-4}$        & 1.34(0.26).10$^{-4}$        & 5.1(1.8).10$^{-5}$ \\ 
\hline
39(1)        & 2.86(0.40).10$^{-3}$        & 2.71(0.38).10$^{-3}$        & 8.4(1.0).10$^{-4}$        & 1.58(0.28).10$^{-4}$        & 5.5(1.5).10$^{-5}$        & 3.2(1.3).10$^{-5}$ \\ 
\hline
\end{tabular}

\begin{tabular}{|c|c|c|c|c|c|}
\hline
$\theta$ & $^{10}$B & $^{11}$B & $^{10}$C & $^{11}$C & $^{12}$C \\ 
(Degrees)  & $d\sigma/d\Omega$ (b.sr$^{-1}$) & $d\sigma/d\Omega$ (b.sr$^{-1}$) & $d\sigma/d\Omega$ (b.sr$^{-1}$) & $d\sigma/d\Omega$ (b.sr$^{-1}$) & $d\sigma/d\Omega$ (b.sr$^{-1}$) \\ 
\hline
\hline
4(1)        & 1.18(0.16)        & 1.62(0.21)        & 2.11(0.46).10$^{-1}$        & 1.35(0.18)        & 1.96(0.26) \\ 
\hline
7(1)        & 2.70(0.37).10$^{-1}$        & 2.42(0.33).10$^{-1}$        & 4.7(1.0).10$^{-2}$        & 1.88(0.25).10$^{-1}$        & 1.56(0.21).10$^{-1}$ \\ 
\hline
9(1)        & 8.3(1.1).10$^{-2}$        & 6.99(0.95).10$^{-2}$        & 1.22(0.27).10$^{-2}$        & 4.60(0.63).10$^{-2}$        & 2.87(0.47).10$^{-2}$ \\ 
\hline
11(1)        & 2.17(0.30).10$^{-2}$        & 1.40(0.19).10$^{-2}$        & 2.85(0.64).10$^{-3}$        & 8.2(1.1).10$^{-3}$        & 3.08(0.83).10$^{-3}$ \\ 
\hline
13(1)        & 8.4(1.2).10$^{-3}$        & 4.63(0.67).10$^{-3}$        & 9.2(2.7).10$^{-4}$        & 2.27(0.40).10$^{-3}$        &  -  \\ 
\hline
15(1)        & 3.42(0.49).10$^{-3}$        & 2.34(0.34).10$^{-3}$        & 2.9(1.3).10$^{-4}$        & 7.8(1.5).10$^{-4}$        &  -  \\ 
\hline
17(1)        & 2.16(0.32).10$^{-3}$        & 1.55(0.24).10$^{-3}$        & 3.7(1.0).10$^{-4}$        & 4.9(1.1).10$^{-4}$        &  -  \\ 
\hline
19(1)        & 1.37(0.21).10$^{-3}$        & 1.15(0.18).10$^{-3}$        & 2.05(0.55).10$^{-4}$        & 3.16(0.66).10$^{-4}$        &  -  \\ 
\hline
21(1)        & 1.18(0.18).10$^{-3}$        & 8.9(1.4).10$^{-4}$        & 1.29(0.43).10$^{-4}$        & 2.69(0.52).10$^{-4}$        &  -  \\ 
\hline
23(1)        & 9.4(1.3).10$^{-4}$        & 7.4(1.1).10$^{-4}$        & 9.3(2.4).10$^{-5}$        & 1.94(0.31).10$^{-4}$        &  -  \\ 
\hline
27(1)        & 5.24(0.91).10$^{-4}$        & 4.06(0.75).10$^{-4}$        & 6.9(2.6).10$^{-5}$        & 1.06(0.29).10$^{-4}$        &  -  \\ 
\hline
31(1)        & 3.15(0.66).10$^{-4}$        & 1.65(0.42).10$^{-4}$        & 1.6(1.2).10$^{-5}$        & 9.4(3.0).10$^{-5}$        &  -  \\ 
\hline
35(1)        & 2.06(0.36).10$^{-4}$        & 1.36(0.27).10$^{-4}$        & 1.07(0.58).10$^{-5}$        & 2.67(0.92).10$^{-5}$        &  -  \\ 
\hline
39(1)        & 1.26(0.26).10$^{-4}$        & 6.7(1.7).10$^{-5}$        & 5.8(4.3).10$^{-6}$        & 1.46(0.68).10$^{-5}$        &  -  \\ 
\hline
\end{tabular}
\end{center}
\caption{$^{12}$C fragmentation cross sections for the aluminum target at different angles. The values in parentheses represent the uncertainties ( 4.55(0.52) is equivalent to 4.55 $\pm$ 0.52)}
\label{Table_cross_section_Al_target}
\end{table}

\section{Cross section for the titanium target}

\begin{table}[H]
\begin{center}
\begin{tabular}{|c|c|c|c|c|c|c|}
\hline
$\theta$ & $^{1}$H & $^{2}$H & $^{3}$H & $^{3}$He & $^{4}$He & $^{6}$He \\ 
(Degrees)  & $d\sigma/d\Omega$ (b.sr$^{-1}$) & $d\sigma/d\Omega$ (b.sr$^{-1}$) & $d\sigma/d\Omega$ (b.sr$^{-1}$) & $d\sigma/d\Omega$ (b.sr$^{-1}$) & $d\sigma/d\Omega$ (b.sr$^{-1}$) & $d\sigma/d\Omega$ (b.sr$^{-1}$) \\ 
\hline
\hline
4(1)        & 8.24(0.75)        & 4.80(0.50)        & 2.33(0.26)        & 3.12(0.67)        & 1.90(0.26).10$^{1}$        & 6.5(3.9).10$^{-1}$ \\ 
\hline
7(1)        & 5.79(0.56)        & 3.25(0.35)        & 1.79(0.21)        & 2.24(0.38)        & 8.3(1.1)        & 6.4(1.8).10$^{-1}$ \\ 
\hline
9(1)        & 5.37(0.52)        & 2.92(0.32)        & 1.48(0.17)        & 1.77(0.29)        & 5.75(0.75)        & 4.1(1.0).10$^{-1}$ \\ 
\hline
11(1)        & 4.35(0.42)        & 2.31(0.25)        & 1.14(0.13)        & 1.24(0.18)        & 3.39(0.41)        & 1.66(0.38).10$^{-1}$ \\ 
\hline
13(1)        & 3.88(0.37)        & 2.03(0.22)        & 9.5(1.1).10$^{-1}$        & 9.7(1.4).10$^{-1}$        & 2.37(0.29)        & 1.15(0.26).10$^{-1}$ \\ 
\hline
15(1)        & 3.13(0.30)        & 1.63(0.18)        & 7.54(0.87).10$^{-1}$        & 6.97(0.97).10$^{-1}$        & 1.56(0.18)        & 5.6(1.2).10$^{-2}$ \\ 
\hline
17(1)        & 2.80(0.27)        & 1.44(0.16)        & 6.36(0.73).10$^{-1}$        & 5.69(0.79).10$^{-1}$        & 1.18(0.13)        & 4.12(0.91).10$^{-2}$ \\ 
\hline
19(1)        & 2.28(0.22)        & 1.17(0.13)        & 5.08(0.58).10$^{-1}$        & 4.16(0.57).10$^{-1}$        & 8.54(0.93).10$^{-1}$        & 2.33(0.52).10$^{-2}$ \\ 
\hline
21(1)        & 2.08(0.20)        & 1.04(0.11)        & 4.25(0.49).10$^{-1}$        & 3.51(0.48).10$^{-1}$        & 6.79(0.74).10$^{-1}$        & 1.75(0.40).10$^{-2}$ \\ 
\hline
23(1)        & 1.73(0.17)        & 8.51(0.93).10$^{-1}$        & 3.46(0.40).10$^{-1}$        & 2.60(0.35).10$^{-1}$        & 5.34(0.57).10$^{-1}$        & 1.17(0.26).10$^{-2}$ \\ 
\hline
27(1)        & 1.36(0.13)        & 6.39(0.70).10$^{-1}$        & 2.47(0.28).10$^{-1}$        & 1.72(0.23).10$^{-1}$        & 3.69(0.38).10$^{-1}$        & 6.7(1.5).10$^{-3}$ \\ 
\hline
31(1)        & 1.12(0.11)        & 4.99(0.55).10$^{-1}$        & 1.83(0.21).10$^{-1}$        & 1.24(0.17).10$^{-1}$        & 2.71(0.28).10$^{-1}$        & 3.77(0.86).10$^{-3}$ \\ 
\hline
35(1)        & 9.66(0.93).10$^{-1}$        & 4.09(0.45).10$^{-1}$        & 1.46(0.17).10$^{-1}$        & 9.4(1.3).10$^{-2}$        & 2.10(0.22).10$^{-1}$        & 2.78(0.62).10$^{-3}$ \\ 
\hline
39(1)        & 8.30(0.80).10$^{-1}$        & 3.31(0.36).10$^{-1}$        & 1.14(0.13).10$^{-1}$        & 7.02(0.95).10$^{-2}$        & 1.60(0.17).10$^{-1}$        & 1.83(0.41).10$^{-3}$ \\ 
\hline
\end{tabular}

\begin{tabular}{|c|c|c|c|c|c|c|}
\hline
$\theta$ & $^{6}$Li & $^{7}$Li & $^{7}$Be & $^{9}$Be & $^{10}$Be & $^{8}$B \\ 
(Degrees)  & $d\sigma/d\Omega$ (b.sr$^{-1}$) & $d\sigma/d\Omega$ (b.sr$^{-1}$) & $d\sigma/d\Omega$ (b.sr$^{-1}$) & $d\sigma/d\Omega$ (b.sr$^{-1}$) & $d\sigma/d\Omega$ (b.sr$^{-1}$) & $d\sigma/d\Omega$ (b.sr$^{-1}$) \\ 
\hline
\hline
4(1)        & 1.37(0.18)        & 1.29(0.17)        & 1.07(0.11)        & 4.98(0.53).10$^{-1}$        & 2.56(0.34).10$^{-1}$        & 1.42(0.38).10$^{-1}$ \\ 
\hline
7(1)        & 4.98(0.68).10$^{-1}$        & 5.13(0.69).10$^{-1}$        & 4.24(0.44).10$^{-1}$        & 1.73(0.19).10$^{-1}$        & 9.8(1.3).10$^{-2}$        & 7.1(1.9).10$^{-2}$ \\ 
\hline
9(1)        & 3.21(0.43).10$^{-1}$        & 3.13(0.42).10$^{-1}$        & 2.26(0.24).10$^{-1}$        & 7.95(0.88).10$^{-2}$        & 3.46(0.47).10$^{-2}$        & 2.69(0.72).10$^{-2}$ \\ 
\hline
11(1)        & 1.68(0.23).10$^{-1}$        & 1.57(0.21).10$^{-1}$        & 9.5(1.0).10$^{-2}$        & 2.45(0.27).10$^{-2}$        & 1.20(0.17).10$^{-2}$        & 8.6(2.3).10$^{-3}$ \\ 
\hline
13(1)        & 1.15(0.16).10$^{-1}$        & 9.6(1.3).10$^{-2}$        & 5.43(0.57).10$^{-2}$        & 1.31(0.15).10$^{-2}$        & 5.91(0.85).10$^{-3}$        & 4.1(1.1).10$^{-3}$ \\ 
\hline
15(1)        & 6.94(0.94).10$^{-2}$        & 6.18(0.84).10$^{-2}$        & 3.01(0.32).10$^{-2}$        & 6.41(0.72).10$^{-3}$        & 3.20(0.45).10$^{-3}$        & 2.27(0.61).10$^{-3}$ \\ 
\hline
17(1)        & 5.25(0.71).10$^{-2}$        & 4.33(0.59).10$^{-2}$        & 2.02(0.21).10$^{-2}$        & 4.45(0.50).10$^{-3}$        & 2.32(0.33).10$^{-3}$        & 1.68(0.45).10$^{-3}$ \\ 
\hline
19(1)        & 3.71(0.50).10$^{-2}$        & 3.29(0.45).10$^{-2}$        & 1.29(0.14).10$^{-2}$        & 3.08(0.36).10$^{-3}$        & 1.60(0.24).10$^{-3}$        & 1.01(0.30).10$^{-3}$ \\ 
\hline
21(1)        & 2.88(0.39).10$^{-2}$        & 2.42(0.33).10$^{-2}$        & 9.5(1.1).10$^{-3}$        & 2.23(0.34).10$^{-3}$        & 1.26(0.24).10$^{-3}$        & 4.7(3.3).10$^{-4}$ \\ 
\hline
23(1)        & 2.24(0.30).10$^{-2}$        & 1.98(0.27).10$^{-2}$        & 7.14(0.76).10$^{-3}$        & 1.89(0.22).10$^{-3}$        & 1.16(0.17).10$^{-3}$        & 3.9(1.1).10$^{-4}$ \\ 
\hline
27(1)        & 1.32(0.18).10$^{-2}$        & 1.21(0.17).10$^{-2}$        & 4.33(0.51).10$^{-3}$        & 1.15(0.18).10$^{-3}$        & 7.6(1.4).10$^{-4}$        & 1.79(0.68).10$^{-4}$ \\ 
\hline
31(1)        & 8.9(1.3).10$^{-3}$        & 7.9(1.1).10$^{-3}$        & 2.93(0.37).10$^{-3}$        & 8.3(1.4).10$^{-4}$        & 6.2(1.2).10$^{-4}$        & 1.54(0.62).10$^{-4}$ \\ 
\hline
35(1)        & 6.77(0.93).10$^{-3}$        & 6.59(0.90).10$^{-3}$        & 1.98(0.22).10$^{-3}$        & 6.05(0.80).10$^{-4}$        & 3.13(0.54).10$^{-4}$        & 7.6(2.6).10$^{-5}$ \\ 
\hline
39(1)        & 4.71(0.65).10$^{-3}$        & 4.62(0.64).10$^{-3}$        & 1.25(0.15).10$^{-3}$        & 3.98(0.56).10$^{-4}$        & 2.64(0.46).10$^{-4}$        & 7.8(2.6).10$^{-5}$ \\ 
\hline
\end{tabular}

\begin{tabular}{|c|c|c|c|c|c|}
\hline
$\theta$ & $^{10}$B & $^{11}$B & $^{10}$C & $^{11}$C & $^{12}$C \\ 
(Degrees)  & $d\sigma/d\Omega$ (b.sr$^{-1}$) & $d\sigma/d\Omega$ (b.sr$^{-1}$) & $d\sigma/d\Omega$ (b.sr$^{-1}$) & $d\sigma/d\Omega$ (b.sr$^{-1}$) & $d\sigma/d\Omega$ (b.sr$^{-1}$) \\ 
\hline
\hline
4(1)        & 1.35(0.18)        & 1.66(0.22)        & 2.45(0.54).10$^{-1}$        & 1.40(0.19)        & 1.86(0.28) \\ 
\hline
7(1)        & 2.59(0.35).10$^{-1}$        & 3.04(0.41).10$^{-1}$        & 4.13(0.91).10$^{-2}$        & 1.77(0.24).10$^{-1}$        & 1.96(0.27).10$^{-1}$ \\ 
\hline
9(1)        & 1.03(0.14).10$^{-1}$        & 7.6(1.0).10$^{-2}$        & 1.46(0.32).10$^{-2}$        & 5.95(0.81).10$^{-2}$        & 4.58(0.62).10$^{-2}$ \\ 
\hline
11(1)        & 2.45(0.34).10$^{-2}$        & 2(0.28).10$^{-2}$        & 3.06(0.70).10$^{-3}$        & 1.03(0.14).10$^{-2}$        & 1.92(0.26).10$^{-2}$ \\ 
\hline
13(1)        & 1.26(0.18).10$^{-2}$        & 6.66(0.95).10$^{-3}$        & 1.74(0.41).10$^{-3}$        & 4.66(0.68).10$^{-3}$        & 1.44(0.20).10$^{-2}$ \\ 
\hline
15(1)        & 4.71(0.65).10$^{-3}$        & 3.80(0.53).10$^{-3}$        & 6.8(1.6).10$^{-4}$        & 1.39(0.20).10$^{-3}$        & 10.0(1.4).10$^{-3}$ \\ 
\hline
17(1)        & 3.58(0.50).10$^{-3}$        & 2.64(0.37).10$^{-3}$        & 6.7(1.5).10$^{-4}$        & 1.28(0.19).10$^{-3}$        &  -  \\ 
\hline
19(1)        & 2.30(0.36).10$^{-3}$        & 2.19(0.33).10$^{-3}$        & 2.8(1.1).10$^{-4}$        & 4.6(1.2).10$^{-4}$        &  -  \\ 
\hline
21(1)        & 1.66(0.31).10$^{-3}$        & 1.18(0.26).10$^{-3}$        & 1.7(1.3).10$^{-4}$        & 5.43(16.0).10$^{-5}$        &  -  \\ 
\hline
23(1)        & 1.43(0.21).10$^{-3}$        & 1.33(0.19).10$^{-3}$        & 1.51(0.41).10$^{-4}$        & 2.77(0.49).10$^{-4}$        &  -  \\ 
\hline
27(1)        & 9.1(1.6).10$^{-4}$        & 4.9(1.0).10$^{-4}$        & 5.1(2.8).10$^{-5}$        & 2.18(0.61).10$^{-4}$        &  -  \\ 
\hline
31(1)        & 4.8(1.0).10$^{-4}$        & 5.5(1.1).10$^{-4}$        & 7.0(3.5).10$^{-5}$        & 5.6(2.9).10$^{-5}$        &  -  \\ 
\hline
35(1)        & 3.23(0.55).10$^{-4}$        & 3.13(0.54).10$^{-4}$        & 2.8(1.1).10$^{-5}$        & 6.5(1.7).10$^{-5}$        &  -  \\ 
\hline
39(1)        & 2.87(0.49).10$^{-4}$        & 1.79(0.34).10$^{-4}$        & 2.9(1.2).10$^{-5}$        & 1.96(0.84).10$^{-5}$        &  -  \\ 
\hline
\end{tabular}
\end{center}
\caption{$^{12}$C fragmentation cross sections for the titanium target at different angles. The values in parentheses represent the uncertainties ( 4.55(0.52) is equivalent to 4.55 $\pm$ 0.52)}
\label{Table_cross_section_Ti_target}
\end{table}

\newpage

\bibliographystyle{unsrt}
\bibliography{myarticleNotes}

\end{document}